\documentclass{article}
\usepackage{qic,epsfig}

\def\0{|0\rangle}
\def\1{|1\rangle}
\def\+{|+\rangle}
\def\cnot{{\sc cnot }}
\def\t{|A_{\pi /4}\rangle }

\begin{document}
%%
%% I commented out this redefinition of length
%\setlength{\textheight}{8.0truein}    %FOR 2ND PAGE ONWARDS
%%

% I commented out the running heads
%\runninghead{Quantum accuracy threshold for concatenated distance-3 code  %$\ldots$
%}
%            {P. Aliferis, D. Gottesman, and J. Preskill %$\ldots$
%}

\normalsize\textlineskip
\thispagestyle{empty}
\setcounter{page}{1}

%% I commented out the copyrightheadings
%\copyrightheading{Vol.}{No.}{Year}{Page Nos.}
%\copyrightheading{0}{0}{2004}{000--000}

\vspace*{0.88truein}

\alphfootnote

\fpage{1}

\centerline{\bf
%%%%%%%%%%%%%%%%%%%%%
%Put in titles here
%%%%%%%%%%%%%%%%%%%%%
QUANTUM ACCURACY THRESHOLD FOR CONCATENATED DISTANCE-3 CODES}
\vspace*{0.37truein}
\centerline{\footnotesize
%%%%%%%%%%%%%%%%%%%%%%%%%%%%%%%%%%%%
%put authors' name and address here
%%%%%%%%%%%%%%%%%%%%%%%%%%%%%%%%%%%%
PANOS ALIFERIS}
\vspace*{0.015truein}
\centerline{\footnotesize\it Institute for Quantum Information, California Institute of Technology }
\baselineskip=10pt
\centerline{\footnotesize\it Pasadena, CA 91125, USA}
\vspace*{10pt}
\centerline{\footnotesize 
DANIEL GOTTESMAN}
\vspace*{0.015truein}
\centerline{\footnotesize\it Perimeter Institute }
\baselineskip=10pt
\centerline{\footnotesize\it Waterloo ON N2V 1Z3, Canada}
\vspace*{10pt}
\centerline{\footnotesize 
JOHN PRESKILL}
\vspace*{0.015truein}
\centerline{\footnotesize\it Institute for Quantum Information, California Institute of Technology }
\baselineskip=10pt
\centerline{\footnotesize\it Pasadena, CA 91125, USA}
\vspace*{0.225truein}
%% I commented out the \publisher line, and added a line dating this draft
%\publisher{(received date)}{(revised date)}
%\centerline{DRAFT --- 26 April 2005}
%
\vspace*{0.21truein}
%
%% \abstracts{first paragraph}{second paragraph}{third paragraph}
%% If there is only one paragraph, just keep the second and third empty 
%% like the following one 
\abstracts{
%%%%%%%%%%%%%%%%%%%%
% put abstract here
%%%%%%%%%%%%%%%%%%%%
We prove a new version of the quantum threshold theorem that applies to concatenation of a quantum code that corrects only one error, and we use this theorem to derive a rigorous lower bound on the quantum accuracy threshold $\varepsilon_0$. Our proof also applies to concatenation of higher-distance codes, and to noise models that allow faults to be correlated in space and in time. The proof uses new criteria for assessing the accuracy of fault-tolerant circuits, which are particularly conducive to the inductive analysis of recursive simulations. Our lower bound on the threshold, $\varepsilon_0 \ge 2.73\times 10^{-5}$ for an adversarial independent stochastic noise model, is derived from a computer-assisted combinatorial analysis; it is the best lower bound that has been rigorously proven so far.   
}{}{}
\vspace*{10pt}
%
%% I commented out the key words, the following space, and the communication line
%\keywords{Quantum error correction, fault tolerance, accuracy threshold}
%\vspace*{3pt}
%\communicate{to be filled by the Editorial}
%%
%%
\vspace*{1pt}\textlineskip    %) USE THIS MEASUREMENT WHEN THERE IS
   %) A SECTION HEADING
%\vspace*{-0.5pt}
%\noindent
%%%%%%%%%%%%%%%%%%%%%%%%%%%%%%%%
%put the text of the paper here
%%%%%%%%%%%%%%%%%%%%%%%%%%%%%%%%%\documentclass[aps,pra,showpacs,preprint]{revtex4}

\section{Introduction}
Our hopes that large-scale quantum computers will be built and operated someday are founded on the theory of quantum fault tolerance \cite{shor_ft}. A centerpiece of this theory is the {\em quantum threshold theorem}, which asserts that an arbitrarily long quantum computation can be executed with high reliability, provided that the noise afflicting the computer's hardware is weaker than a certain critical value, the {\em accuracy threshold} \cite{ben-or,kitaev_threshold,knill, jp_threshold,gottesman_threshold}. In this paper, we will present new proofs of this fundamental theorem, and new rigorous lower bounds on the accuracy threshold.

There have been several very interesting recent developments concerning the accuracy threshold. The original threshold theorem applied to the standard quantum circuit model with nonlocal quantum gates, and for Markovian noise. New threshold theorems have been proved for non-Markovian noise \cite{terhal}, and for the cluster-state model of computation \cite{nielsen,aliferis-leung}, and numerical estimates of the threshold have been carried out for computation using local gates \cite{divincenzo,roychowdhuri}. Furthermore, for computation with nonlocal gates, significantly improved estimates of the threshold have been found \cite{Steane02,knill_detect,reichardt}. These recent developments build on the foundations provided by the threshold theorem first proved by Aharonov and Ben-Or \cite{ben-or}, by Kitaev \cite{kitaev_threshold}, and by Knill, Laflamme, and Zurek \cite{knill}. Our purpose in this paper is to reexamine and strengthen these foundations. 

Our goal is to assess the reliability of a quantum computer whose gates perform imperfectly as described by some specified noise model. Though more general noise models can be analyzed (and will be, in Sec.~\ref{sec:non-markovian} of this paper), it is illuminating to consider the special case of independent stochastic faults. In this noise model, faults are independently and identically distributed at the locations of the noisy quantum circuit; that is, at each location either the gate is executed perfectly (with probability $1-\varepsilon$), or a fault occurs (with probability $\varepsilon$) --- we say that $\varepsilon$ is the {\em fault rate}. (Faults can occur even if the ideal gate is the identity; the ``resting'' qubits are subject to storage errors.) Though the fault locations are chosen probabilistically, once the locations are chosen the action of the faulty gates can be chosen adversarially; that is, we allow the faults to be arbitrary trace-preserving quantum operations. 

The noisy gates are used to build a fault-tolerant simulation of an ideal quantum circuit. In this simulation, the logical qubits processed by the computer are protected from damage using a quantum error-correcting code \cite{shor_9,steane_7}, and the gates acting on the logical qubits are realized by ``gadgets'' that act on the code blocks. The gadgets exploit the redundancy of the quantum error-correcting code to diagnose and remove errors caused by faults; they are carefully designed to minimize propagation of errors among qubits within the same code block. 

The fault-tolerant simulations that we will analyze here are based on {\em concatenated quantum codes} \cite{knill-concatenated}. The code block of a concatenated code is constructed as a hierarchy of codes within codes --- the code block at level $k$ of this hierarchy is built from logical qubits encoded at level $k-1$ of the hierarchy. Likewise, our fault-tolerant gadgets are constructed as a hierarchy of gadgets within gadgets --- the gadgets at level $k$ are built from gate gadgets at level $k-1$. The basic idea of the threshold theorem is very simple: if $\varepsilon$ is below the accuracy threshold $\varepsilon_0$, then the level-1 simulation of the ideal circuit will be more reliable than an unprotected ``level-0'' circuit. Because of the self-similarity of the gadgets, it then follows that the level-2 simulation will be still more reliable, and so on. Thus for $\varepsilon < \varepsilon_0$, our recursive fault-tolerant simulation becomes arbitrarily reliable as we increase the level of concatenation. 

We remark that for the threshold theorem to apply, two features of the simulation are essential: First, quantum gates can be executed in parallel --- otherwise we would be unable to control errors that occur simultaneously in different parts of the computer. Second, qubits can be discarded and replaced by fresh qubits --- otherwise we would be unable to flush from the computer the entropy introduced by noise \cite{aharonov-ancilla}. 

We will discuss two versions of the threshold theorem, each appropriate for a different type of concatenated coding scheme. 
One version applies if each code in the recursive hierarchy can correct two or more errors in the code block (it is a quantum code whose distance is at least 5). This is the scheme considered by Aharonov and Ben-Or \cite{ben-or} and by Kitaev \cite{kitaev_threshold}. The other version applies even if each code in the recursive hierarchy corrects only one error in the code block (it is a distance-3 quantum code). This is the scheme considered by Knill, Laflamme, and Zurek \cite{knill} and in much subsequent work.

These two versions differ  because the fault-tolerant gadgets have different properties depending on which coding scheme is used. For a distance-5 code, level-1 gadgets can be designed such that, if there is one fault in the gadget and no more than one error in each of its input blocks, then there is no more than one error in each of its output blocks. As was shown in \cite{ben-or,kitaev_threshold} using an inductive argument, a similar property can be established at each level of the recursive hierarchy, which suffices to prove the threshold theorem. We will present a new proof that uses the same concepts; this proof actually follows closely the proof due to Aharonov and Ben-Or \cite{ben-or}, but we hope that some readers will find our proof especially clear and accessible.

Distance-3 codes have a smaller block size than distance-5 codes with the same number of logical qubits; hence fault-tolerant gadgets with fewer gates can be constructed using distance-3 codes, and the threshold fault rate $\varepsilon_0$ is expected to be correspondingly larger. Therefore, a threshold theorem that applies to concatenated distance-3 codes is highly desirable. But in contrast to the distance-5 case, since the code can correct only one error, a level-1 gadget with one fault may fail if one of its input blocks has a single error; hence the threshold theorem formulated in \cite{ben-or,kitaev_threshold} does not apply. 

A different (and on the face of it more complex) method of analysis is needed, in which the effectiveness of each gadget is predicated on the performance of the gadgets that immediately precede it in the circuit. Using such reasoning, Knill, Laflamme, and Zurek found a criterion for a level-1 simulation to outperform a level-0 simulation, and asserted without proof that this criterion provides a lower bound on the accuracy threshold \cite{knill}. The main results of this paper are a proof of the threshold theorem for concatenated distance-3 codes that justifies the criterion stated in \cite{knill}, and a rigorous estimate of the accuracy threshold based on this theorem: $\varepsilon_0 \ge 2.73\times 10^{-5}$. Our proof turns out to be remarkably simple, simpler in some ways than our proof for the distance-5 case. As far as we know, no proof of the threshold theorem for concatenated distance-3 codes has previously been published, for either quantum or classical computation.

The threshold estimate in \cite{knill} is done by finding an upper bound on the failure probability of a level-1 gadget. Failure occurs only if two faults occur in an ``extended gadget,'' so it suffices to count the pairs of locations in the extended gadget. Not only do our results put such estimates on a fully rigorous footing; we are also able to improve the threshold estimate, for two reasons. First, our gadgets are more efficient than those constructed in \cite{knill}. Second, many pairs of fault locations in the extended gadget are actually {\em benign}; they do not cause failure. Our new formulation of the threshold theorem allows us to do a more refined count of the {\em malignant} pairs of locations that actually can cause a gadget to fail. 

Another way to estimate the accuracy threshold, described in \cite{jp_threshold,gottesman_threshold,jp_ft}, is to derive and analyze a map (the {\em concatenation flow equation}) that expresses how an effective noise model evolves as the level of concatenation $k$ increases. The accuracy threshold is an unstable fixed point (or fixed surface) of this map. This method applies to concatenated distance-3 codes, but turning such estimates into a rigorous theorem proves to be challenging for several reasons -- for example, it is necessary to control the effects of error correlations that arise because code blocks interact multiple times as a circuit is executed. Two of the authors, whose work on this problem had long been stalled, were delighted to find a much simpler way to prove the threshold theorem for the distance-3 case (following in the footsteps of \cite{knill}) while also obtaining a rigorous threshold estimate that nearly matches the more heuristic result found by analyzing the concatenation flow equations.

We will also present a proof of the quantum threshold theorem that applies to a local non-Markovian noise model, and which goes beyond the result found in \cite{terhal} in two useful ways. First, we make weaker assumptions about the locality of the noise model than assumed in \cite{terhal}; second, our analysis applies to ``extended gadgets'' and therefore to concatenated distance-3 codes. 

Finally we remark that, while our estimate of the quantum accuracy threshold is the best that has so far been rigorously proven, recent studies by Knill have found numerical evidence for a much higher value of the threshold \cite{knill_detect}. For the independent stochastic noise model, Knill calculates that for $\varepsilon < \varepsilon_0'\approx 3 \times 10^{-2}$, it is possible to simulate a universal set of gates with effective fault rate below the threshold $\varepsilon_0$ for a concatenated distance-3 code, thus establishing $\varepsilon_0'$ as an improved lower bound on the quantum accuracy threshold. The theorem proved in this paper, providing a lower bound on $\varepsilon_0$, furnishes a rigorous foundation for one essential part of Knill's analysis. An important goal for future work will be to make the rest of Knill's higher threshold estimate fully rigorous. 

The rest of this paper is organized as follows. In Sec.~\ref{sec:level-1} we explain the essential characteristics of fault-tolerant gadgets built from noisy gates, and relate the {\em goodness} (sparseness of faults) of a level-1 simulation to its {\em correctness} (i.e., accuracy). In Sec.~\ref{sec:recursive} we define appropriate notions of goodness and correctness for level-$k$ simulations, and state our main lemma, which asserts that a good level-$k$ simulation is correct. In Sec.~\ref{sec:threshold-theorem} we invoke the lemma to prove the quantum threshold theorem for independent stochastic noise, and in Sec.~\ref{good-implies-correct} we prove the lemma. In Sec.~\ref{sec:malignant-pairs} we explain how to improve the threshold estimate by counting malignant pairs of fault locations. In Sec.~\ref{sec:ft-general} we discuss in more detail how fault-tolerant gadgets can be constructed, and in Sec.~\ref{sec:explicit} we derive a rigorous lower bound on the quantum accuracy threshold by applying the result of Sec.~\ref{sec:malignant-pairs} to gadgets based on a particular distance-3 code. In Sec.~\ref{sec:higher-distance} we prove a threshold theorem that applies to concatenation of higher-distance codes, and in Sec.~\ref{sec:distance-5} we use a different method to prove a threshold theorem whose content is similar to the result of \cite{ben-or}. In Sec.~\ref{sec:non-markovian} we prove the threshold theorem for local non-Markovian noise, and Sec.~\ref{sec:conclusions} contains our conclusions.

After our work was completed, Reichardt \cite{reichardt-threshold}, using different methods than ours, also proved a quantum threshold theorem for concatenated distance-3 codes.  

\section{Fault-tolerant simulation at level 1}
\label{sec:level-1}
Before proceeding to our proofs of the threshold theorem, we will describe the key properties of the gadgets used in our fault-tolerant simulations. To build intuition, we will at first discuss these properties rather informally; in Sec.~\ref{sec:recursive}, and again in Sec.~\ref{sec:higher-distance}, we will restate the properties in another language to formulate a precise statement of the threshold theorem. Explicit circuits for gadgets will be discussed in Sec.~\ref{sec:ft-general} and \ref{sec:explicit}.

Fault-tolerant simulations of quantum circuits are based on quantum error-correcting codes. Such codes can be constructed for $d$-dimensional quantum systems (``qudits''), but for definiteness we will imagine that our elementary quantum systems, and the encoded systems protected by the code, are qubits ($d=2$). We will denote by $n$ the {\em length} of the code, the number of qubits in the code block. We will focus in this paper on codes with one encoded qubit per block, though our constructions can be generalized to codes with more than one encoded qubit per block.

Suppose that $|\bar \psi\rangle$ denotes a pure quantum state in the code space. The effect on $|\bar \psi\rangle$ of an arbitrary error (a trace-preserving completely positive map acting on the $n$ qubits in the code block) can be expanded in terms of $n$-qubit ``Pauli operators'':
\begin{equation}
\label{general-error}
|\bar \psi\rangle \to \sum_a E_a|\bar \psi\rangle \otimes |a\rangle_E;
\end{equation}
here the states $|a\rangle_E$ are states of the ``environment'', which are not assumed to be normalized or mutually orthogonal, and the $E_a$'s are summed over the $2^{2n}$ operators $\{I,X,Y,Z\}^{\otimes n}$, where $\{I,X,Y,Z\}$ are the single-qubit Pauli operators
\begin{equation}
I= \pmatrix{1 & 0\cr 0& 1\cr}~,\quad X= \pmatrix{0 & 1\cr 1& 0\cr}~,\quad 
Y=\pmatrix{0 & -i\cr i& 0\cr}~,\quad Z= \pmatrix{1 & 0\cr 0& -1\cr}~.
\end{equation}
We denote by $w$ the {\em weight} of a Pauli operator, the number of qubits on which it acts nontrivially (with a Pauli operator other than the identity $I$). Most fault-tolerant protocols are founded on the hypothesis that highly correlated errors that damage many qubits at once should be rare ---  the norm $\parallel |a\rangle_E\parallel$ of the state associated with a high-weight Pauli operator $E_a$ should be quite small. We say that a quantum error-correcting code can correct $t$ errors (for some $t< n$) if the code protects against all Pauli errors with weight $w\le t$. Though uncorrectable errors with weight higher than $t$ might occur, one hopes that these higher weight errors are sufficiently suppressed that the protection against error is still reasonably effective. A code that corrects $t$ errors is said to have {\em distance} $2t+1$, because (at least roughly speaking) Pauli operators that preserve the code space and act nontrivially within the code space have weight at least $2t+1$ \cite{knill-laflamme}. Thus we will say ``distance-3 code'' to indicate a quantum code that can correct one error, and ``distance-5 code'' to indicate a quantum code that can correct two errors. In either case, error recovery proceeds in two steps. First all of the ``check operators'' in a mutually commuting set are measured. This measurement projects the error onto a particular Pauli operator (or a set of Pauli operators that all act on the code space in the same way); furthermore the measurement outcomes constitute an error ``syndrome'' that points to a particular Pauli operator $E_a$ (or to a particular set of equivalent Pauli operators). Then the unitary operator $E_a^\dagger$ is applied to reverse the damage due to the error.

In a fault-tolerant simulation of an ideal quantum circuit, each gate in the ideal circuit is simulated by a gadget constructed from the noisy gates, which acts on the logical qubits that are protected by the code. In describing such simulations, we will make a distinction between an {\em error}, a position in a code block where a qubit has been damaged, and a {\em fault}, a location in a circuit where the operation applied by one of the elementary gates deviates from the ideal operation. Fault-tolerant gadgets are designed to prevent an error introduced by a fault from propagating to become many errors within a single code block. 

Fault-tolerant simulations of quantum computations share many of the same features as fault-tolerant classical simulations, and can be analyzed using similar methods. But there are also some new issues that arise in the quantum case that have no classical analog. First, while in the classical case we need only be concerned about bit flip ($X$) errors, in the quantum case we need to worry about the propagation of phase flip ($Z$) errors as well. (A $Y$ error can be viewed as an $X$ error and a $Z$ error simultaneously afflicting the same qubit.) Second, the construction of a universal set of fault-tolerant gates is especially challenging in the quantum case; the off-line preparation and verification of {\em quantum software} is required to implement some of the gates \cite{gottesman-chuang}. And third, the code states for a quantum code have the property that the qubits in the code block are highly entangled with one another, so that the density operator of each individual qubit is maximally mixed. It can be a subtle matter to speak of an error at a particular position in the block, because the error might have no locally observable effect on the damaged qubit; the damage affects only the quantum correlations of that qubit with other qubits.

In the proof of the threshold theorem we will study the performance of a {\em recursive} simulation. At ``level 0'' of this recursion, the gadget ``simulating'' an ideal gate is just a noisy gate; we call it a 0-gate, or 0-Ga. At level 1, each ideal gate is simulated by a {\em 1-rectangle}, also called a {\em 1-Rec}, which acts on the code blocks (called 1-blocks) of some quantum error-correcting code $C$. At level 2, an ideal gate is simulated by a {\em 2-Rec}, which acts on the code blocks (called 2-blocks) of the concatenated quantum code $C\circ C$. This 2-Rec is constructed by replacing each 0-Ga in the 1-Rec by a 1-Rec. And so on --- a $k$-Rec is constructed by replacing each 0-Ga in the $(k{-}1)$-Rec by a 1-Rec.

In our analysis of the threshold, the level-1 gadgets must have certain properties. We will state the properties here, but we will postpone until Sec.~\ref{sec:ft-general} and \ref{sec:explicit} any detailed discussion of how gadgets with these properties are constructed. Readers who are unfamiliar with the principles of fault-tolerant quantum computing may wish to jump ahead now to Sec.~\ref{sec:ft-general} and \ref{sec:explicit} to see explicit circuits for the gadgets. 

The desired properties are a bit different for codes of distance 5 and higher than for distance-3 codes. In order to proceed as briskly as possible to our proof of the threshold theorem for concatenated distance-3 codes, we will focus here on the distance-3 case, and will return to higher-distance codes in Sec.~\ref{sec:higher-distance} and \ref{sec:distance-5}.

Our 0-Ga's include all of the quantum gates comprising a universal set; in addition there is a 0-preparation, which prepares a qubit in a standard state, and a 0-measurement, which destructively measures a qubit in a standard basis and records the outcome as a classical bit. It will be convenient to suppose that there are two 0-measurements, measurement in the $Z$-eigenstate basis and the $X$-eigenstate basis, and two 0-preparations, the preparation of the $Z$ eigenstate $|0\rangle$ with eigenvalue 1 and of the $X$ eigenstate $|+\rangle$ with eigenvalue 1. For each 0-Ga there is a corresponding 1-Rec, and in the level-1 simulation, each 0-Ga in the ideal circuit is replaced by its corresponding 1-Rec. 

The 1-Recs are constructed using a level-1 error-correction gadget 1-EC, and level-1 gate gadgets, the 1-Ga's. If the 1-EC is executed with no faults, it corrects one error --- it maps the input $E_a|\bar\psi\rangle$ to the output $|\bar \psi\rangle$, where $|\bar \psi\rangle$ is a state in the code space, and $E_a$ is a Pauli operator with weight $w\le 1$. If the distance-3 code is not ``perfect,'' there may be some error syndromes that indicate that more than one qubit has been damaged, and in that case successful error recovery might not be possible. When the 1-EC encounters an ambiguous syndrome, it maps its input to some state in the code space. For most of our discussion, it will not matter exactly how this codeword is chosen. 

For each unitary transformation $U$ in our universal gate set, there is a level-1 gate gadget or 1-Ga; if the 1-Ga is executed with no faults, it applies $U$ to the encoded state. The 1-Rec for the ideal gate $U$ consists of the corresponding 1-Ga followed by a 1-EC acting on each of the output 1-blocks of the 1-Ga, as shown in Fig.~\ref{fig:level-1-sim}. In addition, there is a 1-preparation that prepares a 1-block in the standard code state $|\bar 0\rangle$ (or in the conjugate state $|\bar +\rangle$), and a 1-measurement that destructively measures an encoded qubit in the standard basis $\{|\bar 0\rangle, |\bar 1\rangle\}$ (or in the conjugate basis $\{|\bar +\rangle, |\bar -\rangle\}$), and records the outcome as a classical bit. The preparation 1-Rec consists of 1-preparation followed by 1-EC, and the measurement 1-Rec is the same thing as the 1-measurement.

\begin{figure}
\begin{center}
\vspace{0.5cm}
\begin{picture}(160,54)
\put(0,19){\line(1,0){5}}
\put(0,35){\line(1,0){5}}
\put(5,13){\framebox(20,27){\footnotesize 0-Ga}}
\put(25,19){\line(1,0){5}}
\put(25,35){\line(1,0){5}}
\put(40,21){\makebox(30,12){$\Longrightarrow$}}
\put(78,12){\line(1,0){10}}
\put(78,42){\line(1,0){10}}
\put(88,0){\framebox(28,54){1-Ga}}
\put(116,12){\line(1,0){10}}
\put(116,42){\line(1,0){10}}
\put(126,0){\framebox(24,24){1-EC}}
\put(150,12){\line(1,0){10}}
\put(126,30){\framebox(24,24){1-EC}}
\put(150,42){\line(1,0){10}}
\end{picture}
\vspace{0.3cm}
\end{center}
\fcaption{Level-1 simulation. Each 0-Ga in the ideal circuit is replaced by a 1-Rec, which consists of the 1-Ga that simulates the 0-Ga, followed by a 1-EC acting on each output 1-block of the 1-Ga.}
\label{fig:level-1-sim}
\end{figure}

If $|\psi\rangle$ is an ideal state of a qubit, and $|\bar \psi\rangle$ is the corresponding encoded state of a 1-block, we say that the actual state $\rho$ of the 1-block has at most one error if its purification can be expanded as eq.~(\ref{general-error}), where the sum is restricted to Pauli operators with weight $w\le 1$. By following the principles of quantum fault tolerance, we can construct a 1-EC and 1-Ga's with the following properties:

\begin{description}
\item 0. If a 1-EC contains no fault, it takes any input to an output in the code space.
\item 1. If a 1-EC contains no fault, it takes an input with at most one error to an output with no errors.
\item 2. If a 1-EC contains at most one fault, it takes an input with no errors to an output with at most one error.
\item 3. If a 1-Ga contains no fault, it takes an input with at most one error to an output with at most one error in each output block.
\item 4. If a 1-Ga contains at most one fault, it takes an input with no errors to an output with at most one error in each output block. 
\end{description}

\noindent Property 1 is just the statement that 1-EC is an error recovery circuit for a code that can correct one error, while properties 2--4 express that the gadgets do not propagate errors badly. (In the case of property 3, the input is required to have at most one error {\em all together} acting in all input blocks; this error might propagate to other blocks, but it does not propagate to other qubits in the same block.) Property 0 holds if we adopt a suitable convention for recovering when the error syndrome indicates more than one error. How gadgets satisfying properties 0--4 can be constructed will be discussed in more detail in Sec.~\ref{sec:ft-general} and \ref{sec:explicit}. 

We can also construct a 1-preparation (which has no input) such that a 1-preparation with one fault produces an output with one error; this is a special case of property 4. And we can construct a 1-measurement (which has a classical bit as output) that agrees with an ideal measurement if either its input has one error and the 1-measurement has no faults, or its input has no errors and the 1-measurement has one fault; these can be viewed as special cases of properties 3 and 4. 

Actually, when we assert that a 1-measurement with one fault successfully measures a 1-block with no errors, we are implicitly assuming that either the outcome of the measurement is stored in the block of a classical error-correcting code, or if the outcome is decoded to a single bit, that the classical gates that decode the outcome are perfect. Otherwise, a single fault in the final decoding step could cause an error in the outcome.

We would like to state a criterion that, if satisfied, ensures that the level-1 simulation of the ideal circuit is reliable. For this purpose it is convenient to group each 1-Rec together with the preceding 1-ECs that act on its input blocks; we call this composite object an {\em extended rectangle} or 1-exRec. Note that the extended rectangles can overlap with one another, as in Fig.~\ref{fig:overlapping-exRecs}. Let us refer to the 1-ECs in an exRec that precede the 1-Ga as the {\em leading} error corrections in the exRec and the 1-ECs that follow the 1-Ga as the {\em trailing} error corrections in the exRec. Then a trailing 1-EC of a 1-exRec is also a leading 1-EC of a 1-exRec that follows.

\begin{figure}
\begin{center}
\vspace{0.5cm}
\begin{picture}(201,96)
\put(5,12){\line(1,0){5}}
\put(10,0){\framebox(24,24){1-EC}}
\put(34,12){\line(1,0){14}}
\put(5,44){\line(1,0){5}}
\put(10,32){\framebox(24,24){1-EC}}
\put(34,42){\line(1,0){14}}
\put(48,0){\framebox(28,56){1-Ga}}
\put(76,12){\line(1,0){14}}
\put(76,44){\line(1,0){5}}
\put(85,44){\line(1,0){5}}
\put(90,0){\framebox(24,24){1-EC}}
\put(114,12){\line(1,0){5}}
\put(90,32){\framebox(24,24){1-EC}}
\put(114,44){\line(1,0){5}}
\put(123,44){\line(1,0){5}}
\put(85,76){\line(1,0){5}}
\put(90,64){\framebox(24,24){1-EC}}
\put(114,76){\line(1,0){14}}
\put(128,32){\framebox(28,56){1-Ga}}
\put(156,44){\line(1,0){14}}
\put(156,76){\line(1,0){14}}
\put(170,32){\framebox(24,24){1-EC}}
\put(194,44){\line(1,0){5}}
\put(170,64){\framebox(24,24){1-EC}}
\put(194,76){\line(1,0){5}}
\put(3,-4){\dashbox(118,64){}}
\put(83,28){\dashbox(118,64){}}

\end{picture}
\vspace{0.3cm}
\end{center}
\fcaption{Overlapping extended rectangles. Two consecutive 1-exRecs (indicated by dashed lines) share a 1-EC, which is a trailing 1-EC of the earlier 1-exRec and a leading 1-EC of the later 1-exRec.}
\label{fig:overlapping-exRecs}
\end{figure}

The 1-exRecs have an important property that follows from the properties 0--4 of the 1-gadgets:

\medskip
\noindent {\bf Lemma 1. exRec-Cor at level 1}. {\em Suppose that the level-1 gadgets obey properties 0--4. Then if a 1-exRec contains no more than one fault, and the input to its 1-Rec has no more than one error in each input block, its output has no more than one error in each output block.}
\medskip 

\noindent
Let us say that a 1-exRec is {\em good} if it contains no more than one fault, and that a 1-Rec is {\em correct} if it takes an input with no more than one error per 1-block to an output with no more than one error per 1-block. Then the property exRec-Cor can be stated more succinctly as
\begin{description}
\item {\bf exRec-Cor at level 1}. {\em The 1-Rec contained in a good 1-exRec is correct.}
\end{description}

\medskip
\noindent {\bf Proof of Lemma 1}: First suppose that none of the leading 1-ECs of the 1-exRec contain any faults. Then the output of each leading 1-EC is a codeword by property 0. But if the output of the 1-EC (which is one of the input blocks to the 1-Rec) has at most one error and is also a codeword, then it has no errors. Therefore the input to the 1-Rec actually has no errors. Now, the 1-Rec might contain one fault, which could be in the 1-Ga, or could be in one of the trailing 1-ECs. If the 1-Ga contains a fault, then by property 4 its output has no more than one error in each block, and by property 1 these errors will be corrected by the trailing 1-ECs. If the 1-Ga contains no faults, then its output has no errors by property 3, and since each trailing 1-EC has at most one fault, each output block from the 1-Rec has at most one error by property 2.

On the other hand, suppose that one of the leading 1-ECs in the 1-exRec contains a fault. Then each of the other leading 1-ECs contains no faults, and outputs a codeword by property 0. Therefore, if each input block to the 1-Rec has at most one error, then in fact all but one of the input blocks have no errors. Now there are no faults contained in the 1-Ga or in any of the trailing 1-ECs. Therefore each output block of the 1-Ga has no more than one error by property 3, and the output of each trailing 1-EC has no errors by property 1. 

These arguments also apply to the 1-preparation exRec (which is the same as the 1-preparation Rec), and to the 1-measurement exRec (for which correctness means that the measurement reproduces a perfect measurement of an ideal 1-block).

\rightline{$\square$}

\medskip

If all 1-exRecs are good, then our level-1 simulation will be successful (it produces exactly the same probability distribution for the final readout as the ideal circuit). We simply observe that the initial 1-preparations produce input blocks with at most one error, and that for every 1-Rec that follows, each input block has at most one error so that each output block also has at most one error by exRec-Cor. Finally, the 1-measurements at the end of the circuit simulate the ideal measurements of the output blocks faithfully. The simulation works because the goodness of the exRecs ensures that each error caused by a fault gets corrected before it can be joined by a second error in the same block that would cause the simulation to fail. 

Suppose that the ideal circuit contains $L$ locations, and suppose that stochastic faults occur independently, with probability $\varepsilon$, at each location within the noisy quantum circuit used in our level-1 simulation. (A ``location'' can be a 0-preparation, a 0-measurement, or a gate, including an identity gate acting on a ``resting'' qubit.) For the simulation to fail, there must be at least two faults in at least one 1-exRec. For each specified pair of locations inside a 1-exRec, failure occurs at both of those locations with a probability no larger than $\varepsilon^2$. Therefore, the probability of failure for the level-1 simulation can be bounded as
\begin{equation}
\label{level1-pfail}
P_{\rm fail} \le L A\varepsilon ^2~,
\end{equation}
where $A$ is the number of pairs of locations in the largest 1-exRec. Of course, in this estimate we are being overly pessimistic, because not all pairs of fault locations in the 1-exRec will cause the 1-exRec to be incorrect. We will return to this point in Sec.~\ref{sec:malignant-pairs}.

Thus, for independent stochastic errors, using quantum error correction and fault-tolerant gadgets reduces the probability of failure per gate from $\varepsilon$ to $O(\varepsilon^2)$; quantum coding improves the reliability of the quantum computation if the fault rate $\varepsilon$ is small enough.

\section{Recursive simulation: goodness and correctness}
\label{sec:recursive}

Fault-tolerant simulation at level 1 achieves a modest improvement of the failure probability per gate, from $\varepsilon$ to $O(\varepsilon^2)$. Further improvement to a higher power of $\varepsilon$ can be attained by using codes that correct more errors. But as the codes get more complex, so do the rectangles, and for many families of codes one finds that fault-tolerant protocols are effective only for smaller and smaller values of $\varepsilon$ as the code's distance grows. 

\begin{figure}
\begin{center}
\leavevmode
\epsfysize=2in
\epsfbox{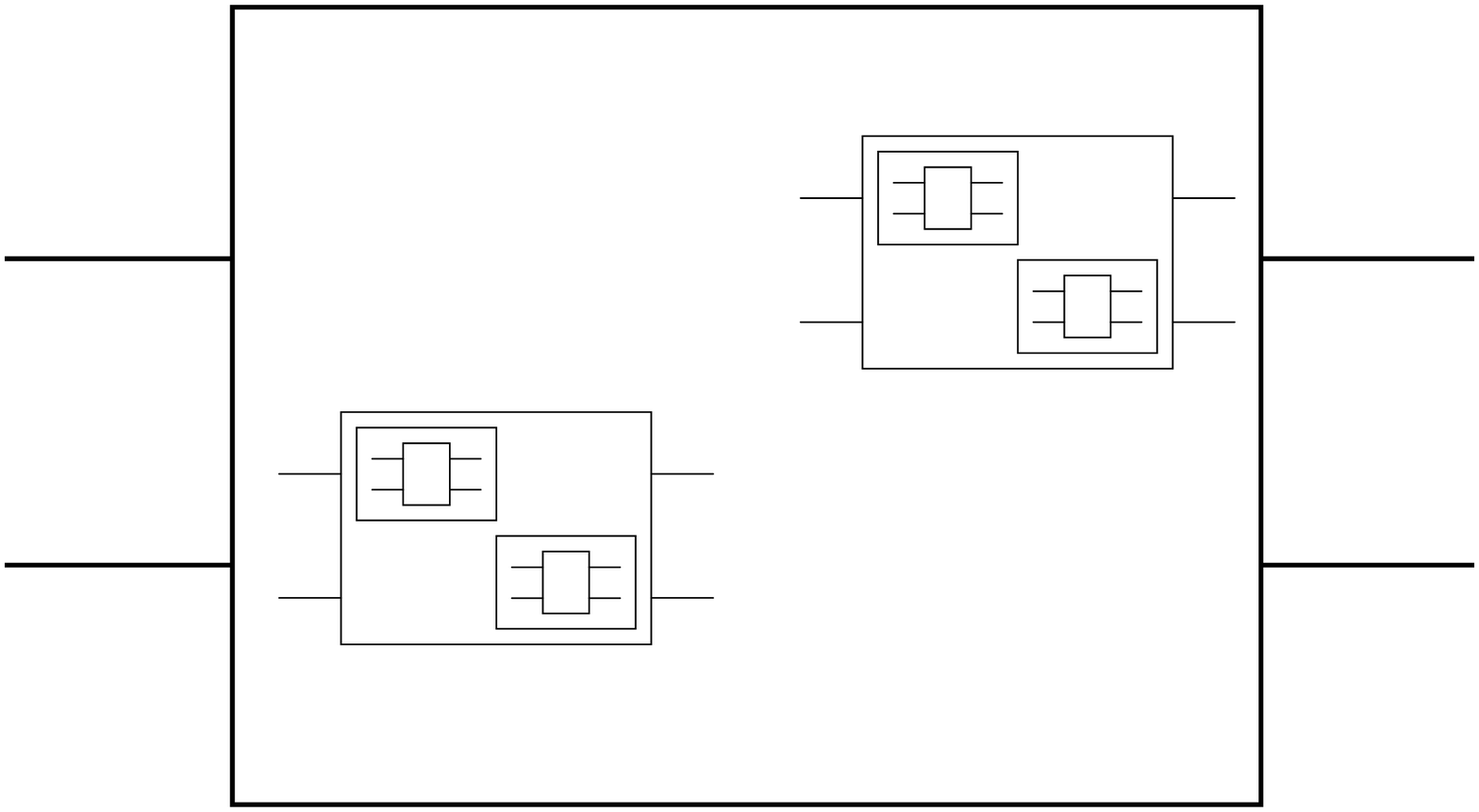}
\end{center}
%\vspace{0.2cm}     
\fcaption{A recursive simulation. A level-$k$ gadget is built from level-$(k{-}1)$ gadgets, which are built from level-$(k{-}2)$ gadgets, and so on.}
\label{fig:recursive}
\end{figure}

If our goal is to compute reliably for as large a value of the fault rate as possible, the best known strategy is to use a recursive simulation, as in Fig.~\ref{fig:recursive}. In this scheme, a fault-tolerant gadget at level $k$ is constructed by replacing each level-0 location in the level-$(k{-}1)$ gadget by the corresponding level-1 rectangle. (This includes the identity gate --- that is, a ``resting'' qubit that is not acted on by any gate in a particular time step is simulated by a 1-EC, which is the 1-Rec for the identity.) Equivalently, we may say that the level-$k$ gadget is constructed by replacing each 0-Ga in the level-1 gadget by the corresponding $(k{-}1)$-Rec. One then hopes to achieve an arbitrarily reliable simulation by observing that, once $\varepsilon$ is small enough, the failure probability per gate declines each time the level of the simulation increases by one. It is almost obvious that this idea is sound, but it will be important to choose our definitions carefully to ensure that the proof can proceed smoothly. To prove the threshold theorem, we wish to extend our observation that ``a good rectangle is correct'' to higher levels than $k=1$. That is, we are to argue that if the faults in a $k$-Rec are sufficiently sparse (the $k$-Rec is good), then it takes accurately encoded inputs to accurately encoded outputs (the $k$-Rec is correct). The key is to find a suitable definition of ``good'' and ``correct'' so that we can easily establish that ``a good rectangle is correct'' by an inductive argument. 

The strategy behind our proofs of the threshold theorem, based on appropriate notions of goodness and correctness, is inspired by the work of Aharonov and Ben-Or \cite{ben-or}. Similar notions were applied earlier in the theory of fault-tolerant classical computation, for example by G\'acs \cite{gacs}.

\subsection{Goodness}
\label{subsec:goodness}
The obvious way to generalize the concept of goodness to higher levels would be to say that a $(k{+}1)$-exRec is good if it contains no more than one bad $k$-exRec; however, that choice would lead to problems, because consecutive exRecs overlap with one another. When we estimate the probability that a $(k{+}1)$-exRec is bad, we wish to regard two bad $k$-exRecs contained in the $(k{+}1)$-exRec as independent events, but consecutive bad $k$-exRecs might not be independent. For example, the $k$-EC shared by two consecutive $k$-exRecs might contain two bad $(k{-}1)$-exRecs, causing both consecutive $k$-exRecs to be bad. Or one bad $(k{-}1)$-exRec in the shared $k$-EC might combine with one bad $(k{-}1)$-exRec in the $k$-Rec that follows the $k$-EC and with one bad $(k{-}1)$-exRec in the $k$-Ga or $k$-EC that precedes it to cause both consecutive $k$-exRecs to be bad.

Fortunately, there is a simple solution to this predicament. We will say that a $(k{+}1)$-exRec is bad only if it contains two bad $k$-exRecs that actually {\em do} fail independently of one another. Thus a bad $(k{+}1)$-exRec is permitted  to contain two overlapping bad $k$-exRecs, but only if the earlier of the two consecutive $k$-exRecs would have been good were it not for the bad $(k{-}1)$-exRecs contained in the $k$-EC it shares with the following $k$-exRec. Nonindependent consecutive pairs of $k$-exRecs are acceptable, because we will be able to argue that they are really no more harmful than a single bad $k$-exRec, at the later of the two consecutive locations. 

Suppose for example, that two consecutive 1-exRecs are both bad, due to a total of three faults --- one in the shared 1-EC, one in the following 1-Rec, and one in the preceding 1-ECGa (we use 1-ECGa to denote leading 1-ECs followed by a 1-Ga), as illustrated in Fig.~\ref{fig:bad-overlapping-exRecs}. These two overlapping 1-exRecs contain two 1-Recs, which simulate a pair of 0-Ga's in the ideal circuit. Ordinarily we would think of the earlier of the two 1-Recs as the gadget that simulates the earlier of the two 0-Ga's.  But in this case we may instead regard the 1-Ga contained in the earlier 1-Rec as the gadget that simulates the earlier 0-Ga, and because the 1-ECGa contains only one fault, the 1-Ga simulates the earlier 0-Ga accurately. We may also regard the later 1-exRec (rather than the 1-Rec it contains) as the gadget that simulates (inaccurately) the later 0-Ga. In this sense the overlapping pair of bad 1-exRecs is really no worse than a single bad 1-exRec. Our inductive proof (see Sec.~\ref{sec:linearity-of-correctness}) will show that this idea also works at higher levels. 

\begin{figure}
\begin{center}
\vspace{0.5cm}
\begin{picture}(250,48)
\put(2,22){\line(1,0){5}}
\put(7,4){\framebox(36,36){1-EC}}
\put(43,22){\line(1,0){14}}
\put(57,4){\framebox(36,36){1-Ga}}
\put(85,26){$\times$}
\put(93,22){\line(1,0){5}}
\put(102,22){\line(1,0){5}}
\put(107,4){\framebox(36,36){1-EC}}
\put(113,6){$\times$}
\put(143,22){\line(1,0){5}}
\put(152,22){\line(1,0){5}}
\put(157,4){\framebox(36,36){1-Ga}}
\put(193,22){\line(1,0){14}}
\put(207,4){\framebox(36,36){1-EC}}
\put(231,32){$\times$}
\put(243,22){\line(1,0){5}}
\put(0,0){\dashbox(150,44){}}
\put(100,-4){\dashbox(150,52){}}
\end{picture}
\vspace{0.3cm}
\end{center}
\fcaption{Two nonindependent consecutive bad 1-exRecs, indicated by dashed lines, with fault locations indicated by $\times$. Because one of the three faults is contained in the shared 1-EC, the bad 1-exRecs are not independent events. In this situation, we may regard the earlier 1-Ga as a gadget that simulates the corresponding ideal 0-Ga accurately, and the later 1-exRec as a gadget that simulates the corresponding ideal 0-Ga inaccurately.}
\label{fig:bad-overlapping-exRecs}
\end{figure}

We note that the opening gates in a gadget (those acting in the first time step) are simulated by Recs whose exRecs are not strictly speaking ``contained in'' the gadget (the leading EC of the exRec is actually part of the preceeding gadget). Nevertheless, in a slight abuse of language, we will say that an opening $(k{-}1)$-exRec in a $k$-gadget is contained in the $k$-gadget. 

When we estimate the probability that a $k$-exRec is bad, we wish to regard the bad $(k{-}1)$-exRecs contained in the $k$-exRec as independent events. Therefore, we define badness as follows:

\medskip
\noindent {\bf Definition. Goodness and Badness}. {\em A 1-exRec is {\em bad} if it contains two faults; if it is not bad it is {\em good}. Two bad 1-exRecs are {\em independent} if they are nonoverlapping or if they overlap and the earlier 1-exRec is still bad when the shared 1-EC is removed. For $k>1$, a $k$-exRec is {\em bad} if it contains two independent bad $(k{-}1)$-exRecs; if it is not bad it is {\em good}. Two bad $k$-exRecs are {\em independent} if they are nonoverlapping or if they overlap and the earlier $k$-exRec is still bad when the shared $k$-EC is removed.}

\medskip
\noindent To ensure that the earlier (truncated) $k$-exRec and the later (complete) $k$-exRec are really independent, it is important that no 0-Ga is contained in both. Thus when we say that ``the shared $k$-EC is removed'' from the earlier $k$-exRec, we mean that the complete $(k{-}1)$-exRecs in the first time step of the later $k$-exRec are excluded from the earlier $k$-exRec. Similarly, the complete $(k{-}2)$-exRecs in the first time step of each of these $(k{-}1)$-exRecs are excluded from the earlier $k$-exRec, and so on. This point will be elaborated in Sec.~\ref{subsubsec:nonindependent-pairs}.

With this definition of badness, bad $k$-exRecs become very unlikely as $k$ increases, if the fault rate is sufficiently small. For independent stochastic faults occuring with probability $\varepsilon$, as in eq.~(\ref{level1-pfail}) the probability $\varepsilon^{(1)}$ that a 1-exRec is bad satisfies
\begin{equation}
\label{pbad-level1}
\varepsilon^{(1)} \le A\varepsilon^2 
\end{equation}
where $A$ is the number of pairs of locations in the largest 1-exRec. Because of the self-similarity of the $k$-exRecs, and because independent bad $(k{-}1)$-exRecs are independent events, the probability $\varepsilon^{(k)}$ that a $k$-exRec is bad satisfies
\begin{equation}
\varepsilon^{(k)} \le  A \left(\varepsilon^{(k{-}1)}\right)^2~,
\end{equation}
which together with eq.~(\ref{pbad-level1}) implies

\medskip
\noindent {\bf Lemma 2. Bad exRecs are rare}. {\em Suppose that stochastic faults occur independently, with probability $\varepsilon$, at each circuit location in a $k$-exRec. Then the probability $\varepsilon^{(k)}$ that the $k$-exRec is bad satisfies
\begin{equation}
\label{p-bad-bound}
\varepsilon^{(k)} \le {\varepsilon_0} \left(\varepsilon/\varepsilon_0\right)^{2^k}~,
\end{equation}
where $\varepsilon_0^{-1}$ is the number of pairs of locations in the largest 1-exRec. }
\medskip

This $\varepsilon_0$ is a lower bound on the accuracy threshold (and in Sec.~\ref{sec:malignant-pairs} we will see that this estimate can be improved); for $\varepsilon < \varepsilon_0$, the probability of badness declines double-exponentially with $k$.

\subsection{Correctness}
How should we define correctness? First we need a notion of what it means for a state to be ``accurately encoded'' at level $k$. Since in our recursive simulation we will read out the result of the computation using a recursive measurement procedure, an accurate encoding should be one that can be successfully decoded recursively. For the purpose of defining our notion of correctness for noisy circuits, we will employ an ideal decoder at level $k$ (a $k$-decoder)--- a conceptual device that maps a level-$k$ encoded block to a single qubit \cite{smith}. The level-1 ideal decoder measures and records the error syndrome, performs recovery as indicated by the syndrome, then decodes the 1-block to a qubit, and finally discards the syndrome. We use the word ``ideal'' to emphasize that this procedure is carried out flawlessly --- there are no faults. The level-$k$ decoder is defined recursively; it is realized by first applying the $(k{-}1)$-decoder to each of the $(k{-}1)$-subblocks, and then applying the 1-decoder to the resulting 1-block. 

When we say that a $k$-Rec is {\em correct}, we mean that it simulates the corresponding ideal gate accurately.

\medskip
\noindent {\bf Definition. Correctness}. {\em A $k$-Rec is {\em correct} if the $k$-Rec followed by the ideal $k$-decoder is equivalent to the ideal $k$-decoder followed by the ideal 0-Ga that the $k$-Rec simulates:}

\vspace{0.5cm}
\begin{picture}(292,24)
\put(0,12){\line(1,0){10}}
\put(10,0){\framebox(48,24){\shortstack{correct\\$k$-Rec}}}
\put(58,12){\line(1,0){10}}
\put(68,0){\framebox(48,24){\shortstack{ideal\\$k$-decoder}}}
\put(116,12){\line(1,0){10}}
\put(126,6){\makebox(20,12){=}}
\put(146,12){\line(1,0){10}}
\put(156,0){\framebox(48,24){\shortstack{ideal\\$k$-decoder}}}
\put(204,12){\line(1,0){10}}
\put(214,0){\framebox(48,24){\shortstack{ideal\\$0$-Ga}}}
\put(262,12){\line(1,0){10}}
\put(272,6){\makebox(20,12){.}}
\end{picture}
\vspace{0.3cm}

\noindent
To be specific, suppose that the ideal 0-Ga applies the unitary transformation $U$ to a qubit or to several qubits. Suppose that the input to the $k$-Rec is a state $\rho$ such that the ideal decoder maps $\rho$ to the ``ideal'' pure state $|\psi\rangle$. According to our criterion, then, if the $k$-Rec is correct, the ideal decoder maps the output of the $k$-Rec to $U|\psi\rangle$. In this sense, our notion of correctness captures the idea that the $k$-Rec maintains the decodability of states. 

There is a similar notion of correctness that applies to the $k$-preparation and the $k$-measurement. A preparation $k$-Rec ($k$-preparation followed by $k$-EC) is correct if the ideal $k$-decoder maps its output to the ideally prepared state, e.g., $|0\rangle$ or $|+\rangle$:

\vspace{0.5cm}
\begin{picture}(272,24)
\put(0,0){\framebox(58,24){\shortstack{correct\\prep. $k$-Rec}}}
\put(58,12){\line(1,0){10}}
\put(68,0){\framebox(48,24){\shortstack{ideal\\$k$-decoder}}}
\put(116,12){\line(1,0){10}}
\put(126,6){\makebox(20,12){=}}

\put(146,0){\framebox(48,24){\shortstack{ideal\\$0$-prep.}}}
\put(194,12){\line(1,0){10}}
\put(204,6){\makebox(20,12){.}}
\end{picture}
\vspace{0.3cm}

\noindent
A $k$-measurement is correct if it realizes the same POVM as the ideal $k$-decoder followed by the ideal 0-measurement:

\vspace{0.5cm}
\begin{picture}(272,24)
\put(0,12){\line(1,0){10}}
\put(10,0){\framebox(48,24){\shortstack{correct\\$k$-meas.}}}
\put(58,6){\makebox(20,12){=}}
\put(78,12){\line(1,0){10}}
\put(88,0){\framebox(48,24){\shortstack{ideal\\$k$-decoder}}}
\put(136,12){\line(1,0){10}}
\put(146,0){\framebox(48,24){\shortstack{ideal\\$0$-meas.}}}
\put(192,6){\makebox(20,12){.}}
\end{picture}
\vspace{0.3cm}

The crucial property of $k$-rectangles used in the proof of the threshold theorem is:

\medskip
\noindent {\bf exRec-Cor}. {\em The $k$-Rec contained in a good $k$-exRec is correct.}
\medskip

\noindent
In other words, good $k$-exRecs satisfy the following identities:

\vspace{0.5cm}
\begin{picture}(378,24)
\put(0,12){\line(1,0){10}}
\put(10,0){\framebox(48,24){$k$-EC}}
\put(58,12){\line(1,0){10}}
\put(68,0){\framebox(48,24){$k$-Rec}}
\put(116,12){\line(1,0){10}}
\put(126,0){\framebox(48,24){\shortstack{ideal\\$k$-decoder}}}
\put(174,12){\line(1,0){10}}
\put(184,6){\makebox(20,12){=}}
\put(204,12){\line(1,0){10}}
\put(214,0){\framebox(48,24){$k$-EC}}
\put(262,12){\line(1,0){10}}
\put(272,0){\framebox(48,24){\shortstack{ideal\\$k$-decoder}}}
\put(320,12){\line(1,0){10}}
\put(330,0){\framebox(48,24){\shortstack{ideal\\$0$-Ga}}}
\put(378,12){\line(1,0){10}}
\put(388,6){\makebox(20,12){,}}
\end{picture}
%\vspace{0.3cm}

\vspace{0.5cm}
\begin{picture}(378,24)
\put(10,0){\framebox(48,24){\shortstack{prep.\\$k$-Rec}}}
\put(58,12){\line(1,0){10}}
\put(68,0){\framebox(48,24){\shortstack{ideal\\$k$-decoder}}}
\put(116,12){\line(1,0){10}}
\put(126,6){\makebox(20,12){=}}
\put(146,0){\framebox(48,24){\shortstack{ideal\\$0$-prep.}}}
\put(194,12){\line(1,0){10}}
\put(204,6){\makebox(20,12){,}}
\end{picture}
%\vspace{0.3cm}

\vspace{0.5cm}
\begin{picture}(378,24)
\put(0,12){\line(1,0){10}}
\put(10,0){\framebox(48,24){$k$-EC}}
\put(58,12){\line(1,0){10}}
\put(68,0){\framebox(48,24){$k$-meas.}}
\put(116,6){\makebox(20,12){=}}
\put(136,12){\line(1,0){10}}
\put(146,0){\framebox(48,24){$k$-EC}}
\put(194,12){\line(1,0){10}}
\put(204,0){\framebox(48,24){\shortstack{ideal\\$k$-decoder}}}
\put(252,12){\line(1,0){10}}
\put(262,0){\framebox(48,24){\shortstack{ideal\\$0$-meas.}}}
\put(308,6){\makebox(20,12){.}}
\end{picture}
\vspace{0.3cm}

\noindent We note that, for $k=1$, this formulation of exRec-Cor is actually somewhat stronger than the formulation used in Sec.~\ref{sec:level-1} --- here the input to the exRec is unrestricted, while the criterion for correctness used in Sec.~\ref{sec:level-1} stipulates an input to the Rec that has at most one error.

In Sec.~\ref{good-implies-correct} we will prove the important

\medskip
\noindent {\bf Lemma 3. Good implies correct}. {\em Suppose that all types of level-1 rectangles satisfy property exRec-Cor. Then exRec-Cor holds for all types of rectangles at each level $k\ge 1$.} 
\medskip

\noindent
It follows from Lemma 3 that our level-$k$ fault-tolerant simulation of an ideal quantum circuit will succeed if every $k$-exRec is good. To reach this conclusion, we first use the correctness of the final $k$-measurements to replace each $k$-measurement by an ideal $k$-decoder followed by an ideal 0-measurement. Then we use the correctness of the $k$-Recs that simulate the quantum gates to move the ideal $k$-decoders to the left through the circuit, transforming each $k$-Rec to an ideal 0-Ga. Finally, using the correctness of the initial $k$-preparations, we replace each $k$-preparation by preparation of a qubit in the ideally prepared state. These steps show that the ideal circuit and its level-$k$ simulation are equivalent --- both produce precisely the same probability distribution of outcomes for the final measurement. Thus we have

\medskip
\noindent {\bf Lemma 4. Good circuits match ideal answers}. {\em Suppose that all level-1 rectangles satisfy property exRec-Cor. Then if every exRec is good in a level-$k$ simulation of an ideal circuit, the probability distribution for the final measurement outcomes in the simulation is exactly the same as the probability distribution for the final measurement outcomes in the ideal circuit.} 
\medskip

It is implicit in the formulation of these lemmas that we assume {\em classical} computation is reliable, or that the final outcome of the quantum computation is protected robustly in a classical code block rather than decoded to a single bit. If the classical gates are noisy and the outcome is decoded, then a single fault in the final decoding step could cause a 1-measurement to fail. Therefore 1-measurement would not satisfy exRec-Cor, and the lemmas would not apply.

The notion of correctness that we have defined using the ideal decoder corresponds fairly closely to the notion of correctness that we used in our discussion of level-1 simulations, but there are also some important differences that should be noted. For one, our new formulation of correctness has been stated as a property of operations, without any explicit reference to states. This feature is advantageous since  $k$-blocks are typically highly entangled with other $k$-blocks, and the goal of a fault-tolerant simulation is to maintain this entanglement; whether this goal is achieved cannot be judged by examining only the local action of each $k$-Rec on its input blocks.

Also, to see that the new version of exRec-Cor is true at level 1, we need to reformulate the properties 0--4 of the level-1 gadgets listed in Sec.~\ref{sec:level-1}. We will postpone discussing the details of this reformulation until Sec.~\ref{sec:higher-distance}. But one point deserves emphasis now: an additional property also must be assumed to enforce exRec-Cor. Let us say that the state of a 1-block is {\em valid} if the purification of the state can be expanded as in eq.~(\ref{general-error}) where each Pauli operator $E_a$ has weight $w\le 1$, and $|\bar \psi\rangle$ is a codeword. The additional necessary property is:
\begin{description}
\item $0'$. If a 1-EC contains one fault, it takes any input to a valid output.
\end{description}
\noindent The property $0'$ is needed to ensure that, if one of the leading 1-ECs of the 1-exRec has a fault, the input to the 1-Rec ``has at most one error'' all together in all input blocks --- that is, to ensure that the simulated gate, followed by ideal decoding, agrees with ideal decoding followed by an ideal gate. If our distance-3 code is ``perfect'' (if every possible error syndrome point to at most one error), then the property $0'$ is automatic. But by following carefully the principles of quantum fault tolerance, we can build a 1-EC that satisfies $0'$ for any distance-3 code; we will return to this point in Sec.~\ref{sec:ft-general}. In any case, for our proof of the threshold theorem, it will not be necessary to explain how gadgets that satisfy exRec-Cor are constructed at level 1; it will suffice just to know that such gadgets exist. 

We have stressed the applicability of our criterion for correctness to distance-3 codes, but we should also point out that it can be generalized to distance-$(2t+1)$ codes that correct $t$ errors. In that case, we can say that a 1-exRec is bad only if it contains faults in at least $t+1$ locations. Then it is possible to construct fault-tolerant gadgets such that exRec-Cor holds at level 1, and by an inductive proof to establish exRec-Cor at level $k$ as well. This generalization to higher-distance codes will be discussed in Sec.~\ref{sec:higher-distance}.

\section{The quantum threshold theorem}
\label{sec:threshold-theorem}

The crux of the quantum threshold theorem is the proof of Lemma 3 asserting the property exRec-Cor (``good implies correct'') for $k$-exRecs. We will postpone this proof until Sec.~\ref{good-implies-correct}. First, we will explain how Lemma 3 is used to complete the proof of our main theorem. 

How accurate is a level-$k$ fault-tolerant simulation? At the conclusion of the computation, some qubits are measured. Let $\{p_i^{(\rm ideal)}\}$ denote the probability distribution for the measurement outcomes for the ideal circuit, and let $\{p_i^{(\rm actual)}\}$ denote the probability distribution for the measurement outcomes for the actual noisy circuit. We define the {\em error} $\delta$ of the noisy computation as the $L^1$ distance between these two probability distributions:
\begin{equation}
\delta = \parallel p^{(\rm actual)}-p^{(\rm ideal)}\parallel \equiv \sum_i | p_i^{(\rm actual)}-p_i^{(\rm ideal)}|~.
\end{equation}
Now, we know that if every $k$-exRec in the simulation is good, then $p_i^{(\rm actual)}=p_i^{(\rm ideal)}$, and furthermore bad $k$-exRecs are rare if the fault rate is low and $k$ is large. Let us say that the $k$-simulation {\em fails} if any $k$-exRec is bad. Then if the ideal circuit has $L$ locations, and stochastic faults occur independently with probability $\varepsilon < \varepsilon_0$ at each location in the noisy circuit, the probability of failure can be bounded as
\begin{equation}
P_{\rm fail}^{(k)} \le  L \varepsilon^{(k)} \le \varepsilon_0 L\left(\varepsilon/\varepsilon_0\right)^{2^k}~,
\end{equation}
using eq.~(\ref{p-bad-bound}).

By averaging over the fault locations in the level-$k$ fault-tolerant circuit, we find
\begin{equation}
p_i^{(\rm actual)} = \left(1-P_{\rm fail}^{(k)}\right) ~p_i^{(\rm ideal)} + P_{\rm fail}^{(k)} ~p_i^{(\rm fail)}~,
\end{equation}
for some distribution $\{p_i^{(\rm fail)}\}$. Therefore, we find an upper bound on the error
\begin{equation}
\label{accuracy-scaling}
\delta = P_{\rm fail}^{(k)}\cdot \sum_i|p_i^{(\rm fail)} - p_i^{(\rm ideal)}| \le ~2 P_{\rm fail}^{(k)}~\le~ 2\varepsilon_0 L\left(\varepsilon/\varepsilon_0\right)^{2^k}~
\end{equation}
(since the maximal $L^1$ distance between any two probability distributions is 2). Rearranging, we see that error $\delta$ or better can be achieved by choosing
\begin{equation}
\label{2k-bound}
2^k \ge \left({\log\left(2\varepsilon_0L/\delta\right)\over \log\left(\varepsilon_0/\varepsilon\right)}\right)~.
\end{equation}

Suppose that the largest size (number of locations) of any 1-Rec is $\ell$ and that the largest depth (number of time steps) of any 1-Rec is $d$. Then because of the self-similarity of the $k$-Recs, no $k$-Rec can have size larger than $\ell^k=\left(2^k\right)^{\log_2 \ell}$ and no $k$-Rec can have depth larger than $d^k=\left(2^k\right)^{\log_2 d}$. From eq.~(\ref{2k-bound}) we therefore obtain

\medskip
\noindent {\bf Theorem 1. Quantum accuracy threshold for independent stochastic noise}. {\em Suppose that fault-tolerant gadgets can be constructed such that all 1-exRecs obey the property exRec-Cor, and such that $\ell$ is the maximal number of locations in a 1-Rec, $d$ is the maximal depth of a 1-Rec, and $\varepsilon_0^{-1}$ is the maximal number of pairs of locations in a 1-exRec. Suppose that independent stochastic faults occur with probability $\varepsilon < \varepsilon_0$ at each location in a noisy quantum circuit. Then for any fixed $\delta$, any ideal circuit with $L$ locations and depth $D$ can be simulated with error $\delta$ or better by a noisy circuit with $L^*$ locations and depth $D^*$, where
\begin{equation}
L^*=O\left(L(\log L)^{\log_2 \ell}\right)~,\quad D^*= O\left(D(\log L)^{\log_2 d}\right)~.
\end{equation}
}
\medskip

\noindent Theorem 1 is our main result. What is new is the connection between the accuracy threshold and properties of extended rectangles, which are satisfied by suitably designed level-1 gadgets for {\em distance-3} codes. 

An implicit assumption in the statement of the theorem is that either {\em classical} computation is reliable or else the final output of the computation is protected in a classical code block. If classical gates are noisy, and we wish to decode the final outcome to a single bit, then a single fault in the final decoding step could cause an error in the output. For this reason, it would not be possible to construct a 1-measurement that obeys exRec-Cor, and the theorem would not apply. There is still a quantum accuracy threshold if the classical computation is noisy (or if all gates are noisy quantum gates), but the error $\delta$ cannot be arbitrarily small in that case; rather it is limited to $\delta=O(\varepsilon)$. We are also taking it for granted that all input qubits are prepared in standard known states such as $|0\rangle$. Only then can we construct a 1-preparation with property exRec-Cor. If the input includes qubits in unknown states, a single fault in the first encoding step could cause an error in the computation.

%{\em [Should this caveat be stated after each of the theorems that we formulate?]}

As we will discuss in Sec.~\ref{sec:higher-distance}, Theorem 1 can be generalized to codes of higher distance (for which the threshold estimate $\varepsilon_0$ may be lower but the overhead may scale more favorably). For distance-3 codes, our estimate of the accuracy threshold can be further improved; see Sec.~\ref{sec:malignant-pairs}.

\section{Good implies correct}
\label{good-implies-correct}

\subsection{The threshold dance}
To complete the proof of the threshold theorem, it only remains to prove Lemma 3: ``good implies correct.'' The proof will be by induction on the level $k$ of the simulation. We assume that level-1 gadgets have been constructed so that the property exRec-Cor is satisfied. Then we must show that if exRec-Cor holds at level $k$ it also holds at level $k+1$. 

The idea underlying the inductive step is quite clear. We may regard each $(k{+}1)$-gadget as a simulation of the corresponding 1-gadget, where each gate is replaced by a $k$-Rec. Then we invoke the induction hypothesis at level $k$ to justify that this circuit composed of $k$-Recs simulates the 1-gadget accurately, and invoke property exRec-Cor at level 1 to complete the induction step. We have chosen our definitions and properties so that this idea can be realized relatively easily.

Specifically, our proof exploits the recursive construction of the ideal decoder --- decoding of a $(k{+}1)$-block is achieved by first decoding each $k$-subblock to a qubit, and then decoding the resulting 1-block. To show that a $(k{+}1)$-Rec is correct, we view the $(k{+}1)$-decoder as a $k$-decoder acting on each $k$-subblock, followed by the 1-decoder. Repeatedly using the property exRec-Cor at level $k$, we steadily move the $k$-decoders to the left, one $k$-Rec at a time, until they reach the front of the $(k{+}1)$-exRec; thereby we transform the $(k{+}1)$-exRec to a 1-exRec. If the original $(k{+}1)$-exRec is good, so is the resulting 1-exRec (see Sec.~\ref{sec:linearity-of-correctness}); therefore using exRec-Cor at level 1, we can move the 1-decoder to the front of the 1-Rec, transforming it to the corresponding ideal 0-Ga. Finally, we can sweep the $k$-decoders back to the right to the front of the $(k{+}1)$-Rec, replacing the leading 1-EC of the 1-exRec by the original $(k{+}1)$-EC, and reuniting the $k$-decoders with the 1-decoder to reassemble the $(k{+}1)$-decoder.  This completes the demonstration of exRec-Cor at level $(k{+}1)$.

We affectionately refer to this maneuver, in which the $k$-decoders first sweep forward to the front of the $(k{+}1)$-exRec, the 1-decoder follows to the front of the resulting 1-Rec, and then the $k$-decoders sweep backward to rejoin the 1-decoder, as the {\em threshold dance}. See Fig.~\ref{fig:dance}.

\begin{figure}
\begin{center}
\begin{picture}(378,24)
\put(0,6){\makebox(20,12){}}
\put(20,12){\line(1,0){10}}
\put(30,0){\framebox(54,24){$(k{+}1)$-EC}}
\put(84,12){\line(1,0){10}}
\put(94,0){\framebox(54,24){$(k{+}1)$-Rec}}
\put(148,12){\line(1,0){10}}
\put(158,0){\framebox(72,24){$(k{+}1)$-decoder}}
\put(230,12){\line(1,0){10}}
\end{picture}

\vspace{0.5cm}
\begin{picture}(378,24)
\put(0,6){\makebox(20,12){=}}
\put(20,12){\line(1,0){10}}
\put(30,0){\framebox(54,24){$(k{+}1)$-EC}}
\put(84,12){\line(1,0){10}}
\put(94,0){\framebox(54,24){$(k{+}1)$-Rec}}
\put(148,12){\line(1,0){10}}
\put(158,0){\framebox(54,24){$k$-decoders}}
\put(212,12){\line(1,0){10}}
\put(222,0){\framebox(54,24){$1$-decoder}}
\put(276,12){\line(1,0){10}}
\end{picture}

\vspace{0.5cm}
\begin{picture}(378,24)
\put(0,6){\makebox(20,12){=}}
\put(20,12){\line(1,0){10}}
\put(30,0){\framebox(54,24){$k$-decoders}}
\put(84,12){\line(1,0){10}}
\put(94,0){\framebox(54,24){1-EC}}
\put(148,12){\line(1,0){10}}
\put(158,0){\framebox(54,24){1-Rec}}
\put(212,12){\line(1,0){10}}
\put(222,0){\framebox(54,24){$1$-decoder}}
\put(276,12){\line(1,0){10}}
\end{picture}
%\vspace{0.3cm}
%

\vspace{0.5cm}
\begin{picture}(378,24)
\put(0,6){\makebox(20,12){=}}
\put(20,12){\line(1,0){10}}
\put(30,0){\framebox(54,24){$k$-decoders}}
\put(84,12){\line(1,0){10}}
\put(94,0){\framebox(54,24){1-EC}}
\put(148,12){\line(1,0){10}}
\put(158,0){\framebox(54,24){1-decoder}}
\put(212,12){\line(1,0){10}}
\put(222,0){\framebox(54,24){0-Ga}}
\put(276,12){\line(1,0){10}}
\end{picture}

\vspace{0.5cm}
\begin{picture}(378,24)
\put(0,6){\makebox(20,12){=}}
\put(20,12){\line(1,0){10}}
\put(30,0){\framebox(54,24){$(k{+}1)$-EC}}
\put(84,12){\line(1,0){10}}
\put(94,0){\framebox(54,24){$k$-decoders}}
\put(148,12){\line(1,0){10}}
\put(158,0){\framebox(54,24){1-decoder}}
\put(212,12){\line(1,0){10}}
\put(222,0){\framebox(54,24){0-Ga}}
\put(276,12){\line(1,0){10}}
\end{picture}

\vspace{0.5cm}
\begin{picture}(378,24)
\put(0,6){\makebox(20,12){=}}
\put(20,12){\line(1,0){10}}
\put(30,0){\framebox(54,24){$(k{+}1)$-EC}}
\put(84,12){\line(1,0){10}}
\put(94,0){\framebox(72,24){$(k{+}1)$-decoder}}
\put(166,12){\line(1,0){10}}
\put(176,0){\framebox(54,24){0-Ga}}
\put(230,12){\line(1,0){10}}
\end{picture}
\vspace{0.3cm}
\end{center}
%\vspace*{0.1truein}
\fcaption{The {\em threshold dance}, shown schematically here, is the pivotal maneuver in the inductive proof that a good exRec is correct.}
\label{fig:dance}
\end{figure}

\subsection{Bad rectangles as simulated faults}
\label{sec:linearity-of-correctness}

The idea behind the argument sketched above is that, by moving the $k$-decoders to the left through a good $(k{+}1)$-exRec, we transform it to a good 1-exRec. It is clear that the property exRec-Cor at level $k$ allows us to move a $k$-decoder through the $k$-Rec contained in a good $k$-exRec, transforming the $k$-Rec to an ideal 0-Ga.  But a good $(k{+}1)$-exRec might also contain a bad $k$-exRec, or a consecutive pair of nonindependent bad $k$-exRecs. Intuitively, we should be able to regard each $k$-Rec contained in a bad $k$-exRec as a simulation of a faulty 0-Ga contained in the 1-exRec, and as suggested in Sec.~\ref{subsec:goodness}, we expect that a consecutive pair of nonindependent bad $k$-exRecs can be regarded as a simulation of a pair of 0-Ga's such that only the later of the two is faulty. Then the 1-exRec becomes a circuit with a single fault, and is good.

To complete the argument, then, we need to explain what happens as the $k$-decoder sweeps past a bad $k$-exRec or a nonindependent pair of consecutive $k$-exRecs. Let us first discuss how we treat the case of two nonindependent bad $k$-exRecs.

\subsubsection{Nonindependent pairs of bad $k$-exRecs}
\label{subsubsec:nonindependent-pairs}

By definition, if two bad $k$-exRecs comprise a nonindependent consecutive pair, the earlier of the two consecutive $k$-exRecs is good when the shared $k$-EC is removed. Therefore, when the $k$-decoders, as they migrate to the left, reach the consecutive pair of bad $k$-exRecs, we may proceed as follows. The $k$-decoders first leap past the later of the two bad $k$-exRecs --- past the entire $k$-exRec, rather than past the $k$-Rec that this $k$-exRec contains. As explain in Sec.~\ref{subsubsec:transforming} below, the bad $k$-exRec is thereby effectively replaced by a faulty 0-Ga. Next we are to move the $k$-decoders another step to the left, past the $k$-Rec contained in the earlier of the two bad $k$-exRecs. Because the $k$-decoder leapt past the shared $k$-EC in the previous step, the earlier $k$-Rec has been truncated --- it is now missing the trailing $k$-EC that it shares with the later bad $k$-exRec. 

Actually, to enforce the independence of the bad $k$-exRec and the truncated $k$-exRec that precedes it, we must ensure that no level-0 gate is contained in both; therefore we should specify carefully where the boundary lies between the later untruncated $k$-exRec and the earlier truncated $k$-exRec. We define this boundary by excluding from the earlier truncated $k$-exRec every 0-Ga that is ``contained in'' the later $k$-exRec. For example, the opening $(k{-}1)$-exRecs ``contained in'' the later $k$-exRec have leading $(k{-}1)$-ECs that are considered to be part of the later $k$-exRec, and are not included in the earlier truncated $k$-exRec. Similarly, at the next level down, the opening $(k{-}2)$-exRecs ``contained in'' the opening $(k{-}1)$-exRecs of the later $k$-Rec have leading $(k{-}2)$-ECs that are considered to be part of the later $k$-exRec and are not included in the earlier truncated $k$-exRec. And so on at all lower levels. 

Now we need to consider what happens when we move the $k$-decoders further to the left, past the earlier truncated $k$-exRec. 
For this purpose, we  can use an obvious identity that follows from the definition of the $j$-decoder: 

\vspace{0.5cm}
\begin{picture}(292,24)
\put(0,12){\line(1,0){10}}
\put(10,0){\framebox(48,24){\shortstack{ideal\\$j$-decoder}}}
\put(58,12){\line(1,0){10}}
\put(68,6){\makebox(20,12){=}}
\put(88,12){\line(1,0){10}}
\put(98,0){\framebox(48,24){\shortstack{ideal\\$j$-EC}}}
\put(146,12){\line(1,0){10}}
\put(156,0){\framebox(48,24){\shortstack{ideal\\$j$-decoder}}}
\put(204,12){\line(1,0){10}}
\put(214,6){\makebox(20,12){.}}
\end{picture}
\vspace{0.3cm}

\noindent
Invoking this identity, we may replace the $j$-ECs that have been amputated ($j=1,2,3,\dots,k$) by ideal $j$-ECs (ones with no faults). This restoration of the amputated ECs proceeds level by level: first the final truncated 1-exRecs are augmented by adding ideal trailing 1-ECs, then the final truncated 2-exRecs are augmented by adding ideal trailing 2-ECs, and so on at all higher levels. At each step, we can justify adding the ideal $j$-ECs by observing that the $k$-decoder can be realized as $j$-decoders applied to all $j$-subblocks, followed by a $(k{-}j)$-decoder.

At this point, the $k$-decoders are behind the complete $k$-exRec with restored $k$-ECs. Furthermore, if the truncated $k$-exRec is good (does not contain two nonindependent bad $(k{-}1)$-exRecs), then the completed $k$-exRec is also good, because there are no bad $(j{-}1)$-exRecs in the $j$-ECs that we inserted ($j=1,2,3,\dots,k$). Therefore, using exRec-Cor at level $k$, we may move the $k$-decoders to the left of the $k$-Ga, transforming it to an ideal 0-Ga. We conclude, in other words, that the property exRec-Cor for complete $k$-exRecs implies that exRec-Cor is also true for truncated $k$-exRecs. 

Thus we have shown, as desired, that as the $k$-decoders move to the left, a nonindependent pair of bad $k$-Recs is transformed to an ideal 0-Ga followed by a faulty 0-Ga. If, on the other hand, the truncated $k$-exRec is bad, we move the $k$-decoders past the entire truncated $k$-exRec --- the pair of independently bad $k$-exRecs is transformed to a pair of faulty 0-Ga's, and the $k$-exRecs that immediately precede the bad pair become truncated.

\subsubsection{Transforming a bad $k$-exRec to a faulty 0-Ga}
\label{subsubsec:transforming}

It still remains to justify moving the $k$-decoders to the left through a bad $k$-exRec (or a bad truncated $k$-exRec), transforming the $k$-exRec to a faulty implementation of the ideal 0-Ga. Here an annoying technical point arises. Our induction hypothesis prescribes no special properties of the bad rectangles, which therefore should be regarded as arbitrary operations (trace-preserving completely positive maps). Thus a bad $k$-exRec cannot necessarily be regarded as a level-$k$ simulation of a well-defined (faulty) level-0 gate, because the bad $k$-exRec might map distinct valid encodings of the same ideal input to valid encodings of two distinct ideal outputs. (We say that a state $\rho$ of a $k$-block is a {\em valid} encoding of an ideal pure state $|\psi\rangle$ of a qubit if the ideal decoder maps $\rho$ to $|\psi\rangle$.) If we attempt to move the ideal $k$-decoder from behind a bad $k$-exRec to in front of it, and in so doing transform the bad $k$-exRec to a faulty 0-Ga, the snag is that the particular faulty 0-Ga we obtain may depend on the syndrome that the ideal decoder measures. For example, at level $k=1$, errors due to faults in the 1-exRec might combine differently with an input error on the first qubit in the 1-block than with an input error on the second qubit in the 1-block; then applying the ideal 1-decoder to the output of the 1-exRec in these two cases might yield different output states, even though applying the ideal 1-decoder to the input yields the same state in both cases.

This observation indicates that if we are to transform a bad $k$-exRec to a level-0 fault, in doing so we cannot completely disregard the error syndromes of the incoming $k$-blocks. Nevertheless the transformation is possible, if we allow the syndrome to assume the role of an ``environment'' that interacts with the level-0 data whenever a 0-fault occurs. 

Up until now, we have prescribed that the ideal $k$-decoder discards the error syndrome after performing the error-recovery operation, but for analyzing the action of bad $k$-exRecs on the data, it will be convenient to consider the error syndrome to be part of the output from the ideal decoder. For clarity, we will refer to the $k$-decoder that retains the syndrome information as the $k$-$^*$decoder. Let $|\bar \psi\rangle$ denote the ideal encoding in a level-$k$ concatenated code of the single-qubit state $|\psi\rangle$, and let $\{E_i\}$ denote the set of correctable Pauli errors acting on the $k$-block; then the action of the $k$-$^*$decoder (denoted ${\cal D}$) is
\begin{equation}
\label{decoder*-action}
 {\cal D}: E_i|\bar\psi\rangle \mapsto |\psi\rangle\otimes |i\rangle~,
\end{equation}
where $|i\rangle$ denotes the state of the register that records the syndrome. For a perfect code, (or for a perfect code of the CSS type), the states $\{E_i|\bar\psi\rangle\}$ are a complete basis, and eq.~(\ref{decoder*-action}) completely characterizes the action of the $k$-$^*$decoder on the $k$-block. If the code is not perfect, this action can be extended to a larger set of Pauli errors that includes some non-correctable errors, such that the states $\{E_i|\bar\psi\rangle\}$  are mutually orthogonal and complete. Then ${\cal D}$ is invertible, and its inverse ${\cal D}^{-1}$ is a $k$-$^*$encoder, which takes the input qubit $|\psi\rangle$ and input syndrome $|i\rangle$ to the encoded state $|\bar \psi\rangle$ with error $E_i$. An ideal $k$-decoder is just the $k$-$^*$decoder ${\cal D}$ followed by disposal of the output syndrome:

%%% decoder and star-decoder
\vspace{0.5cm}
\begin{picture}(152,48)
\put(0,36){\line(1,0){10}}
\put(10,24){\framebox(48,24){\shortstack{ideal\\$k$-decoder}}}
\put(58,36){\line(1,0){10}}
\put(68,30){\makebox(20,12){=}}
\put(88,36){\line(1,0){10}}
\put(98,24){\framebox(24,24){${\cal D}$}}
\put(122,36){\line(1,0){10}}
\put(110,24){\line(0,-1){15}}
\put(110,9){\line(1,0){10}}
\put(114,3){\makebox(20,12){$\times$}}
\put(132,30){\makebox(20,12){.}}
\end{picture}
\vspace{0.3cm}

Suppose we assume the induction hypothesis, that the property exRec-Cor holds at level $k$. Therefore, if a $k$-exRec is good, we can move a $k$-decoder that follows the $k$-exRec to the left, past the $k$-Rec contained in the good $k$-exRec, converting the $k$-Rec to the corresponding ideal 0-Ga. The same property still holds when we move the $k$-$^*$decoder to the left instead. As far as the data is concerned, there is no difference between the $k$-decoder (which discards the syndrome that it measures) and the $k$-$^*$decoder (which retains the syndrome). The only question is: what happens to the syndrome when the $k$-$^*$decoder moves left?

For simplicity, consider a good $k$-exRec that acts on a single input $k$-block. Moving the $k$-$^*$ decoder ${\cal D}$ past the noisy $k$-Rec ${\cal M}$ generates an operation ${\cal D} {\cal M}{\cal D}^{-1}$ whose input consists of a single qubit and a syndrome (here our operator ordering convention is that the operator furthest to the right acts first), and the action of ${\cal D} {\cal M}{\cal D}^{-1}$ on the reduced state of the qubit alone, for any input, is the ideal unitary gate ${\cal M}_{\rm ideal}$. Therefore, the action of ${\cal D} {\cal M}{\cal D}^{-1}$ on the qubit must be uncorrelated with its action on the syndrome, for otherwise the qubit would decohere. That is, ${\cal D} {\cal M}{\cal D}^{-1}$ is a tensor product
\begin{equation}
{\cal D} {\cal M}{\cal D}^{-1}= {\cal M}_{\rm ideal}\otimes {\cal M}_{\rm syndrome}
\end{equation}
where ${\cal M}_{\rm syndrome}$ is a trace-preserving operation acting on the syndrome alone that depends on the details of the noise in the $k$-Rec:

%%% exRec-Cor with syndrome propagation
\vspace{0.5cm}
\begin{picture}(324,48)
\put(0,36){\line(1,0){10}}
\put(10,24){\framebox(30,24){$k$-EC}}
\put(40,36){\line(1,0){10}}
\put(50,24){\framebox(48,24){$k$-Rec}}
\put(98,36){\line(1,0){10}}
\put(108,24){\framebox(24,24){${\cal D}$}}
\put(132,36){\line(1,0){10}}
\put(120,24){\line(0,-1){15}}
\put(120,9){\line(1,0){22}}
\put(142,30){\makebox(20,12){=}}
\put(162,36){\line(1,0){10}}
\put(172,24){\framebox(30,24){$k$-EC}}
\put(202,36){\line(1,0){10}}
\put(212,24){\framebox(24,24){${\cal D}$}}
\put(236,36){\line(1,0){10}}
\put(246,24){\framebox(48,24){\shortstack{ideal\\0-Ga}}}
\put(294,36){\line(1,0){10}}
\put(224,24){\line(0,-1){15}}
\put(224,9){\line(1,0){22}}
\put(246,0){\framebox(48,18){${\cal M}_{\rm syndrome}$}}
\put(294,9){\line(1,0){10}}
\put(304,30){\makebox(20,12){.}}
\end{picture}
\vspace{0.3cm}

\noindent
In effect, then, the $k$-Recs contained in good $k$-exRecs perform two independent quantum computations in parallel: the ideal processing of the encoded data, and the noisy processing of the syndrome. For our purposes, the details of how the syndrome is processed are not relevant; all that matters is that a good noisy circuit processes the data and the syndrome independently, so that in principle we can propagate the syndrome through the good noisy circuit without interfering with the ideal evolution of the data. Since good truncated $k$-exRecs are also correct, as explained in Sec.~\ref{subsubsec:nonindependent-pairs}, they too process the encoded data and the syndrome independently.

For a bad $k$-exRec, on the other hand, the processing of the encoded data and of the syndrome are not independent. If ${\cal N}$ denotes the action of the bad $k$-exRec on the data and syndrome, then we may write ${\cal D}{\cal N} = {\cal F}{\cal D}$, where 
\begin{equation}
{\cal F}= {\cal D}{\cal N}{\cal D}^{-1} 
\end{equation}
can be regarded as the level-0 fault simulated by the bad $k$-exRec; diagrammatically,

%%% bad exRec identity
\vspace{0.5cm}
\begin{picture}(312,48)
\put(0,36){\line(1,0){10}}
\put(10,24){\framebox(48,24){\shortstack{bad\\$k$-exRec}}}
\put(58,36){\line(1,0){10}}
\put(68,24){\framebox(24,24){${\cal D}$}}
\put(92,36){\line(1,0){10}}
\put(80,24){\line(0,-1){15}}
\put(80,9){\line(1,0){22}}
\put(102,30){\makebox(20,12){=}}
\put(122,36){\line(1,0){10}}
\put(132,24){\framebox(24,24){${\cal D}$}}
\put(156,36){\line(1,0){10}}
\put(144,24){\line(0,-1){15}}
\put(144,9){\line(1,0){34}}
\put(178,24){\line(0,-1){15}}
\put(166,24){\framebox(48,24){\shortstack{faulty\\0-Ga}}}
\put(214,36){\line(1,0){10}}
\put(202,24){\line(0,-1){15}}
\put(202,9){\line(1,0){22}}
\put(224,30){\makebox(20,12){,}}
\end{picture}
\vspace{0.3cm}

\noindent where

%%% definition of faulty 0-Ga
\vspace{0.5cm}
\begin{picture}(244,48)
\put(0,36){\line(1,0){10}}
\put(10,24){\framebox(48,24){\shortstack{faulty\\0-Ga}}}
\put(58,36){\line(1,0){10}}
\put(0,9){\line(1,0){22}}
\put(22,24){\line(0,-1){15}}
\put(46,24){\line(0,-1){15}}
\put(46,9){\line(1,0){22}}
\put(68,30){\makebox(20,12){=}}
%
%\put(122,36){\line(1,0){10}}
\put(88,36){\line(1,0){10}}
\put(98,24){\framebox(24,24){${\cal D}^{-1}$}}
\put(88,9){\line(1,0){22}}
\put(110,24){\line(0,-1){15}}
\put(122,36){\line(1,0){10}}
\put(132,24){\framebox(48,24){\shortstack{bad\\$k$-exRec}}}
\put(180,36){\line(1,0){10}}
\put(190,24){\framebox(24,24){${\cal D}$}}
\put(214,36){\line(1,0){10}}
\put(202,24){\line(0,-1){15}}
\put(202,9){\line(1,0){22}}
\put(224,30){\makebox(20,12){.}}
\end{picture}
\vspace{0.3cm}

\noindent
In contrast with the ideal level-0 gate simulated by the $k$-Rec contained in a good $k$-exRec, the faulty level-0 operation simulated by a bad $k$-exRec depends on the syndrome that is input to the $k$-$^*$encoder. Moving the $k$-$^*$decoder left past a bad $k$-exRec transforms the $k$-exRec to an operation that actually acts collectively on the level-0 data and the syndrome. In effect, the syndrome functions as an ``environment'' that interacts with the data at the locations where level-0 faults occur. In a level-$k$ circuit that contains $s$ independent bad $k$-exRecs, moving the $k$-$^*$decoder left through the circuit transforms it to a circuit of 0-Ga's, with $s$ faults. Schematically, two bad $k$-exRecs, with an intervening good $k$-circuit, become transformed like this: 

%%% good between two bad circuits
\vspace{0.5cm}
\begin{picture}(444,48)
\put(0,36){\line(1,0){10}}
\put(10,24){\framebox(40,24){\shortstack{bad\\$k$-exRec}}}
\put(50,36){\line(1,0){10}}
\put(60,24){\framebox(52,24){\shortstack{good\\$k$-circuit}}}
\put(112,36){\line(1,0){10}}
\put(122,24){\framebox(40,24){\shortstack{bad\\$k$-exRec}}}
\put(162,36){\line(1,0){10}}
\put(172,24){\framebox(24,24){${\cal D}$}}
\put(196,36){\line(1,0){10}}
\put(184,24){\line(0,-1){15}}
\put(184,9){\line(1,0){22}}
\put(206,30){\makebox(20,12){=}}
\put(226,36){\line(1,0){10}}
\put(236,24){\framebox(24,24){${\cal D}$}}
\put(260,36){\line(1,0){10}}
\put(248,24){\line(0,-1){15}}
\put(248,9){\line(1,0){34}}
\put(282,24){\line(0,-1){15}}
\put(270,24){\framebox(36,24){\shortstack{faulty\\0-Ga}}}
\put(306,36){\line(1,0){10}}
\put(294,24){\line(0,-1){15}}
\put(294,9){\line(1,0){22}}
\put(316,0){\framebox(52,18){${\cal M}_{\rm syndrome}$}}
\put(368,36){\line(1,0){10}}
\put(316,24){\framebox(52,24){\shortstack{ideal\\0-circuit}}}
\put(368,9){\line(1,0){18}}
\put(386,24){\line(0,-1){15}}
\put(378,24){\framebox(36,24){\shortstack{faulty\\0-Ga}}}
\put(414,36){\line(1,0){10}}
\put(402,24){\line(0,-1){15}}
\put(402,9){\line(1,0){22}}
\put(424,30){\makebox(20,12){.}}
\end{picture}
\vspace{0.3cm}

Under this transformation, the syndrome becomes an effective environment that remains isolated from the data during the ideal 0-Ga's, but interacts with the data during 0-faults. Faults that share an environment can be strongly correlated with one another, but fortunately this feature does not interfere with the successful execution of the threshold dance. In the property exRec-Cor, the goodness of an exRec is determined only by the locations of the faults in the exRec ---  once the fault locations are determined, the action of the faults at those locations can be arbitrary. In particular, a good 1-exRec is correct even if the 0-faults contained in the 1-exRec share a quantum memory.

Now, for the inductive step, we are to assume the property exRec-Cor at level $k$, and we are to prove exRec-Cor at level $k{+}1$. A $(k{+}1)$-decoder following a $(k{+}1)$-exRec can be realized as $k$-decoders acting on $k$-blocks, followed by a 1-decoder, and each $k$-decoder can be regarded as a $k$-$^*$decoder whose output syndrome is discarded. When the $k$-$^*$decoders sweep left to the front of a good $(k{+}1)$-exRec, they transform it to a good 1-exRec.  The 0-faults in this 1-exRec share access to an environment (the syndrome), but otherwise the evolution of the environment is independent of the evolution of the data. The good 1-exRec is correct, so we can move the 1-decoder left past the 1-Rec, transforming it to the corresponding ideal 0-Ga. Finally, we move the $k$-$^*$decoders back to the right to join the 1-decoder. The outputs of these $k$-$^*$ decoders are discarded, so they are $k$-decoders which together with the 1-decoder reconstitute the $(k{+}1)$-decoder; this completes the proof of the inductive step.

Straightforward modifications of this argument apply to $(k{+}1)$-preparation (for which there are no leading ECs) and $(k{+}1)$-measurement (for which there is no output block).

We have considered $(k{+}1$)-exRecs that contain several embedded bad $k$-exRecs in order to emphasize that this argument can be applied to codes of any odd distance $2t+1$, where a good $(k{+}1)$-ex-Rec is defined to contain no more than $t$ independent bad $k$-exRecs. Higher-distance codes will be further discussed in Sec.~\ref{sec:higher-distance}.

\section{Improving the threshold estimate}
\label{sec:malignant-pairs}

\subsection{Benign and malignant sets of locations}
In the threshold theorem as formulated in Sec.~\ref{sec:threshold-theorem}, we have estimated the accuracy threshold by counting all pairs of locations in the largest 1-exRec. But that estimate is too pessimistic, because there are many pairs of locations in the 1-exRec such that arbitrary faults at both locations do not cause the 1-Rec to be incorrect. Can we derive a sharper estimate using this observation?

Let us say that a set of locations in a 1-exRec is {\em benign} if the 1-Rec contained in the 1-exRec is correct for arbitrary faults occuring at the locations in the set. If the set of locations is not benign it is {\em malignant}. Using a distance-3 code and fault-tolerant gadgets, we can ensure that any set containing only one location is benign. We used this property to prove the threshold theorem in Sec.~\ref{sec:threshold-theorem}. But there are many other benign sets of locations, a fact we can exploit to obtain improved rigorous estimates of the accuracy threshold.

Our analysis must take into account that overlapping bad $k$-rectangles need not be independent for $k\ge 1$. We may relax our definition of goodness: 

\medskip
\noindent {\bf Definition. Goodness and Badness (revised)}. {\em A 1-exRec is {\em bad} if it contains faults at a malignant set of locations; if it is not bad it is {\em good}. For $k>1$, a $k$-exRec is {\em bad} if it contains independent bad $(k{-}1)$-exRecs at a malignant set of locations; if it is not bad it is {\em good}.}

\medskip

\noindent Then exRec-Cor is true at level-1 by definition. Furthermore, with this definition of goodness, the inductive proof of exRec-Cor at level $k+1$ (via the threshold dance) proceeds simply. What makes this definition useful is that the arguments in Sec.~\ref{sec:linearity-of-correctness} show that when $k$-$^*$decoders sweep from behind a good $(k{+}1)$-exRec to in front of it, the $(k{+}1)$-exRec is transformed to a good 1-exRec.

\subsection{Counting malignant pairs}
Enumerating all malignant sets of locations in a 1-exRec would be a combinatoric challenge. But it is a worthwhile and manageable task to count all malignant {\em pairs} of locations. 

We will outline how this counting can be done for a perfect distance-3 code of the CSS type, like the 7-qubit Steane code; further details are provided in Sec.~\ref{sec:explicit}. Suppose we pick a pair of locations in the 1-exRec, and we wish to test whether this pair of locations is malignant. First we note that although in our error model we allow the faults at specified locations to be chosen adversarially, it suffices to test pairs of Pauli faults to check whether a pair of locations is benign, since an arbitrary fault can be expanded in terms of Pauli operators. We may therefore consider replacing a faulty single-qubit 0-Ga by one of the four Pauli operators $\{I,X,Y,Z\}$, or replacing a faulty two-qubit 0-Ga by one of the sixteen tensor products of Pauli operators in $\{I,X,Y,Z\}\otimes \{I,X,Y,Z\}$. (Equivalently, we may consider inserting Pauli operators other than the identity right after or right before the gates.)

Furthermore, we can re-express the criterion for correctness of a 1-Rec in terms of the propagation of Pauli errors through the 1-Rec. For a perfect distance-3 CSS code, we can choose a basis for a 1-block, such that each element of the basis deviates from the code space by at most one $X$ error and at most one $Z$ error. With the Pauli faults fixed at a particular pair of locations, we consider Pauli errors afflicting the input to the 1-exRec, with at most one $X$ and at most one $Z$ acting at arbitrary positions in each input 1-block. We propagate this input error through the leading 1-ECs of the 1-exRec, to find the Pauli error for the output of the 1-ECs, which is the input to the 1-Rec. By applying a logical $X$ and/or logical $Z$ as needed, the error in the input to the 1-Rec can also be expressed as at most one $X$ and at most one $Z$. Then we propagate this error acting on its input through the 1-Rec to find the corresponding error acting on the output of the 1-Rec. If, within any output block, there are two or more $X$ errors acting on the block, or two or more $Z$ errors, then the pair of locations we are testing has been found to be potentially malignant. (If the 1-Rec contains a 0-Ga that is not in the Clifford group, then the output error might not be a Pauli error, but we can check whether the error has weight higher than 1.)

With the fault locations fixed, we conduct this test for each possible choice of the error acting on the input to the 1-exRec, and for each possible choice of the Pauli faults acting at the fixed fault locations. If for each output block there is no more than one $X$ error and no more than one $Z$ error acting on the block for all such choices, then the tested pair of locations is benign.

This analysis can be usefully partitioned into consideration of various cases. It is obvious that the pair is benign if both fault locations are in one of the leading 1-ECs, since then the 1-Rec has no faults. If each of two different leading 1-ECs has a fault (for a 1-Rec that acts on two or more input blocks), then again the 1-Rec has no faults, but we must check whether the 1-Ga propagates an error from one block to the other. The most delicate case is when there is one fault in one of the leading 1-ECs and one in the 1-Rec. This case can be further divided into subtasks. The location of the first fault, together with the input error, determines the error in the output from the leading 1-EC. Then, with the error in the input to the 1-Rec fixed, we can determine whether a second fault at a particular location in the 1-Rec causes multiple errors in the output.

\subsection{Refined calculation of the failure probability}
Now we should consider how the likelihood of a bad $k$-exRec is affected by our revised definition of goodness. We denote the probability that a $k$-exRec is bad by $\varepsilon^{(k)}$. For a $k$-exRec to be bad, either independent bad $(k{-}1)$-exRecs occur at a malignant pair of locations, or else there must be independent bad $(k{-}1)$-exRecs at three or more locations (in the latter case it might be that no two of the locations form a malignant pair). Therefore, since the independent bad $(k{-}1)$-exRecs may be regarded as statistically independent, an upper bound on $\varepsilon^{(k)}$ is
\begin{equation}
\label{malignant-recursion}
\varepsilon^{(k)} \le A \left(\varepsilon^{(k{-}1)}\right)^2 + B\left(\varepsilon^{(k{-}1)}\right)^3~,
\end{equation} 
where $A$ is the number of malignant pairs of locations, and $B$ is the total number of ways to choose three locations (where it is not required that at least two of the three form a malignant pair). 

It follows from eq.~(\ref{malignant-recursion}) that
\begin{equation}
\varepsilon^{(k)} \le A' \left(\varepsilon^{(k{-}1)}\right)^2~,
\end{equation} 
where
\begin{equation}
\varepsilon \le \varepsilon_0 = \left(A'\right)^{-1}~,
\end{equation}
is our threshold estimate and 
\begin{equation}
\label{better-threshold}
A'= A + B\varepsilon_0=A +B/A'\quad  \Rightarrow \quad A'=\frac{1}{2}A\left(1 +\sqrt{1+4B/A^2}\right)~.
\end{equation}
Hence we have

\medskip
\noindent {\bf Theorem 2. Quantum accuracy threshold for independent stochastic noise (revised)}. {\em Suppose that fault-tolerant gadgets can be constructed such that all 1-exRecs obey the property exRec-Cor, and such that $\ell$ is the maximal number of locations in a 1-Rec, and $d$ is the maximal depth of a 1-Rec. Let $\varepsilon_0$ be the minimal value of $A'^{-1}$ for any 1-exRec, where $A'$ is given by eq.~(\ref{better-threshold}), A is the number of malignant pairs of locations in the 1-exRec, and B is the total number of ways to choose three locations in the 1-exRec. Suppose that independent stochastic faults occur with probability $\varepsilon < \varepsilon_0$ at each location in a noisy quantum circuit. Then for any fixed $\delta$, any ideal circuit with $L$ locations and depth $D$ can be simulated with error $\delta$ or better by a noisy circuit with $L^*$ locations and depth $D^*$, where
\begin{equation}
L^*=O\left(L(\log L)^{\log_2 \ell}\right)~,\quad D^*= O\left(D(\log L)^{\log_2 d}\right)~.
\end{equation}
}
\medskip

Sometimes, it is convenient to design gadgets that are {\em nondeterministic} --- the gadgets include subroutines in which certain ancilla states are prepared and verified, where the ancilla is discarded if the verification test fails. In that case, strictly speaking a 1-exRec contains many ancilla preparations performed in parallel, where just one of these ancillas is accepted and used in the circuit that follows. It would be overly pessimistic to include all the locations in the preparation and verification of rejected ancillas in the estimate of $B$ used in eq.~(\ref{better-threshold}), and it is possible to perform a refined count from which many sets of three locations can be excluded. Alternatively, by estimating the probability that an ancilla is accepted and by using Bayes' rule, we can bound the probability of failure for each type of exRec, given that all the ancillas used in that exRec are successfully verified. In Sec.~\ref{subsec:results} we will use the latter method in conjunction with Theorem 2 to obtain an explicit lower bound on the quantum accuracy threshold: $\varepsilon_0 \ge 2.73\times 10^{-5}$.

%--------------------------------------------------------------------------------------------%
\section{Construction of fault-tolerant gadgets: generalities}
\label{sec:ft-general}

The threshold theorem proved in Sec.~\ref{sec:threshold-theorem} is premised on the existence of level-1 gadgets (based on a distance-3 code) that obey the property exRec-Cor. In this section we will explain how such gadgets can be constructed. Then in Sec.~\ref{sec:explicit} we will use one such construction to obtain an explicit estimate of the quantum accuracy threshold.

We have seen in Sec.~\ref{sec:level-1} and \ref{sec:recursive} that exRec-Cor at level 1 holds for level-1 error correction and gate gadgets with the following properties:
\begin{description}
\item 0. If a 1-EC contains no fault, it takes any input to an output in the code space.
\item $0'$. If a 1-EC contains one fault, it takes any input to a valid output. (The state of a level-1 block is ``valid'' if it deviates from the code space by the action of a weight-1 operator.) 
\item 1. If a 1-EC contains no fault, it takes an input with at most one error to an output with no errors.
\item 2. If a 1-EC contains at most one fault, it takes an input with no errors to an output with at most one error.
\item 3. If a 1-Ga contains no fault, it takes an input with at most one error to an output with at most one error in each output block.
\item 4. If a 1-Ga contains at most one fault, it takes an input with no errors to an output with at most one error in each output block. 
\end{description}

\noindent
Therefore, it will suffice to verify that the 1-gadgets satisfy properties 0--4.

%----------------------------------%
\subsection{Stabilizer codes}

Before proceeding to explicit gadget constructions, we briefly review the theory of binary stabilizer codes \cite{att,gott_stab}, which are well suited for applications to fault-tolerant quantum computing. The code space of a stabilizer code of length $n$ is the simultaneous eigenspace of a set $\mathcal{G}$ of commuting Pauli operators; these operators generate the code's \emph{stabilizer} $\mathcal{S}$, an abelian subgroup of the $n$-qubit Pauli group $\mathcal{C}_1^{(n)}= \{\{\pm 1,\pm i\}\cdot \{I,X,Y,Z\}^{\otimes n}\}$. Up to an overall phase, an element $P_a$ of $\mathcal{C}_1^{(n)}$ can be labeled by a length-$2n$ binary vector $a$, where  
\begin{equation}
     P_a = \bigotimes_{i=1}^n \; X^{a_i} \bigotimes_{i=1}^n Z^{a_{i{+}n}} \, ,
\end{equation}
and $a_i$ denotes the $i$th component of $a$. 
Two elements $P_a$ and $P_b$ of $\mathcal{C}_1^{(n)}$ obey the commutation relation
\begin{equation}
\label{commrel}
P_a P_b = (-1)^{a\Lambda b^T} P_b P_a \, ,
\end{equation}
\noindent where 
\begin{equation}
\Lambda = \left( \begin{array}{cc}
                                      \mathbf{0}_n & \mathbf{I}_n \\
                                      \mathbf{I}_n & \mathbf{0}_n
                     \end{array}
                     \right) \, ;
\end{equation}
\noindent here $\mathbf{0}_n $ and $\mathbf{I}_n$ are the zero and identity $n \times n$ matrices respectively. 

For a length-$n$ stabilizer code with $k$ encoded qubits, the generating set $\mathcal{G}=\{G(1),G(2),\dots G(n-k)\}$ of the stabilizer $\mathcal{S}$ contains $n-k$ independent commuting elements of $\mathcal{C}_1^{(n)}$. The set $\mathcal{G}$ can be represented by the $(n-k)\times 2n$ binary matrix 
\begin{equation}
 G \equiv \left( \begin{array}{c}
                                      g(1)\\
                                      g(2)\\
                                      \vdots\\
                                      g(n-k)
                     \end{array}
                     \right) \, ;
\label{genmat}
\end{equation}
here $g(i)$ is the binary vector labeling the Pauli operator $G(i)$.
Each $G(i)$ squares to the identity, and has eigenvalues $\pm 1$ with equal degeneracy; thus the $2^n$-dimensional $n$-qubit Hilbert space decomposes into $2^{n-k}$ disjoint eigenspaces, each of dimension $2^{k}$. Each subspace can be labeled by a $(n-k)$-bit binary vector $e$, where the eigenvalue of $G(i)$ is $(-1)^{e_i}$, for $i=1,\dots, n-k$. This vector $e$ is called the {\em syndrome} of the subspace; the code space, which is the simultaneous eigenspace with eigenvalue one of all the generators, has syndrome $e=(0 0 \dots 0)$. Using Eq.$\,$(\ref{commrel}), we see that the action of an Pauli operator $P_a$ on the code space changes the syndrome to the value $e=a \Lambda G^T$.

For a {\em nondegenerate} stabilizer code that corrects $t$ errors, all $P_a$ with weight $w\le t$ take the code space to mutually orthogonal subspaces with distinct syndromes; thus, under the assumption that no more than $t$ errors occured, each syndrome $e$ points to a unique Pauli operator. In error correction, the syndrome $e$ is measured, and then $P_a^\dagger$ is applied, where $P_a$ is the unique Pauli operator with weight no more than $t$ such that $e=a \Lambda G^T$. If the code is {\em degenerate}, then $P_a$ may not be unique, but each $P_a$ of weight up to $t$ satisfying $e=a \Lambda G^T$ is equally effective in correcting the error.

%---------------------------------------------------%
\subsection{Fault-tolerant error correction}

We wish to construct a level-1 error correction gadget (1-EC) that satisfies properties 0--2. The key goal guiding the construction is property 2 --- we must assure that an 1-EC with a single fault, when applied to an input with no errors, produces an output with only one error.

\subsubsection{Syndrome measurement using cat states}
\label{sec:cat-state-method}
The error correction consists of syndrome measurement followed by a recovery step. How can the syndrome be measured? Putting aside for the moment any concerns about fault tolerance, there is a general method for constructing a quantum circuit for measuring an $n$-qubit unitary operator $U$ that has eigenvalues $\pm 1$. We may prepare a single ancilla qubit in the state $|+\rangle\equiv(|0\rangle + |1\rangle)/\sqrt{2}$, and then apply $U$ to a block of $n$-qubits, conditioned on the value of the ancilla (that is, $I$ is applied if the state of the ancilla is $|0\rangle$ and $U$ is applied if the state of the ancilla is $|1\rangle$). Then we measure the ancilla qubit in the basis $\{|\pm\rangle\equiv (|0\rangle \pm |1\rangle)/\sqrt{2}\}$. The outcome $|+\rangle$ for the ancilla measurement indicates that the eigenvalue of $U$ is $+1$, and the outcome $|-\rangle$ for the ancilla measurement indicates that the eigenvalue of $U$ is $-1$.

For the case of weight-$w$ Pauli operator, this circuit consists of the ancilla preparation, a sequence of $w$ two-qubit gates (each a conditional Pauli operator acting on the ancilla and one of the qubits in the code block), and a final measurement of the ancilla qubit. This procedure is {\em not} fault tolerant, because the ancilla qubit interacts with multiple qubits in the same code block. A single fault could damage the ancilla qubit, and that error could propagate, infecting several of the qubits in the block.

A better procedure is to replace the single ancilla qubit by an encoded ancilla block. For example, to measure a weight-$w$ Pauli operator, we could prepare the ancilla in the ``cat'' state $|+\rangle_{\rm rep}={(\0 ^{\otimes w} + \1 ^{\otimes w}) / \sqrt{2}}$ of the $w$-qubit quantum repetition code \cite{shor_ft,DiVincenzo96}. Now each qubit in the ancilla block interacts with just one qubit in the data block, so that if both the cat state and the input data block have no errors, a single fault in the circuit results in just one error in the data block. The final destructive measurement of the ancilla block in the basis $\{|\pm\rangle_{\rm rep}\}$ is equivalent to a measurement of $X^{\otimes w}$; it can be achieved by measuring $X$ on each of the $w$ ancilla qubits and evaluating the parity of the outcomes. 

\begin{figure}[thb]
\begin{center}
  \begin{tabular}{c}
                \epsfig{file=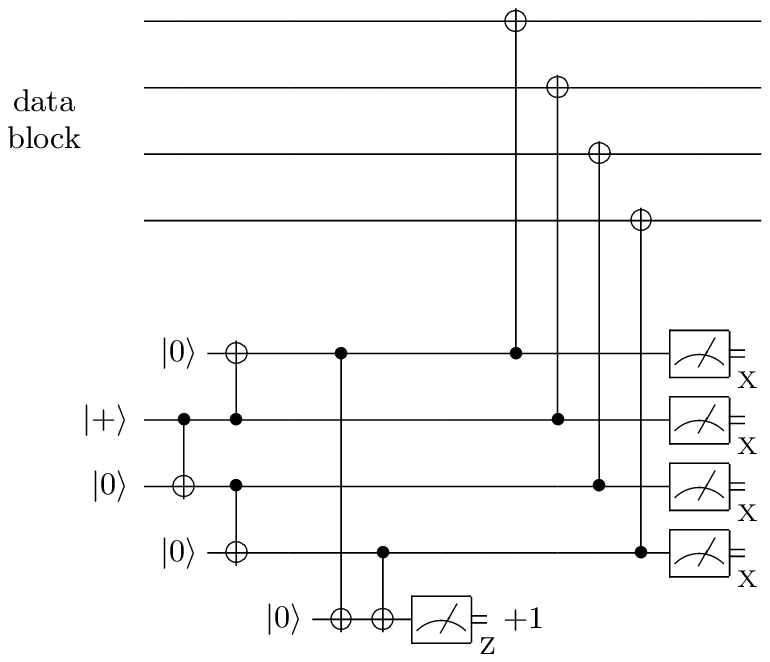}  
  \end{tabular}
\end{center}  
  \vspace{0.2cm}
\fcaption{
           Cat-state method for fault-tolerant measurement of a Pauli operator, in this case $X^{\otimes 4}$. A four-qubit cat state $(|0\rangle^{\otimes 4} + |1\rangle^{\otimes 4})/\sqrt{2}$ is prepared and verified, then controls the application of the Pauli operator to the data block. All four ancilla qubits are measured in the $X$ basis, and the parity of the measurement outcomes determines the eigenvalue of the measured Pauli operator. 
           }
\label{cat_method}
\end{figure}

But our procedure must also include the fault-tolerant preparation of $|+\rangle_{\rm rep}$. A complete procedure for measuring one bit of the syndrome is shown in Fig.$\,$\ref{cat_method} (where the Pauli operator being measured is $X^{\otimes 4}$).  To prevent a single fault in the encoding circuit from causing multiple errors in the output data block, a verification step is included. If the outcome of the verification measurement is $Z=-1$, then the cat state might have multiple errors --- the state is discarded before it ever comes into contact with the data, and the preparation is repeated. If the verification succeeds, then at least two faults are required to introduce two or more errors into the data block.

%\vspace{0.2cm}

To see why the procedure is fault-tolerant, it is important to understand how errors are propagated by \cnot gates. A \cnot propagates an $X$ error ``forward'' from its control qubit to its target qubit, and it propagates a $Z$ error ``backward'' from its target qubit to its control qubit. Thus while a $Z$ error in the cat state might result in a faulty syndrome measurement, it is the $X$ errors in the cat state that are especially dangerous, for these are the errors that might propagate into the data block. Since the ideal cat state is actually an eigenstate with eigenvalue one of $X^{\otimes 4}$, it cannot be afflicted with more than two $X$ errors; when two $X$ errors occur, only one of the two qubits that participate in the verification test will have been affected, so that the test (if performed without faults) will detect the damage. It is easy to adapt the principles underlying the construction of this circuit to achieve a fault-tolerant measurement of any Pauli operator. 

Finally, after the complete syndrome $e$ is measured, a weight-1 Pauli operator is applied to correct the error. Since a single fault in the circuit might cause both an error in the data block and an error in a syndrome bit, the syndrome measurement should be repeated to ensure accuracy. If the same syndrome is found twice in a row, it is safe to accept the syndrome and recover from the error accordingly. (As we will explain in Sec.~\ref{sec:explicit}, for some fault-tolerant constructions, the error recovery step is not necessary at all; rather the propagation of errors through subsequent gates can be tracked by a classical computation.)  Hence the full procedure obeys property 2.

\subsubsection{Syndrome measurement using encoded blocks}
\label{sec:steane-ec}
With the cat state method, a separate encoded ancilla is used to measure each of the code's stabilizer generators. Steane \cite{steane_anc} proposed another fault-tolerant method for measuring the syndrome that is more highly parallelized. Steane's method requires fewer gates than the cat state method, and is especially advantageous if the identity gate and nontrivial gates have comparable fault rates (that is, if storage errors are about as likely as gate errors). 

Steane's method applies to CSS codes \cite{cal_shor,steane_css}, for which each generator in $\mathcal{G}$ can be chosen to be either a tensor product of $X$s and $I$s (an $X$-type generator) or a tensor product of $Z$s and $I$s (a $Z$-type generator). For any CSS code with $k=1$ encoded qubit, a \cnot gate can be implemented {\em transversally} --- an encoded \cnot is realized by performing (in parallel) a \cnot from each qubit in the control block to each qubit in the corresponding position in the target block \cite{shor_ft,gott_ft}. 

In Steane's method, the ancillas for the syndrome measurement are encoded using the same CSS quantum error-correcting code as protects the data. To measure all of the $Z$-type generators, the ancilla is prepared in the encoded state $|\bar +\rangle\equiv (|\bar 0\rangle+\bar 1\rangle)/\sqrt{2}$, and an encoded \cnot is applied with the data block as control and the ancilla block as target. Since $|+\rangle$ is an eigenstate with eigenvalue one of $X$={\sc not}, the \cnot has no affect on the encoded state of the ancilla or the data, but the \cnot gates propagate each $X$ error in the data block to the corresponding position of the ancilla block. Assuming, then, that the ancilla block had no errors of its own initially, and that none of the \cnot gates are faulty, the $Z$-type syndrome (which detects the $X$ errors) can be extracted by measuring all of the ancilla qubits in the $Z$ basis, and applying a classical parity check matrix to the outcomes. Specifically, for the $Z$-type stabilizer generator $Z(b)\equiv\otimes_{i=1}^n \; Z^{b_i}$, if the outcome of the $Z$ measurement is $z=(z_1,z_2,\dots , z_n)$, then the measured eigenvalue of $Z(b)$ is $b\cdot z$ (mod 2). Likewise, the $X$-type syndrome (which detects the $Z$ errors) can be extracted by preparing the ancilla in the encoded state $|\bar 0\rangle$, applying an encoded \cnot with the ancilla block as control and the data block as target, measuring each qubit in the ancilla qubit in the $X$ basis, and applying a classical parity check to the measurement outcomes. For the $X$-type stabilizer generator $X(a)\equiv\otimes_{i=1}^n \; X^{a_i}$, if the outcome of the $X$ measurement is $x=(x_1,x_2,\dots , x_n)$, then the measured eigenvalue of $X(a)$ is $a\cdot x$ (mod 2). 

The encoded $|\bar 0\rangle$ and $|\bar +\rangle$ are prepared using encoding circuits that are {\em not} fault tolerant, and therefore a single fault during encoding might result in more than one error in the encoded state; as for the cat state method, a verification step is needed to enforce property 2. For the encoded $|\bar 0\rangle$, it is the $X$ errors that might propagate from ancilla to data, and the ancilla should be rejected if the verification detects $X$ errors. One way to conduct the verification is to prepare two blocks in the state $|\bar 0\rangle$, the ancilla block and a verifier block. A \cnot is applied with the ancilla block as control and the verifier block as target, the qubits in the verifier block are measured in the $Z$ basis, and a parity check is applied to the measurement outcomes. In this case, not just the $Z$-type stabilizer generators are extracted; after a classical error correction step, the eigenvalue of the encoded $\bar Z$ (another $Z$-type Pauli operator) is also found. The ancilla block is rejected if the measurement of the verifier block detects a nontrivial syndrome or if the eigenvalue of $\bar Z$ is $-1$. The verification of the $|\bar +\rangle$ ancilla can be conducted similarly.

%\vspace{0.2cm}
\begin{figure}[htb]
\begin{center}
  \begin{tabular}{c}
             \epsfig{file=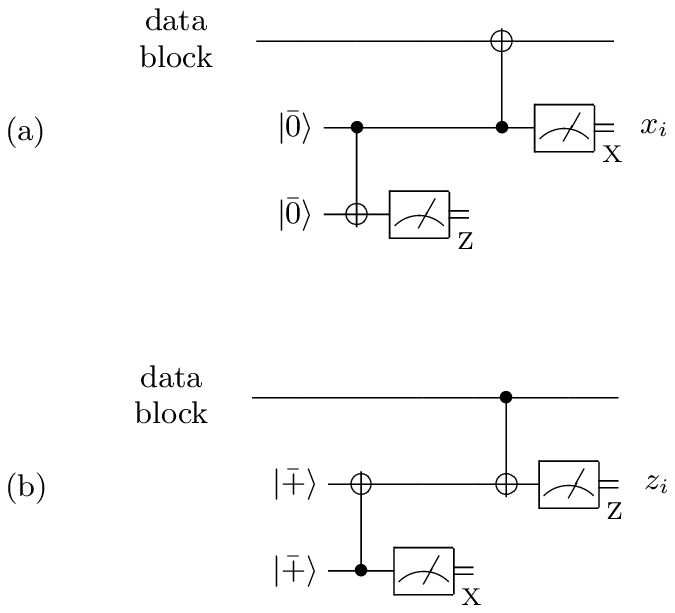}
  \end{tabular}
\end{center}  
  \vspace{0.2cm}
 \fcaption{
Steane's method for measuring the syndrome of a CSS code using encoded ancilla blocks. 
(a) For measurement of the $X$-type stabilizer generators, an ancilla block and a verifier block are prepared in the state $|\bar 0\rangle$. After a transversal \cnot from ancilla block to verifier block, the verifier block is measured and a classical parity check is computed to test the ancilla for $X$ errors. If the ancilla is accepted, a \cnot is applied from ancilla block to data block, the ancilla qubits are measured in the $X$ basis, and a classical parity check is computed to extract the measured eigenvalues of the $X$-type generators. (b) A similar procedure is used to measure the $Z$-type generators, with the ancilla and verifier prepared in the state $|\bar +\rangle$, with \cnot gates acting in the opposite direction, and with qubits measured in the conjugate bases. 
           }
\label{steane_method}
\end{figure}

The syndrome measurement circuit, including the verification of the encoded ancillas, is shown in Fig.$\,$\ref{steane_method}. A single fault during encoding of the ancilla block (not shown in Fig.$\,$\ref{steane_method}) may propagate badly within that block. But if the ancilla is badly damaged, and there is only one fault in the complete syndrome measurement circuit, then the verification is perfect --- the ancilla will be rejected, preventing the errors from propagating to the data.

The syndrome measurement is followed by a recovery step, in which at most one $X$ and at most one $Z$ are applied to the data block. The recovery operation can be safely applied even if the syndrome is measured only once. A single fault during syndrome measurement might result in both an incorrect syndrome and an error in the data; however, using Steane's method, the incorrect syndrome affects the recovery operation only at position in the data block that is already damaged by the fault. Therefore, the faulty syndrome together with the error caused by the fault cannot combine to produce errors at two distinct positions in the data block, and the error correction procedure satisfies property 2.

Note that this syndrome measurement procedure, like the cat state procedure described earlier, is nondeterministic --- there are probabilistic fluctuations  in the number of encoded blocks that must be prepared before an ancilla is successfully verified. There are some advantages to replacing the procedure with a deterministic one, in which a fixed number of encoded blocks are verified; for example, in that case we can make more definite statements about the overhead cost of the procedure. If we don't mind paying the price of making the accuracy threshold slightly worse, we can replace the nondeterministic procedure by a deterministic one that uses more ancilla blocks. But since we are more interested here in  obtaining a good estimate of the threshold rather than in minimizing the (worst-case) overhead, we will stick with the nondeterministic procedure in our analysis. The fact that ancillas are sometimes rejected has an impact on our calculation of the threshold, as will be explained in Sec.~\ref{subsec:results}.

\subsubsection{Syndrome measurement using encoded Bell pairs}
\begin{figure}[tb]
\vspace{0.5cm}
\begin{center}
   \begin{tabular}{c}
                \epsfig{file=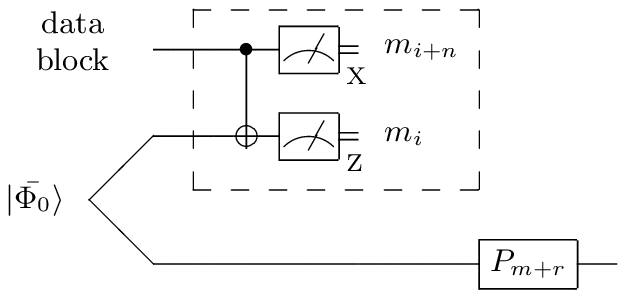}
                            
  \end{tabular}
\end{center}  
  \vspace{0.2cm}
\fcaption{
           Error correction by teleportation. A pair of encoded ancilla blocks is prepared in the Bell state $|\Phi_0\rangle$; then a transversal \cnot and measurements in the $X$ and $Z$ bases are used to perform a destructive Bell measurement on one of the ancilla blocks and the data block, yielding the outcome $(P_m\otimes I)|\Phi_0\rangle^{\otimes n}$, where $P_m$ is a Pauli operator.  This outcome determines an error syndrome pointing to the recovery Pauli operator $P_r$. Thus applying $P_{m+r}$ to the unmeasured ancilla block completes the teleportation of the data while also correcting errors. 
           }
\label{knill_method}
\end{figure}

Knill \cite{knill_detect} proposed another fault-tolerant method for measuring the syndrome, based on quantum teleportation, that also uses encoded ancilla blocks. In this method, depicted in Fig.\,\ref{knill_method}, a pair of ancilla blocks is prepared in the encoded Bell state $ |\bar \Phi_0\rangle = (|\bar 0\bar 0\rangle + |\bar 1 \bar 1\rangle) / \sqrt{2} \, $. Then the data block together with one of the ancilla blocks is measured transversally in the Bell basis, and a Pauli operator is applied to the other ancilla block to complete the ``teleportation'' of the data.

The outcome of the destructive Bell measurement is $(P_m\otimes I) |\Phi_0\rangle^{\otimes n}$, where $P_m$ is an $n$-qubit Pauli operator. If the ancilla has no errors, this outcome indicates an error syndrome $e=m\Lambda G^T$ for the data, which points to the recovery Pauli operator $P_r$. Thus applying the operator $P_m\cdot P_r=P_{m+r}$ to the unmeasured ancilla block implements the encoded Pauli operator needed for teleportation of the data block, and at the same time corrects the errors in the data. To ensure fault tolerance, the encoded Bell pair must be verified.

While Steane's method applies only to CSS codes, Knill's method works for any stabilizer code. Furthermore, it provides good protection against leakage errors, since leaked qubits are replaced when teleportation is carried out \cite{carlos}. And  Knill has shown \cite{knill_detect} that this method, combined with other tricks, seems to yield especially favorable estimates of the threshold. It is an important open problem to derive a rigorous lower bound on the quantum accuracy threshold based on Knill's circuitry, but we will use Steane's method in our analysis in this paper.

\subsubsection{Properties $0$, $0'$,and $1$}
\label{sec:prop00'1}
Property $1$ merely says that a 1-EC with no faults can correct one error successfully, which is true for the procedures explained above.

Property $0$ says that a 1-EC with no faults takes any input to the code space. For a perfect code, each value of the syndrome points to a unique correctable error. But even if the code is not perfect, we may by convention assign to each syndrome one particular Pauli operator that returns the subspace labeled by the syndrome to the code space. Therefore, if $|\bar\psi\rangle$ denotes a state in the code space, then for each Pauli error $E_a$, the ideal 1-EC maps $E_a|\bar \psi\rangle$ to ${\cal O}_a|\bar\psi\rangle$, where ${\cal O}_a$ is some encoded operation that may depend on $a$. Any state of a data block can be purified, expressed as a pure state $|\tilde\psi\rangle$ of the data and its environment $E$, and the ideal 1-EC can be expressed as a unitary transformation acting on the data and an ancilla $A$ according to

\begin{equation}
\label{purification}
  |\tilde{\psi}\rangle =  \sum_{a} E_a |\bar{\psi}\rangle \otimes |a\rangle_E \otimes |0\rangle_A 
                 \stackrel{\rm{EC}}{\longrightarrow}  \sum_{a} {\cal O}_a |\bar{\psi}\rangle \otimes |a\rangle_E \otimes |a\rangle_A \, ,
\end{equation}
\noindent 
where the states $\{|a\rangle_E\}$ and $\{|a\rangle_A\}$ of environment and ancilla are not assumed to be normalized or mutually orthogonal. After tracing over the data and ancilla, the state of the data block will be, in general, a mixture of pure states in the code space. 

Property $0'$ says that a 1-EC with one fault takes any input to a {\em valid} output (a superposition of states that each deviate from the code space by the action of a weight-1 Pauli operator). Though property $0'$ holds for other error correction schemes as well, it is most easily explained for Steane's method. 

Suppose, for now, that the fault in the 1-EC circuit occurs during the syndrome measurement, not during the final recovery step. Because the circuit is fault tolerant, the fault affects the output data block in at most one position, and it also affects the outcome of the ancilla measurement in at most one position. Furthermore, when we consider all the possible fault locations in the 1-EC, and all the possible ways for a Pauli error introduced by the fault to propagate through the 1-EC circuit, we conclude that the damage due to the fault occurs at the {\em same} position in the data block and in the ancilla block. Call this position $i$.

Therefore, if the input data block has Pauli error $E_a$, the error in the output data block is $E_b^{(i)} E_a$, where $E_b^{(i)}$ is a Pauli operator with support at the position $i$. The syndrome measured by the decoder will be the syndrome associated with $ E_c^{(i)}E_a$, where $E_c^{(i)}$ has support at position $i$, but it might be interpreted as the syndrome of another error $E_d$ with recovery operator $E_d^\dagger$. However, since $E_d$ and $E_c^{(i)}E_a$ have the same syndrome, $E_d^\dagger E_c^{(i)}E_a\equiv {\cal O}_a$ preserves the code space, and we may write  $E_d= E_c^{(i)} \tilde E_a$ where $\tilde E_a^\dagger E_a={\cal O}_a$. 

Therefore, after the recovery operation is applied, the combined effect of the input error, the fault, and the recovery step on the data block is 
\begin{equation}
\left({E_c}^{(i)}\tilde E_a\right)^\dagger \left(E_b^{(i)} E_a\right)
=\left(\tilde E_a^\dagger \left({E_c}^{(i)\dagger} E_b^{(i)} \right)\tilde E_a\right)\left(\tilde E_a^\dagger E_a\right)
=E_e^{(i)}{\cal O}_a~,
\end{equation}
where $E_e^{(i)}$ is also a Pauli operator with support at position $i$. Thus, the output of the 1-EC deviates from the code space by the action of the weight-one Pauli operator $E_e^{(i)}$.

We should also consider what happens if the fault occurs in the final recovery step. Then if the fault affects the Pauli gate that is intended to correct the error, the error might not be corrected successfully, but no other error will appear. Otherwise the fault is a storage error affecting one of the resting qubits in the final step. Then the fault does not interfere with the recovery Pauli operator, and the only error is the new one introduced by the fault. 

Finally, since any input state can be expanded in terms of states with Pauli errors, and any fault can be expanded in terms of Pauli faults, the 1-EC maps any input to a valid output, proving property $0'$. A similar argument shows that for a code that corrects $t$ errors, a 1-EC with $s$ faults maps any input to an output that deviates from the code space by the action of weight-$s$ Pauli operators. 

This observation completes our general discussion of how error correction gadgets satisfying properties 0--2 are constructed.

%---------------------------------------------------------%
\subsection{Fault-tolerant encoded gates}

Next we discuss how to construct a universal set of level-1 gate gadgets obeying properties 3 and 4. For an efficient recursive simulation, we should choose a universal set of 0-Ga's that enables us to build a relatively simple fault-tolerant 1-Ga to simulate each 0-Ga in the set. 

The quest for a universal set of fault-tolerant gates can be guided by a helpful classification of $n$-qubit unitary transformations \cite{gottesman-chuang}. We have already discussed the $n$-qubit Pauli group $\mathcal{C}_1^{(n)}$. The Clifford group $\mathcal{C}_2^{(n)}$ contains the $n$-qubit unitaries whose action by conjugation maps Pauli operators to Pauli operators. The Clifford group is generated by the Hadamard gate $H\equiv (X+Z)/\sqrt{2}$, the phase gate $S \equiv U_z(\pi/2)=\exp(-i {\pi \over 4} Z)$, and the controlled-not gate \cnot$\equiv \Lambda (X)$. The set $\mathcal{C}_r^{(n)}$ (which is not a group for $r\ge 3$) contains the $n$-qubit unitaries whose action by conjugation maps Pauli operators to the set $\mathcal{C}_{r-1}^{(n)}$. Important elements of $\mathcal{C}_{3}$ that are not in $\mathcal{C}_{2}$ include the Toffoli gate $\Lambda^2(X)$, the conditional Hadamard gate $\Lambda(H)$, the conditional phase gate $\Lambda(S)$, and the single-qubit rotation $T\equiv U_z(\pi/4)=\exp\left(-i\frac{\pi}{8}Z\right)$. The Clifford group $\mathcal{C}_2^{(n)}$ is not dense in $U(2^n)$ --- in fact quantum computation using Clifford group gates, Pauli operator measurements, and Pauli operator eigenstate preparations can be simulated efficiently with a classical computer \cite{gott_heisenberg}. But by adding to the generators of $\mathcal{C}_2$ an element of $\mathcal{C}_3$ that is not contained in $\mathcal{C}_2$, we  can obtain a universal gate set.

Let us suppose that our 0-Ga's include the gates $H$, $S$, and \cnot that generate $\mathcal{C}_2$, as well as the preparation of a qubit in the state $|0\rangle$ and a $Z$ measurement. With these tools, for any stabilizer code we can build fault-tolerant level-1 gadgets that realize the encoded Pauli operators, the preparation of the encoded state $|\bar 0\rangle$, and the measurement of encoded Pauli operators. (The measurement of encoded Pauli operators includes a classical parity computation which we assume to be reliable.) Furthermore, preparation of $|0\rangle$, Pauli operators, and Pauli measurements are adequate for realizing each of $H$, $S$, and {\sc cnot}. By this route level-1 $\mathcal{C}_2$ generators satisfying properties 3 and 4 can be constructed for any stabilizer code \cite{gott_ft}. 

For some codes the construction of fault-tolerant $\mathcal{C}_2$ gates is particularly simple. We say that a 1-Ga is {\em transversal} if it is realized in a single time step by 0-Ga's acting in parallel, such that all multiple-qubit 0-Ga's act on qubits at corresponding positions in distinct blocks. Transversal gates are fault tolerant. Property 3 is satisfied, because although an ideal  1-Ga might propagate an error from one block to another, it cannot propagate an error from one position in a block to another position in the same block. And property 4 is also satisfied, because a faulty 0-Ga cannot cause two errors in the same block.

Thus, in fault-tolerant simulations, it is advisable to use quantum error-correcting codes for which transversal gates can be constructed. For any CSS code with $k=1$ encoded qubit, the encoded \cnot gate can be implemented transversally. And if the CSS code is constructed from a punctured doubly-even self-dual classical code, the encoded $H$ and $S$ gates are also transversal \cite{shor_ft}. An example of a code with these properties is Steane's $[[7,1,3]]$ quantum code \cite{steane_7}; we will describe the details of the level-1 gadgets for this code in Sec.~\ref{sec:explicit}.

But these generators of the Clifford group do not comprise a universal gate set, and indeed it does not appear to be possible, for any code, to construct a universal set of transversal 1-Ga's. Fortunately, though, there is a general scheme for using the Clifford 1-Ga's and a 0-Ga in $\mathcal{C}_3$ to build a fault-tolerant 1-Ga that simulates the $\mathcal{C}_3$ gate \cite{shor_ft,gottesman-chuang,zhou}. This scheme involves the off-line preparation and verification of {\em quantum software} that is then consumed during the execution of the gate.

\begin{figure}[htb]
\begin{center}
              \begin{tabular}{c}
               \epsfig{file=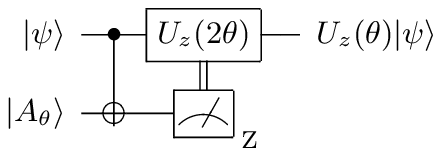}           
               \end{tabular}
\end{center} 
\vspace{0.2cm}               
\fcaption{
        Implementation of the gate $U_z(\theta)$. An ancilla is prepared in the state $|A_\theta\rangle=U_z(\theta)|+\rangle$, and a \cnot gate is executed with the data as control and the ancilla as target; then the ancilla is measured in the basis $\{|0\rangle$, $|1\rangle\}$. The gate $U_z(2\theta)$ is applied to the data conditioned on the measurement outcome.
        }
\label{t_circuit}
\end{figure}

The version of this scheme that we will use in our analysis is based on the circuit shown in Fig.\ \ref{t_circuit}. To realize the single-qubit rotation $U_z(\theta)=\exp\left(-i\frac{\theta}{2}Z\right)$, an ancilla qubit is prepared in the state $|A_\theta\rangle=U_z(\theta)|+\rangle$. A \cnot gate is applied with the data qubit as control and the ancilla qubit as target, and then the ancilla qubit is measured in the $Z$ basis. If the outcome of the measurement is $|0\rangle$, then $U_z(\theta)$ has been successfully applied to the data. If the measurement outcome is $|1\rangle$, then $U_z(-\theta)$ has been applied instead; in the latter case, the implementation of $U_z(\theta)$ can be completed by applying $U_z(2\theta)$.

%\vspace{1cm}

Why is this circuit useful? Suppose that we have fault-tolerant 1-Ga's for the gates $H$, $S$, and \cnot that generate $\mathcal{C}_2$, and that we seek a fault-tolerant level-1 gadget for the $\mathcal{C}_3$ gate $T\equiv U_z(\pi/4)$ to complete a universal set. For $\theta=\pi/4$, $U_z(2\theta)=S$ is a $\mathcal{C}_2$ gate, so in that case, all of the operations in Fig.\ \ref{t_circuit} can be implemented fault tolerantly. To simulate the $T$ gate at level 1, we replace the {\sc cnot}, the $S$ gate, and the $Z$ measurement by the corresponding 1-Recs. What is needed to complete the simulation is an additional 1-Rec that reliably prepares (off-line) some suitably quantum software, namely the encoded state $|\bar A_{\pi/4}\rangle$. 

To enforce property 4, we must ensure that a single fault in the preparation circuit for $|\bar A_{\pi/4}\rangle$ will not cause more than one error in the circuit's output. It is helpful to observe that $|A_{\theta}\rangle$ is an eigenstate with eigenvalue one of the operator $U_z(\theta)XU_z(\theta)^\dagger= U_z(2\theta)\cdot X$, which for $\theta=\pi/4$ is the $\mathcal{C}_2$ operator $SX$.

Thus the fault-tolerant realization of the $T$ gate reduces to fault-tolerantly measuring the $\mathcal{C}_2$ operator $SX$, an operator for which a fault-tolerant 1-Ga can be constructed. This measurement can be accomplished using the cat state method. That is, we can measure an encoded operator $U$ by preparing an ancilla in the (verified) state $|+\rangle_{\rm rep}\equiv (|0\rangle_{\rm rep} +| 1\rangle_{\rm rep})/\sqrt{2}$ of the quantum repetition code, applying $U$ to the data block conditioned on the ancilla being in the state $| 1\rangle_{\rm rep}$, measuring the ancilla qubits in the $X$ basis, and then computing the parity of the outcomes. We used this method previously in Sec.~\ref{sec:cat-state-method} to perform fault-tolerant measurements of encoded $\mathcal{C}_1$ operators, and it can also be used to measure encoded $\mathcal{C}_2$ operators, once we have constructed the fault-tolerant gadgets for the $\mathcal{C}_2$ operators. 

A single fault in the measurement circuit causes only one error in the measured block, but it might also cause an error in the measurement outcome. Hence, after an error correction, the measurement should be repeated. If the same result is obtained twice in a row, the ancilla block can be accepted safely. Finally, we have assembled all the elements of a 1-Ga for the $T$ gate that satisfies properties 3 and 4.

%---------------------------------------------------------%
\subsection{Fault-tolerant preparation and measurement}
\label{sec:ft-prep-and-meas}

For (destructive) measurements of encoded Pauli operators, properties 3 and 4 can be restated as:

\begin{description}
\item 3. ~A 1-measurement with no faults applied to an input with one error agrees with an ideal measurement.
\item 4. A 1-measurement with at most one fault applied to an input with no errors agrees with an ideal measurement. 
\end{description}

\noindent
Although the measurement is allowed to destroy the input block, for general stabilizer codes measurements must be repeated to enforce property 4, and therefore nondestructive measurements should be used. As discussed in Sec.~\ref{sec:cat-state-method}, fault-tolerant nondestructive measurements of encoded operators can be executed by the cat state method; one fault in the measurement circuit causes just one error in the output block. But a single such measurement obeys neither property 3 nor property 4. Therefore, we extend the procedure --- the first measurement is followed by the 1-EC, then a second measurement, another 1-EC, and finally a third measurement. The outcome of the 1-measurement is the majority of the results of the three individual measurements.

If there are no faults, then the first 1-EC corrects the error in the input, and the second and third measurements both agree with the ideal outcome. If there is one fault, it could occur in one of the measurements or in one of the 1-ECs. But no matter where it is located, the fault can disturb the outcome of only one of the three measurements. Therefore properties 3 and 4 are satisfied.

For CSS codes, an especially efficient destructive 1-measurement can be constructed, at least for measurement of the encoded operations $\bar X$ and $\bar Z$. For example, to measure $\bar Z$, which is a $Z$-type Pauli operator, all qubits are measured in the $Z$ basis, and the results are postprocessed classically. A single erroneous outcome can be diagnosed by applying a classical parity check, the error can be corrected, and the eigenvalue of the encoded $\bar Z$ then evaluated. The error could be due to an error in the input, or due to a fault in one of the measurements --- hence properties 3 and 4 are satisfied, if we assume that the classical processing is flawless. The same procedure can also be used to measure $\bar X$.

Since as far as the quantum processing is concerned the 1-measurement is built from only 0-measurements, the recursively defined $k$-measurement is also very simple. All of the qubits can be measured simultaneously in a single time step. Then the classical processing is done recursively. First each 1-block is (classically) decoded and a one-bit result recorded, then the process is repeated altogether $k$ times to extract the final one-bit outcome of the measurement of the $k$-block. 

For a 1-preparation, which has no input, we need only worry about property 4, which can be restated:
\begin{description}
\item 4. A 1-preparation with at most one fault produces an output with at most one error. 
\end{description}
\noindent
While encoding circuits typically propagate errors badly, we can build a fault-tolerant 1-preparation of the encoded $|\bar 0\rangle$ for any stabilizer code, using the cat state method for measuring encoded Pauli operators. Starting with an arbitrary state (such as a product of $|0\rangle$'s), the 1-EC can be executed to obtain a state in the code space if there are no faults (property 0), or a valid state if there is one fault (property $0'$). Then the encoded Pauli operator $\bar Z$ can be measured nondestructively three times as described above. If a majority vote on the results finds the result $|\bar 0\rangle$, the preparation is finished, and if the result $|\bar 1\rangle$ is found, an encoded $\bar X$ is applied to complete the procedure. 

For CSS codes, there are more efficient 1-preparation procedures. The 1-preparation satisfies property 4 in the CSS sense if its output has no more than one $X$ error and no more than one $Z$ error; we can enforce this property by subjecting an encoded block to a verification test, and rejecting the block when multiple errors are detected. This procedure is especially simple for a perfect CSS code like the [[7,1,3]] code, since in that case, we do not need to worry about more than one $Z$ error in the encoded state $|\bar 0\rangle$ --- for the [[7,1,3]] code, two $Z$ errors are equivalent to an encoded $\bar Z$ and a single $Z$ error, and the state $|\bar 0\rangle$ is an eigenstate of $\bar Z$ with eigenvalue one.  We can check for multiple $X$ errors just as in the verification test included in the Steane error correction circuit. We encode both the data block and an ancilla block in the state $|\bar 0\rangle$, then execute a \cnot gate from data to ancilla, and measure the ancilla in the $Z$ basis. Applying classical parity checks to the measurement outcomes, we extract the data's $X$ error syndrome, and after a classical error correction (if necessary), the eigenvalue of $\bar Z$. If the outcome of the encoded $\bar Z$ measurement is $|\bar 0\rangle$, then the data block is accepted. Otherwise it is rejected. 

\begin{figure}[htb]
\begin{center}
              \begin{tabular}{c}
               \epsfig{file=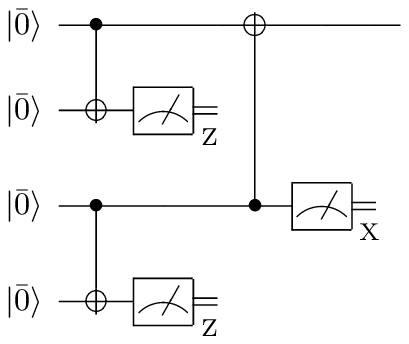}           
               \end{tabular}
\end{center} 
\vspace{0.2cm}               
\fcaption{
        1-preparation of the encoded state $|\bar 0\rangle$, for CSS codes. Three ancilla blocks are used to verify that the data block contains no more than one $X$ error and no more than one $Z$ error. For the perfect [[7,1,3]] code, there is no need to check for multiple $Z$ errors, and only one ancilla block is needed.
        }
\label{fig:CSS_prep}
\end{figure}

However, for a CSS code that is not perfect, we need to check for multiple $Z$ errors as well as multiple $X$ errors; this can be achieved with the circuit in Fig.~\ref{fig:CSS_prep}, in which three encoded ancilla blocks are used for the verification. The first ancilla block detects multiple $X$ errors in the data, as before. A second ancilla block is used to check for multiple $Z$ errors. But if there is a fault in the encoding of the second ancilla block, it might have multiple $X$ errors that could propagate to the data; to prevent this the second ancilla block is also verified, using the third ancilla block.

%\vspace{1cm}

%--------------------------------------------%
\section{Threshold estimate for the [[7,1,3]] code}
\label{sec:explicit}

Now we will discuss explicit constructions of the level-1 gadgets for a particular quantum error-correcting code, Steane's [[7,1,3]] code \cite{steane_7}, the distance-3 CSS code with the smallest block size. By counting the malignant pairs of locations in the 1-exRecs as outlined in Sec.~\ref{sec:malignant-pairs}, we will obtain a lower bound on the quantum accuracy threshold $\varepsilon_0$, assuming an independent stochastic noise model. In our analysis, we will assume that classical processing of measurement outcomes is flawless and instantaneous. 

Our goal is to design gadgets that are as simple as possible, and so to obtain a good threshold estimate. We will be less attentive to overhead requirements, and in particular, the overhead of some of the gadgets is nondeterministic; encoded states are verified and are discarded if the verification fails. Thus there is a small probability that the verification step needs to be repeated many times before the state is accepted. We could modify the gadgets so that the overhead is deterministic, but the gadgets then become a bit more complicated and the threshold estimate worsens, so we will not discuss such modifications here.

The [[7,1,3]] code has the generator matrix
\begin{equation}
   G = \left( \begin{array}{cc}
                                      \mathbf{H}_X & \mathbf{0}\\
                                      \mathbf{0} & \mathbf{H}_Z
                     \end{array}
                     \right) \, ,
\end{equation}
\noindent where
\begin{equation}
   \mathbf{H}_Z = \mathbf{H}_X = \left( \begin{array}{ccccccc}
                                      0 & 0 & 0 & 1 & 1 & 1 & 1\\
                                      0 & 1 & 1 & 0 & 0 & 1 & 1\\
                                      1 & 0 & 1 & 0 & 1 & 0 & 1
                     \end{array}
                     \right) \, .
\label{7qubit-gen}
\end{equation}
The encoded $\bar X$ and $\bar Z$ can be realized by the weight-3 Pauli operators $XXXIIII$ and $ZZZIIII$. The encoded generators of $\mathcal{C}_2$ can all be realized transversally --- $\bar H$ by applying $H$ to each of the seven qubits in the block, $\bar S$ by applying $S^\dagger$ to each qubit, and the encoded \cnot by applying \cnot gates from each qubit in the control block to the corresponding qubit in the target block. A universal gate set can be completed by adding $\bar T$, realized via fault-tolerant preparation of the state $|\bar A_{\pi / 4}\rangle$, followed by execution of the circuit shown in Fig.$\,$\ref{t_circuit}.

%--------------------------------------------------------%
\subsection{Fault-tolerant gadgets}

We will suppose that our quantum computer is equipped with the following 0-Ga's: preparation of $|0\rangle$ and $|+\rangle$, the gates $\{H,T,T^\dagger,~${\sc cnot}$\}$, and measurement of $Z$ and $X$. There is some redundancy in this list --- we could prepare $|+\rangle$ by first preparing $|0\rangle$ and then applying $H$, and we could measure $X$ by first applying $H$ and then measuring $Z$. But in our threshold analysis, we regard each listed 0-Ga as a primitive location in our quantum circuit, where a fault can occur with probability $\varepsilon$. (Note that $T^2=S$, so the set is universal.) In a recursive simulation, we are to use these 0-Ga's to realize the same set of primitive objects as 1-Ga's.

The 1-preparation and 1-measurement are realized following the CSS constructions described in Sec.~\ref{sec:ft-prep-and-meas}, and error correction is performed using the Steane method depicted in Fig.$\,$\ref{steane_method}. The encoded ancilla is rejected whenever a non-trivial syndrome or the incorrect value of the encoded $\bar Z$ or $\bar X$ is obtained in the verification step. We will assume that many encoding and verification circuits are executed in parallel, so that a verified ancilla is always available when needed. 

As discussed in Sec.~\ref{sec:steane-ec}, although a single fault can cause both an error in the data block and an error in the syndrome measurement, the syndrome measurement need not be repeated to ensure fault tolerance. Furthermore, we will not need to include a recovery step in which a weight-1 Pauli error is applied to correct the error. Instead, Pauli errors will be recorded in a classical register, and propagated through subsequent $\mathcal{C}_2$ gates by an efficient classical computation. No corrections are needed until encoded blocks are measured; at that stage, the recorded $X$ errors are consulted to properly decode the outcome of a $\bar Z$ measurement, and the recorded $Z$ errors are consulted to properly decode the outcome of a $\bar X$ measurement. This procedure works because the $\mathcal{C}_3$ gate $T$ is realized through the off-line preparation and verification of quantum software; only $\mathcal{C}_2$ gates, which propagate Pauli operators to Pauli operators, act directly on the data blocks.

\begin{figure}[htb]
\begin{center}
        \begin{tabular}{c}
               \epsfig{file=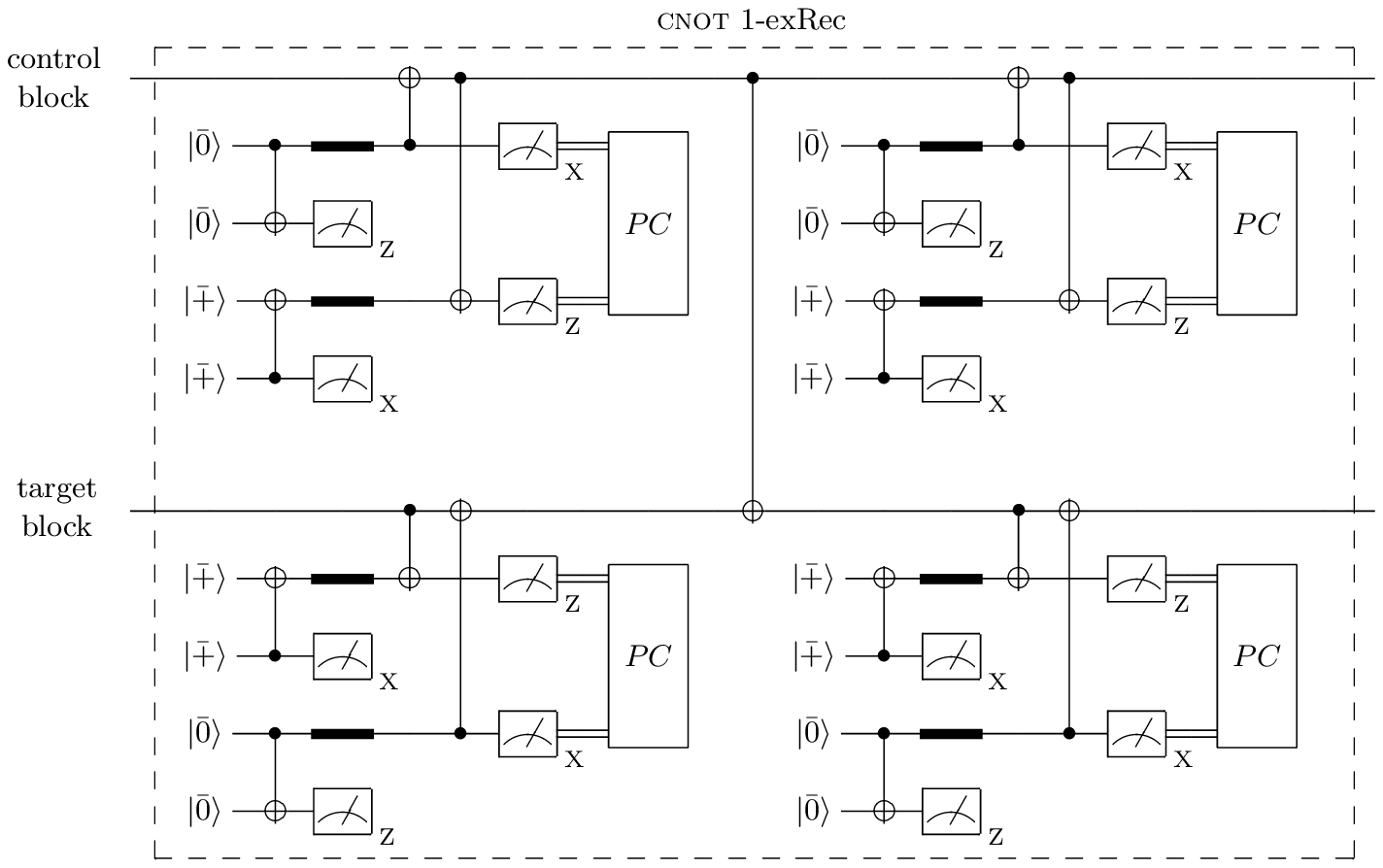}
        \end{tabular}
\end{center} 
\vspace{0.2cm}
\fcaption{
         The \cnot extended rectangle for a CSS code. The encoding circuits that prepare $|\bar 0\rangle$ and $|\bar +\rangle$ are suppressed here (the $|\bar 0\rangle$ encoder is shown explicitly in Fig.\,\ref{encoded_0}). Classical parity checks (PC), which are assumed to be instantaneous and flawless, are performed on the measurement outcomes to diagnose errors in the data blocks. Diagnosed Pauli errors are not explicitly corrected; rather they are stored in a classical register and propagated through subsequent $\mathcal{C}_2$ gates by an efficient classical computation. Locations where storage faults can occur are indicated by thickened wires.
         }
%\vspace{0.2cm}  
\label{cnot_rect}
\end{figure}

\begin{figure}[htb]
\begin{center}
             \begin{tabular}{c}
             \epsfig{file=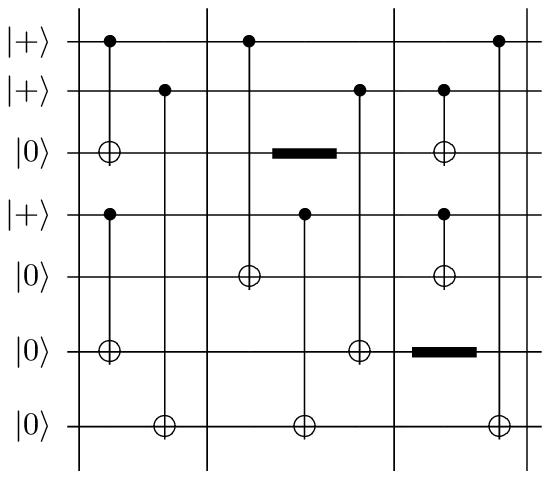}
             \end{tabular}
\end{center} 
\vspace{0.2cm}
\fcaption{
        The level-1 $|\bar 0\rangle$ encoding circuit for the [[7,1,3]] code. The $|\bar +\rangle$ encoder is identical, except all \cnot gates are reversed in direction and the input states $|0\rangle$ and $|+\rangle$ are interchanged. Vertical lines separate successive time steps, and locations where storage faults can occur are indicated by thickened wires. 
(There is no storage fault in the first time step because one 0-preparation occurs a step behind the others.)        }
%\vspace{0.2cm}  
\label{encoded_0}
\end{figure}

The level-1 $\mathcal{C}_2$ gate with the largest 1-exRec is the {\sc cnot}, whose 1-exRec is shown schematically in Fig.$\,$\ref{cnot_rect}. Contained within each 1-EC, but not shown explicitly in Fig.$\,$\ref{cnot_rect}, are encoding circuits for the $|\bar 0\rangle$ and $|\bar +\rangle$ states; the $|\bar 0\rangle$ encoder is shown in Fig.$\,$\ref{encoded_0}. Locations in the circuits where the qubits are ``resting'' and subject to storage faults are indicated by a thickening of the wires. 

%\vspace{1cm}

\begin{figure}[htb]
\begin{center}
              \begin{tabular}{c}
               \epsfig{file=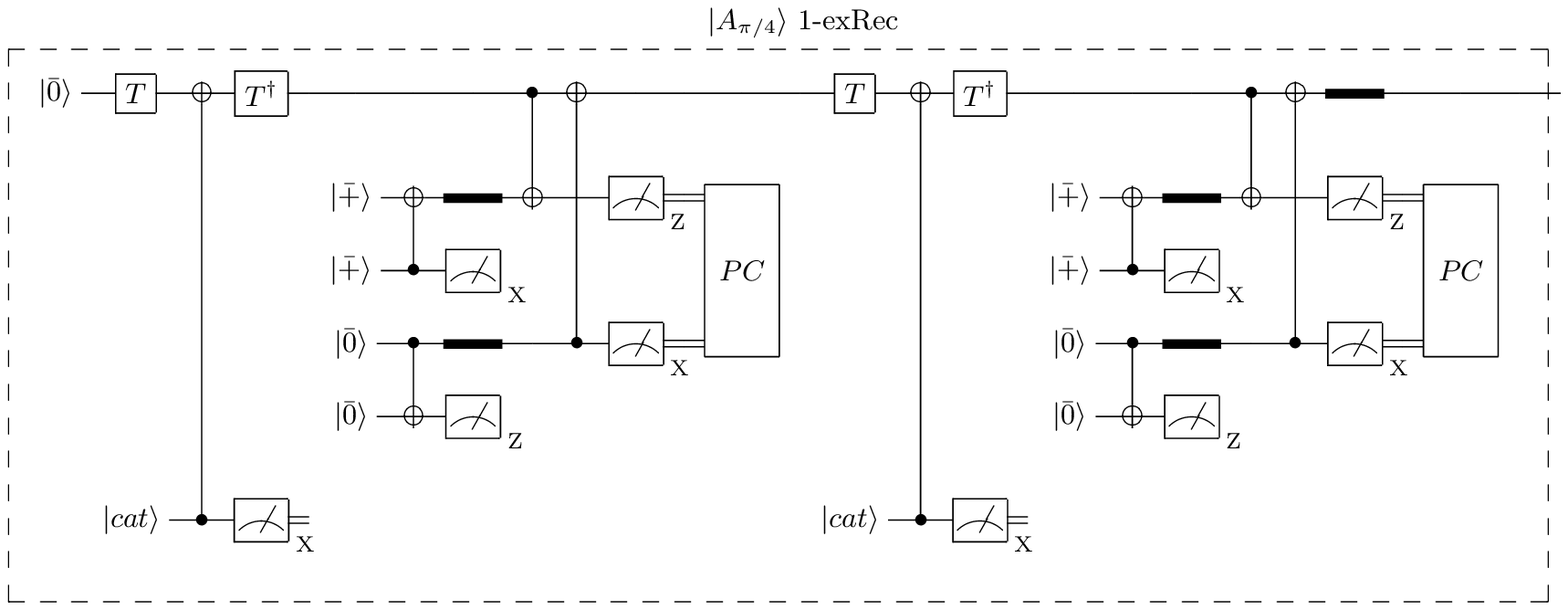}           
               \end{tabular}
\end{center} 
\vspace{0.2cm}               
\fcaption{\label{t_ancilla}
        Preparation of an ancilla in the encoded state $|\bar A_{\pi / 4}\rangle$, which is needed for the execution of the encoded $\mathcal{C}_3$ gate $\bar T$. The ancilla is prepared in the state $|\bar 0\rangle$, then a seven-qubit cat state is prepared and verified, and the cat state is used to measure $\bar T \bar X \bar T^{\dagger}$. The error syndrome is measured and then the procedure is repeated. The ancilla is accepted if both measurements of $\bar T \bar X \bar T^{\dagger}$ yield the same eigenvalue and both error syndromes are trivial. In this circuit $T$ denotes, not the 1-Ga $\bar T$, but rather the 0-Ga $T$ applied transversally to each qubit in the code block. The preparation and verification of the cat states, suppressed here, is shown in Fig.$\,$\ref{cat}. 
        }
\end{figure}

\begin{figure}[htb]
\begin{center}
             \begin{tabular}{c}
             \epsfig{file=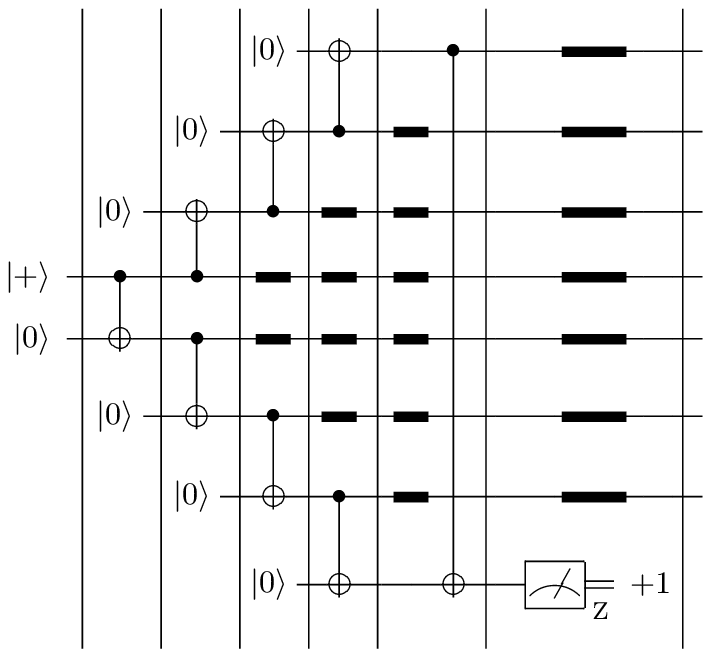}
             \end{tabular}
\end{center}
\vspace{0.2cm}             
\fcaption{\label{cat}
        Encoding and verification of the seven-qubit cat state, needed for the preparation of the encoded ancilla state $|\bar A_{\pi/4}\rangle$. The state is rejected if the outcome of the measurement of the verifier qubit is $|1\rangle$, indicating multiple $X$ errors that could have resulted from a single fault in the encoder.
        } 
\end{figure}

To realize the gate $\bar T$ needed to complete our universal set, we prepare an ancilla in the encoded state $|\bar A_{\pi /4}\rangle$ using the circuit shown in Fig.$\,$\ref{t_ancilla}. This circuit employs the cat state method to twice measure the $\mathcal{C}_2$ operator $\bar T \bar X \bar T^{\dagger}= \bar S\bar X$, which in the [[7,1,3]] code can be implemented transversally by applying $S^\dagger X=T^{\dagger} X T$ to each qubit in the block. First an ancilla block is prepared in the state $|\bar 0\rangle$, and a seven-qubit cat state $\left(|0\rangle^{\otimes 7} +|1\rangle^{\otimes 7}\right)/\sqrt{2}$ is encoded and verified using the circuit in Fig.$\,$\ref{cat}. Then the transversal $T^\dagger XT$ is applied to the ancilla controlled by the cat state, all qubits of the cat state are measured in the $X$ basis, and the parity of the measurement outcomes is evaluated classically to determine the outcome of the measurement of  $\bar T \bar X \bar T^{\dagger}$. Then the error syndrome is measured, and the ancilla is discarded unless the syndrome is trivial (indicates no errors). Finally the whole procedure is repeated --- $\bar T \bar X \bar T^{\dagger}$ and then the error syndrome are measured a second time. The ancilla is accepted only if both measurements of $\bar T \bar X \bar T^{\dagger}$ find the same eigenvalue, and if both measured syndromes are trivial. Repeating the measurement ensures the fault-tolerance of the ancilla measurement, and rejecting states with nontrivial error syndromes improves the fidelity of the preparation. (If the measured eigenvalue of $\bar T \bar X \bar T^{\dagger}$ is $-1$, we can flip the eigenvalue by applying $\bar Z$, or we can incorporate this $\bar Z$ into the Pauli error that is recorded and propagated classically.)

The encoding of $|\bar 0\rangle$ is included in the $|\bar A_{\pi/4}\rangle$ preparation 1-exRec to ensure that, if there are no faults during encoding, the input to the encoded measurement is in the code space. It is not necessary to verify the encoded $|\bar 0\rangle$, since any other state in the code space would serve as well. A fault during encoding may propagate badly, but if there are no other faults, the resulting deviation from the code space will be detected subsequently and the prepared state will be rejected.

%\vspace{1cm}

%---------------------------------------------------%
\subsection{Counting of malignant pairs: Procedure}
\label{sec:malignant-procedure}

In Sec.~\ref{sec:malignant-pairs}, we described how to estimate the quantum accuracy threshold by counting the number of malignant pairs of locations within an extended rectangle. This counting is entirely combinatorial, and we have used a computer program written in Matlab to carry it out exhaustively.

Two noise models were analyzed: The first is (adversarial) independent stochastic noise, where fault locations are independently and identically distributed, and once the fault locations are chosen the operations at those locations are arbitrary trace-preserving completely positive maps. Since all faults have a Pauli expansion, this model can be analyzed by testing Pauli faults at a pair of locations, and declaring the pair of locations to be benign if and only if the 1-Rec contained in the 1-exRec is correct for all possible Pauli faults at those locations. For this model, we can apply Theorem 2 to derive a rigorous lower bound on the quantum accuracy threshold $\varepsilon_0$. 

The second noise model inserts the depolarizing channel at each location. Again the fault locations are independently and identically distributed, but now the fault at a bad location is assumed to be a Pauli operator, chosen equiprobably from among $\{X,Y,Z \}$ for single-qubit gates, and equiprobably among the 15 nontrivial Pauli operators for two-qubit gates. For this model, we can estimate a critical noise rate $\varepsilon_0^{(1)}$, such that for noise rate $\varepsilon < \varepsilon_0^{(1)}$ a level-1 simulation outperforms an unprotected level-0 simulation. This estimate allows us to make a heuristic comparison between the adversarial and depolarizing noise models. But because the effective noise model that governs whether a $k$-Rec is bad for $k>1$ is not exactly self-similar, this critical noise rate cannot be identified as a rigorously established accuracy threshold.

In both models, faults at a bad location are inserted right after the ideal implementation of the 0-Ga. For a faulty 0-preparation,  an $X$ or $Z$ is inserted right after the ideal preparation of  $|0\rangle$ or $|+\rangle$, and for a faulty 0-measurement, an $X$ or $Z$ is inserted right before an ideal measurement of $Z$ or $X$. 

The investigation of whether a specified pair of locations is benign is conducted differently for the \cnot 1-exRec, which contains only $\mathcal{C}_2$ 0-Ga's, than for the $|A_{\pi/4}\rangle$ 1-exRec, which contains the $\mathcal{C}_3$ gates $T$ and $T^\dagger$.
Since $\mathcal{C}_2$ gates propagate Pauli operators to Pauli operators, for the \cnot $1$-exRec, input Pauli errors and Pauli errors caused by faults can be propagated through the circuit without ever leaving the Pauli group. When an ancilla qubit with an $X$ error (say) is measured in the $Z$ basis, a nontrivial syndrome bit is recorded. Finally, the Pauli error acting on the output of the 1-Rec is compared with the Pauli error acting on the input to test for encoded errors. 

The \cnot 1-Rec is certainly correct if both faults in the 1-exRec are contained in one of the leading 1-ECs; therefore we may assume that each leading 1-EC has no more than one fault. This observation leads to a further simplification. We have already seen in Sec.~\ref{sec:prop00'1} that a 1-EC with one fault takes an input with one error to a valid output, deviating from the code space due to an error at the position in the code block where the fault acted. Therefore, to investigate the consequences of a fault at a single location in one of the leading 1-ECs, we may disregard the input error while allowing an arbitrary fault at the specified location.

We say that the $|A_{\pi/4}\rangle$ preparation 1-exRec {\em succeeds} if its output has no more than one error relative to the eigenstate found in the measurement of $\bar T \bar X \bar T^\dagger$; otherwise it {\em fails}.  A pair of locations is benign if the 1-exRec succeeds for arbitrary faults at those locations. To investigate whether a specified pair of locations is benign we consider arbitrary Pauli faults at the specified locations, propagate the errors to the output, and check whether the 1-exRec succeeds. However, an $X$ error propagated through a $T$ gate becomes $TXT^\dagger=(X+Y)/\sqrt{2}=X\cdot (I+iZ)/\sqrt{2}$, which is not a Pauli operator. In our analysis, we pessimistically assume that when an $X$ error propagates through a $T$ or $T^\dagger$ gate, the result is an $X$ error accompanied by a potential $Z$ error that can be turned on or off adversarially.  

A pair of locations in the $|A_{\pi/4}\rangle$ 1-exRec is benign if both locations are in the first half of the circuit (the first $\bar{T} \bar{X} \bar{T}^{\dagger}$ measurement and the first 1-EC). In that case the second half of the circuit has no faults, so the state is a codeword if the second 1-EC detects no error, and the eigenvalue found by the second $\bar{T} \bar{X} \bar{T}^{\dagger}$ measurement is correct. A pair of locations is also benign if exactly one of the locations is in the second 1-EC. In that case, since there is only one fault at a location preceding the second 1-EC, the input to the second 1-EC has at most one error, and therefore, if no error is detected by the 1-EC, its output also has at most one error; furthermore, at least one of the two $\bar{T} \bar{X} \bar{T}^{\dagger}$ measurements must be correct. By similar reasoning, we see that no $T$ or $T^\dagger$ gates belong to any malignant pair. A fault in one of these gates can cause a single error, but the error will be detected if the following 1-EC has no faults. On the other hand, if the following 1-EC does have a fault, then, if no error is detected, its output has at most one error.

It may be that both fault locations are in the final 1-EC; in that case the Pauli errors can be propagated through the circuit and the output tested for an encoded error. All other malignant pairs contain at least one location in the encoding or measurement of a cat state.

%------------------------------------------------% 
\subsection{Counting of malignant pairs: Results}
\label{subsec:results}

In general, distinct 0-Ga's are expected to have distinct fault rates. Let us designate the types of 0-Ga's by the labels 1--8, as follows:
\begin{description}
\item 1. rest during a gate cycle.
\item 2. rest during a measurement cycle.
\item 3. preparation of $|0\rangle$.
\item 4. preparation of $|+\rangle$.
\item 5. measurement of $X$.
\item 6. measurement of $Z$.
\item 7. \cnot gate.
\item 8. $T$ or $T^\dagger$ gate.
\end{description} 
We denote the probability of a fault at a location of type $i$ by $\varepsilon_i$. Since the time needed to execute a gate might differ from the time needed for a measurement, we allow the rate $\varepsilon_1$ of storage faults during gate cycles  to differ from the rate $\varepsilon_2$ of storage faults during  measurement cycles. (We will not explicitly discuss the $H$ 0-Ga, which appears in the transversal $\bar H$ 1-Ga, but not in any other gadgets.)

The 1-EC gadget contains 142 locations, which can be enumerated as follows: There are four encoders (18 locations each), four 1-measurements (7 locations each), two rests (7 locations each) and four encoded \cnot gates (7 locations each).
Thus the \cnot 1-exRec contains 575 locations: four 1-ECs (142 locations each) and one \cnot (7 locations). This is the largest 1-exRec used in the level-1 simulation. 

To use the method of Sec.~\ref{sec:malignant-pairs} to estimate the threshold, we need to be careful to take into account that ancilla blocks are used for syndrome extraction only after successfully passing a verification test. Therefore, in a recursive estimate of the failure probability for a $k$-exRec, we should upper bound the failure probability {\em conditioned on the acceptance of all ancilla blocks used in the $k$-exRec}. Furthermore, this bound should be expressed in terms of failure probabilities for the $(k{-}1)$-exRecs, also conditioned on the acceptance of all ancilla blocks used in the $(k{-}1)$-exRecs. In our enumeration of malignant pairs, it is convenient to count the pairs of locations such that suitable faults at those locations lead to the acceptance of all ancilla blocks, and cause failure of the $k$-exRec. Thus we obtain an upper bound on the {\em joint} probability of acceptance of all ancillas and failure of the $k$-exRec. After an appropriate adjustment, we can then find an upper bound on the {\em conditional} probability of failure, given acceptance of all ancillas. 

Let  $\alpha_{ij}$ denote the number of pairs of locations of types $i$ and $j$ where faults can cause the \cnot 1-exRec to fail, while all ancilla blocks contained in the 1-exRec are accepted. Then, arguing as in Sec.~\ref{sec:malignant-pairs}, we find that the {\em joint} probability $\epsilon_{7,~{\rm joint}}^{(k)}$ of failure for the \cnot $k$-exRec and acceptance of all ancillas can be bounded in terms of the failure rates $\{\varepsilon_i^{(k{-}1)}\}$ for $(k{-}1)$-exRecs ({\em conditioned} on acceptance of all ancillas contained in these $(k{-}1)$-exRecs) as
\begin{equation}
\varepsilon_{7,~{\rm joint}}^{(k)} \leq \sum_{j\leq i=1}^{7} \alpha_{ij}\, \varepsilon_i^{(k{-}1)} \varepsilon_j^{(k{-}1)} + B_{\rm CNOT}\, (\varepsilon_{\rm{max}}^{(k{-}1)})^3 ,
\end{equation}
\noindent where $\varepsilon^{(k{-}1)}_{\rm {max}}$ is the largest of the (conditional) failure rates in $\{\varepsilon^{(k{-}1)}_i\}$, and 
\begin{equation}
  B_{\rm CNOT} \equiv \left( \begin{array}{c}  575 \\  3  \end{array} \right) = 31,519,775 \, 
\end{equation}
\noindent is the number of ways to choose three locations in the \cnot 1-exRec. We have computed $\alpha_{ij}$ using the procedure outlined in Sec.~\ref{sec:malignant-procedure}. Represented as a lower triangular $7\times 7$ matrix (the $T$ and $T^\dagger$ gates do not appear in the 1-exRec), the result is:
\begin{equation}
\alpha = \left( \begin{array}{ccccccc}
              64    & \\
              624	& 630 \\
              160	& 468	& 96 \\
              160	& 468	& 0	    & 96 \\
              192	& 546	& 0	    & 288	& 168 \\
              192	& 546	& 288	& 0	    & 0	    & 168 \\
              2560	& 5924	& 1888	& 1888	& 2288	& 2288	& 13245
              \end{array} \right) \, .
\end{equation}
\noindent To treat the case where the fault probability is $\varepsilon$ at every location, we add all entries in $\alpha_{ij}$ with equal weight, obtaining  that the joint probability of failure and acceptance of all ancillas can be bounded as
\begin{equation}
\varepsilon^{(k)}_{7,~{\rm joint}} ~\leq~ A_{\rm CNOT}\left(\varepsilon^{(k{-}1)}\right)^2 + B_{\rm CNOT}\left(\varepsilon^{(k{-}1)}\right)^3 ~\le~ A'_{\rm CNOT}\left(\varepsilon^{(k{-}1)}\right)^2\, ,
\end{equation}
where 
$A_{\rm CNOT}=35,235$ malignant pairs, $\varepsilon^{(k{-}1)} < \left(A_{\rm CNOT}'\right)^{-1}$, and $A'_{\rm CNOT} \approx 36,108 $, as in eq.~(\ref{better-threshold}). Now, to obtain an upper bound on the probability of failure $\varepsilon^{(k)}_{7}$ for the \cnot $k$-exRec, conditioned on the acceptance of all ancillas, we can use Bayes' rule. Let $P^{(k)}_{|\bar 0\rangle, ~{\rm accept}}$ denote the probability that a level-$k$ encoded $|\bar 0\rangle$ or $|\bar +\rangle$ block passes the verification test. Then, since eight ancilla blocks, prepared independently, are used in the four $k$-ECs contained in the \cnot $k$-exRec, we have
\begin{equation}
\varepsilon^{(k)}_{7} =\left(P^{(k)}_{|\bar 0\rangle, ~{\rm accept}}\right)^{-8}\varepsilon^{(k)}_{7,~{\rm joint}}~.
\end{equation}
To obtain a lower bound on $P^{(k)}_{|\bar 0\rangle, ~{\rm accept}}$, and hence an upper bound on $\varepsilon^{(k)}_{7}$, we observe that for the ancilla to be rejected the encoding and verification circuit must contain at least one bad $(k{-}1)$-exRec. This circuit contains $C=50$ locations (18 for each of two encoders, 7 for the {\sc cnot}, and 7 for the measurement of the verifier block); therefore rejection occurs with probability no larger than $C\varepsilon^{(k{-}1)}$, and
\begin{equation}
P^{(k)}_{|\bar 0\rangle, ~{\rm accept}}\ge 1- C\varepsilon^{(k{-}1)}~.
\end{equation}
Therefore, assuming $\varepsilon^{(k{-}1)}< \left(A'_{\rm CNOT}\right)^{-1}$ we have
\begin{equation}
\varepsilon^{(k)}_{7} \le \left(1-C/A'_{\rm CNOT}\right)^{-8}A'_{\rm CNOT}\left(\varepsilon^{(k{-}1)}\right)^2 \le A''_{\rm CNOT}\left(\varepsilon^{(k{-}1)}\right)^2\le \varepsilon_0\left(\varepsilon/\varepsilon_0\right)^{2^k}~,
\end{equation}
where $A''_{\rm CNOT}\approx (1.0111)A'_{\rm CNOT}\approx 36,511$, and $\varepsilon_0= \left(A''_{\rm CNOT}\right)^{-1}$. Thus we obtain a rigorous lower bound on the accuracy threshold (assuming that the \cnot exRec dominates the threshold estimate): 
\begin{equation}
\varepsilon_0\ge 2.739 \times 10^{-5}\, .
\end{equation} 

If we assume that storage faults are negligible, we do not need to count the rests in the circuit, reducing the total number of locations in the \cnot 1-exRec to 487, so that $B_{\rm CNOT}=19,131,795$; furthermore, we can set $\varepsilon_1=\varepsilon_2=0$, eliminating the entries in the first two rows and columns of $\alpha$, and the number of locations in the ancilla encoding/verification circuit is reduced to $C=46$. We then find $A_{\rm CNOT}=22,701$, $A_{\rm CNOT}'\approx 23,515$, $A''_{\rm CNOT}\approx 23,887$, and $\varepsilon_0\approx 4.186 \times 10^{-5}$.

The matrix $\alpha_{ij}$ is informative about how fault locations inside the \cnot exRec combine to lead to a failure. For example, the malignant pairs of \cnot faults (counted in $\alpha_{77}$) are $37\%$ of the total, and the malignant pairs containing at least one \cnot gate (the last row of $\alpha$) are $85\%$ of the total. In contrast, storage faults belong to only $36\%$ of the total number of malignant pairs, reflecting the highly parallelized nature of Steane's error correction method. 

We need to check that the \cnot 1-exRec really dominates the accuracy threshold. Its strongest competitor is the $|A_{\pi /4}\rangle$ preparation 1-exRec shown in Fig.$\,$\ref{t_ancilla}. To count the locations in this exRec, we must keep in mind that at level 1 and above the $T$ gate is executed using the circuit in Fig.~\ref{t_circuit}, which should be regarded as four locations (the $|A_{\pi /4}\rangle$ preparation, the {\sc cnot}, the measurement, and (if necessary) the $S$ gate). Therefore the $|A_{\pi /4}\rangle$ preparation exRec has 521 locations: there are two 1-ECs (142 locations each), two circuits for cat state encoding and verification (36 locations each), two $\bar T$ gates and two $\bar T^\dagger$ gates (28 locations each), two \cnot gates from cat state to data block (7 locations each), two cat state measurements (7 locations each), a $|\bar 0\rangle$ encoding circuit (18 locations), and a rest at the end of the 1-exRec during the final measurement that determines whether the ancilla is accepted (7 locations). 

The $|A_{\pi /4}\rangle$ preparation exRec uses two cat states and four encoded ancillas, all of which must pass verification tests, and furthermore the output of the exRec is accepted only if the same measured eigenvalue is obtained twice in a row. As in our analysis of the \cnot exRec, the counting of malignant pairs in the $|A_{\pi /4}\rangle$ preparation exRec provides an upper bound on the {\em joint} probability of failure of the exRec and a successful outcome in all verification tests. Let $\beta_{ij}$ denote the number of pairs of locations of type $i$ and $j$ where faults can cause the $|A_{\pi /4}\rangle$ preparation 1-exRec to fail, while all verification tests succeed. (This matrix is $7\times 7$, because it turns out that the $T$ and $T^\dagger$ locations do not belong to any malignant pair.) Then, the joint probability $\varepsilon_{8,~{\rm joint}}^{(k)}$ of failure for the $|A_{\pi /4}\rangle$ $k$-exRec and success in all verification tests can be bounded in terms of the failure rates $\{\varepsilon_i^{(k{-}1)}\}$ for the $(k{-}1)$-exRecs as
\begin{equation}
\epsilon_{8,~{\rm joint}}^{(k)} \leq \sum \limits_{j\leq i=1}^{7} \beta_{ij}\, \varepsilon_i^{(k{-}1)} \varepsilon_j^{(k{-}1)} + B_{\t}\, (\epsilon_{\rm{max}}^{(k{-}1)})^3 ,
\end{equation}
\noindent where $\varepsilon_{\rm max}^{(k{-}1)}$ is the largest of the failure rates in $\{\varepsilon_i^{(k{-}1)}\}$, and
\begin{equation}
   B_{\t} \equiv \left( \begin{array}{c}   521 \\   3  \end{array} \right) =23,434,580
\end{equation}
is the number of ways to choose three locations in this 1-exRec. Represented as a lower triangular $7\times 7 $ matrix, the result of our computation of $\beta_{ij}$ is:

\begin{equation}
\beta = \left( \begin{array}{cccccccc}
      144 \\
      168 &  133 \\      
        0 &    0 &      2 \\
       24 &   14 &      0 &      1 \\
      168 &   98 &      0 &     14 &     49 \\
        0 &    0 &      1 &      0 &      0 &      0  \\
      360 &  462 &      8 &     30 &    210 &      2 &    442

              \end{array} \right) \,.
\end{equation} 
\noindent  To treat the case where the fault probability is $\varepsilon$ at every location, we add all entries of $\beta_{ij}$ with equal weight, finding $A_{\t}=2,330$ and  $A'_{\t} \approx 6,144$. In order for any verification to fail, there must be at least one bad $(k{-}1)$-exRec in the $k$-exRec. Therefore, all verifications succeed with probability 
\begin{equation}
P_{\rm succeed}^{(k)} \ge 1- D\varepsilon^{(k{-}1)}~,
\end{equation}
where $D=521$ is the number of locations in the exRec, and we have
\begin{equation}
\varepsilon_8^{(k)} \le \left(1-D/A'_{\t}\right)^{-1}A'_{\t}\left(\varepsilon^{(k{-}1)}\right)^2\le A''_{\t}\left(\varepsilon^{(k{-}1)}\right)^2~,
\end{equation}
where $A''_{\t}=(1.0927)A'_{\t}\approx 6,713$. Since $A''_{\t} < A_{\rm CNOT}''\approx 36,511$, the \cnot 1-exRec does indeed determine our lower bound on the quantum accuracy threshold. Therefore, we have proved:

\medskip
\noindent {\bf Theorem 3. Lower bound on the quantum accuracy threshold}. {\em Suppose that independent stochastic faults occur with probability $\varepsilon$ at each location in a noisy quantum circuit. Then for any fixed $\delta$, any ideal circuit with $L$ locations can be simulated with error $\delta$ or better by a noisy circuit with $L^*=L({\rm polylog}~L)$ locations, provided that
\begin{equation}
\varepsilon < \varepsilon_0= 2.73 \times 10^{-5}\, .
\end{equation}
}
\medskip

\noindent We note that if we merely counted the number of pairs of locations in the \cnot 1-exRec and used Theorem 1, we would estimate $\varepsilon_0={575\choose 2}^{-1}\approx 6.06 \times 10^{-6}$ (times a small correction arising from the acceptance probability of the ancillas). By counting malignant pairs and using Theorem 2 instead, we have improved the lower bound on the accuracy threshold by a factor of about 4.5. 

We have also analyzed the performance of our level-1 gadgets against depolarizing noise. For a specified malignant pair of fault locations, there are some pairs of Pauli errors that do not cause failure. In the analysis, we weight each of the three Pauli errors at a single-qubit location of type $i$ by $\varepsilon_i/3$, and we weight each of the fifteen Pauli errors at a \cnot location by $\varepsilon_7/15$. Preparation and measurement errors occur with probabilities, $\frac{2}{3}\varepsilon_3,\frac{2}{3}\varepsilon_4,\frac{2}{3}\varepsilon_5,\frac{2}{3}\varepsilon_6$. We then find that the joint probability that the \cnot 1-exRec fails and all ancillas are accepted is
\begin{equation}
\varepsilon_{7,~{\rm joint}}^{(1)} \le \sum_{j\le i=1}^7 \left(\alpha_{\rm depol}\right)_{ij}\varepsilon_i\varepsilon_j + B_{\rm CNOT}\varepsilon_{\rm max}^3~,
\end{equation}
where
\begin{equation}
\alpha_{\rm {depol}}  = \left( \begin{array}{cccccccc}
7.1	\\
138.7 &	350	\\
35.6  &	208	   &   42.7 \\
35.6  &	208	   &    0 	  & 42.7 \\
42.7  &	242.7  &	0	  & 128	   &  74.7 \\
42.7  &	242.7  &	128	  & 0      &	0   &  74.7 \\
287.3 &	1462.4 &	423.8 &	423.8  &   512	&   512	 & 1517.2
                            \end{array} \right) \,.
\end{equation} 
\noindent Again, we can treat the case where the fault probability is $\varepsilon$ at every location by adding all entries in $\alpha_{ij}$ with equal weight, obtaining $A_{\rm CNOT}\approx 7,183$ malignant pairs, $A'_{\rm CNOT} \approx 10,256$, and $A''_{\rm CNOT} \approx 10,665$. The value $A''_{\t} \approx 6,713$ for the $|A_{\pi/4}\rangle$ preparation with adversarial noise is already smaller than our value of $A'_{\rm CNOT}$ for depolarizing noise, so there is no need to reanalyze the $|A_{\pi/4}\rangle$ preparation for depolarizing noise.

Thus we find that $\varepsilon_i^{(1)} < \varepsilon$ for $\varepsilon < (A''_{\rm CNOT})^{-1}\equiv \varepsilon_{0,{\rm depol}}^{(1)}$, or
\begin{equation}
\varepsilon_{0,{\rm depol}}^{(1)} \approx 9.376 \times 10^{-5}~,
\end{equation}
an improvement over our estimate of $\varepsilon_0$ by about a factor of 3.4. We emphasize, though, that in contrast to our calculation of $\varepsilon_0$ for adversarial independent stochastic noise reported in Theorem 3, this calculation of $\varepsilon_{0,{\rm depol}}^{(1)}$ is not a rigorous lower bound on the accuracy threshold for depolarizing noise. Rather, $\varepsilon_{0,{\rm depol}}^{(1)}$ is a lower bound on the critical fault rate for which the level-1 simulation is more reliable than the unprotected level-0 circuit, and at best a rough indication of how effectively a recursive simulation protects against depolarizing noise compared to adversarial noise. (If we assume that storage faults are negligible ($\varepsilon_1=\varepsilon_2=0$ and $C=46$), then $A_{\rm CNOT}\approx 3,880$, $A_{\rm CNOT}'\approx 6,725$, $A''_{\rm CNOT}\approx 7,105$, and $\varepsilon_{0,{\rm depol}}^{(1)} \approx 1.407 \times 10^{-4}$.)

Higher estimates of the accuracy threshold have been reported based on numerical and heuristic studies of fault-tolerant gadgets for the [[7,1,3]] code \cite{Zalka00,Steane02,divincenzo, reichardt,knill_detect}. We emphasize again that our lower bound on the quantum accuracy threshold (Theorem 3), in contrast to those previous estimates, has been rigorously proven.

\section{Higher-distance codes}
\label{sec:higher-distance}

As we have formulated it, the quantum threshold theorem is premised on the existence of a universal set of level-1 gadgets that satisfy the property exRec-Cor. Our proof shows that exRec-Cor is also satisfied at higher levels, and that bad $k$-exRecs are very rare. 

In Sec.~\ref{sec:level-1} and \ref{sec:recursive}, we stated conditions on the level-1 gadgets that suffice to ensure that exRec-Cor is satisfied, and those conditions guided our explicit gadget constructions in Sec.~\ref{sec:ft-general} and \ref{sec:explicit}. Now we will revisit these conditions and state them in a different language.

Our motivation is two-fold. First, the original statement of the conditions employs the somewhat vague notion that an input or output block has ``at most one error.'' But as we have emphasized previously, since the qubits comprising an encoded block are highly entangled, an error damages not the local state of the qubit but rather its correlations with other qubits. Therefore, it is preferable to characterize gadgets {\em syntactically} rather than {\em semantically} --- that is, to speak of the properties of operators rather than the properties of states. Our restated criteria are syntactic. Secondly, it will be especially easy to apply these criteria to quantum codes that can correct multiple errors in a block.

For a code that can correct $t$ errors, we will say that a 1-exRec is good if it contains no more than $t$ faults. As before, a 1-Rec is correct if the 1-Rec followed by the ideal 1-decoder is equivalent to the ideal 1-decoder followed by the corresponding ideal 0-Ga. At level 1, exRec-Cor says that the 1-Rec contained in a good 1-exRec is correct; that is, for a good 1-exRec:

\vspace{0.5cm}
\begin{picture}(446,24)
\put(0,12){\line(1,0){10}}
\put(10,0){\framebox(48,24){1-EC}}
\put(58,12){\line(1,0){10}}
\put(68,0){\framebox(48,24){1-Ga}}
\put(116,12){\line(1,0){10}}
\put(126,0){\framebox(48,24){1-EC}}
\put(174,12){\line(1,0){10}}
\put(184,0){\framebox(48,24){\shortstack{ideal\\$1$-decoder}}}
\put(232,12){\line(1,0){10}}

\put(242,6){\makebox(20,12){=}}
\put(262,12){\line(1,0){10}}
\put(272,0){\framebox(48,24){1-EC}}
\put(320,12){\line(1,0){10}}
\put(330,0){\framebox(48,24){\shortstack{ideal\\$1$-decoder}}}
\put(378,12){\line(1,0){10}}
\put(388,0){\framebox(48,24){0-Ga}}
\put(436,12){\line(1,0){10}}
\put(446,6){\makebox(20,12){.}}
\end{picture}
\vspace{0.3cm}

We will state some (syntactic) properties of 1-gadgets, and show that these properties imply exRec-Cor at level 1. These properties do invoke the condition that the input or output of a gadget is not too badly damaged, but in a form that can be stated syntactically. Namely, it is enough to characterize how a state deviates from the code space, rather than its deviation from an ideal state in the code space. For this purpose, we introduce the concept of a {\em filter}. An $s$-filter is the orthogonal projection onto the space spanned by all states that can be obtained by acting on a codeword with a Pauli operator of weight no larger than $s$. We will say that a gadget is ``$r$-good'' if it contains no more than $r$ faults. 

The properties satisfied by the 1-gadgets are more precise formulations of the properties that we stated earlier for distance-3 codes ($t=1$), and they also generalize those properties to larger values of $t$. These properties are as follows.

\medskip
\noindent {\em Property $0$}:

\vspace{0.5cm}
\begin{picture}(368,24)
\put(0,12){\line(1,0){10}}
\put(10,0){\framebox(48,24){\shortstack{$r$-good\\1-EC}}}
\put(58,12){\line(1,0){10}}
\put(68,6){\makebox(20,12){=}}
\put(88,12){\line(1,0){10}}
\put(98,0){\framebox(48,24){\shortstack{$r$-good\\1-EC}}}
\put(146,12){\line(1,0){10}}
\put(156,0){\framebox(48,24){$r$-filter}}
\put(204,12){\line(1,0){10}}
\put(300,0){\makebox(68,24){$(r \le t)$}}
\end{picture}
\vspace{0.3cm}

\noindent {\em Property $1$}:

\vspace{0.5cm}
\begin{picture}(436,24)
\put(0,12){\line(1,0){10}}
\put(10,0){\framebox(48,24){$s$-filter}}
\put(58,12){\line(1,0){10}}
\put(68,0){\framebox(48,24){\shortstack{$r$-good\\1-EC}}}
\put(116,12){\line(1,0){10}}
\put(126,0){\framebox(48,24){\shortstack{ideal\\$1$-decoder}}}
\put(174,12){\line(1,0){10}}

\put(184,6){\makebox(20,12){=}}
\put(204,12){\line(1,0){10}}
\put(214,0){\framebox(48,24){$s$-filter}}
\put(262,12){\line(1,0){10}}
\put(272,0){\framebox(48,24){\shortstack{ideal\\$1$-decoder}}}
\put(320,12){\line(1,0){10}}
\put(368,0){\makebox(68,24){$(r+s\le t)$}}
\end{picture}
\vspace{0.3cm}

\noindent {\em Property $2$}:

\vspace{0.5cm}
\begin{picture}(456,24)
\put(0,12){\line(1,0){10}}
\put(10,0){\framebox(48,24){$\{s_i\}$-filters}}
\put(58,12){\line(1,0){10}}
\put(68,0){\framebox(48,24){\shortstack{$r$-good\\1-Ga}}}
\put(116,12){\line(1,0){10}}
\put(126,6){\makebox(20,12){=}}
\put(146,12){\line(1,0){10}}
\put(156,0){\framebox(48,24){$\{s_i\}$-filters}}
\put(204,12){\line(1,0){10}}
\put(214,0){\framebox(48,24){\shortstack{$r$-good\\1-Ga}}}
\put(262,12){\line(1,0){10}}
\put(272,0){\framebox(48,24){$s$-filter}}
\put(320,12){\line(1,0){10}}
\put(350,0){\makebox(90,24){$(s=r+\sum_i s_i \le t)$}}
\end{picture}
\vspace{0.3cm}

\noindent
{\em Property $3$}:

\vspace{0.5cm}
\begin{picture}(472,24)
\put(0,12){\line(1,0){10}}
\put(10,0){\framebox(48,24){$\{s_i\}$-filters}}
\put(58,12){\line(1,0){10}}
\put(68,0){\framebox(48,24){\shortstack{$r$-good\\1-Ga}}}
\put(116,12){\line(1,0){10}}
\put(126,0){\framebox(48,24){\shortstack{ideal\\$1$-decoder}}}
\put(174,12){\line(1,0){10}}

\put(184,6){\makebox(20,12){=}}
\put(204,12){\line(1,0){10}}
\put(214,0){\framebox(48,24){$\{s_i\}$-filters}}
\put(262,12){\line(1,0){10}}
\put(272,0){\framebox(48,24){\shortstack{ideal\\$1$-decoder}}}
\put(320,12){\line(1,0){10}}

\put(330,0){\framebox(48,24){\shortstack{ideal\\$0$-Ga}}}
\put(378,12){\line(1,0){10}}
\put(398,0){\makebox(68,24){$(r+\sum_i s_i\le t)$}}
\end{picture}
\vspace{0.3cm}

\noindent For properties 2 and 3, in the case where the 1-Ga has more than one input block, by $\{s_i\}$-filters we mean that the $s_i$-filter is applied to the $i$th input block, while the $s$-filter on the right means that {\em each} output block deviates from the code space by at most distance $s$. 

There are also special versions of the properties that hold for the 1-preparation and the 1-measurement. For the 1-preparation, property 2 becomes:

\vspace{0.5cm}
\begin{picture}(194,24)

\put(0,0){\framebox(48,24){\shortstack{$r$-good\\1-prep.}}}
\put(48,12){\line(1,0){10}}
\put(58,6){\makebox(20,12){=}}

\put(78,0){\framebox(48,24){\shortstack{$r$-good\\1-prep.}}}
\put(126,12){\line(1,0){10}}
\put(136,0){\framebox(48,24){$r$-filter}}
\put(184,12){\line(1,0){10}}
\end{picture}
\vspace{0.3cm}

\noindent
and property 3 is:

\vspace{0.5cm}
\begin{picture}(268,24)

\put(0,0){\framebox(48,24){\shortstack{$r$-good\\1-prep.}}}
\put(48,12){\line(1,0){10}}
\put(58,0){\framebox(48,24){\shortstack{ideal\\$1$-decoder}}}
\put(106,12){\line(1,0){10}}

\put(116,6){\makebox(20,12){=}}
\put(136,0){\framebox(48,24){\shortstack{ideal\\$0$-prep.}}}
\put(184,12){\line(1,0){10}}
\put(194,0){\makebox(68,24){$(r\le t)$}}
\end{picture}
\vspace{0.3cm}

\noindent
For the 1-measurement, which has only a classical output, there is no analog of property 2, but property 3 becomes:

\vspace{0.5cm}
\begin{picture}(378,24)
\put(0,12){\line(1,0){10}}
\put(10,0){\framebox(48,24){$s$-filter}}
\put(58,12){\line(1,0){10}}
\put(68,0){\framebox(48,24){\shortstack{$r$-good\\1-meas.}}}

\put(116,6){\makebox(20,12){=}}
\put(136,12){\line(1,0){10}}
\put(146,0){\framebox(48,24){$s$-filter}}
\put(194,12){\line(1,0){10}}
\put(204,0){\framebox(48,24){\shortstack{ideal\\$1$-decoder}}}
\put(252,12){\line(1,0){10}}

\put(262,0){\framebox(48,24){\shortstack{ideal\\$0$-meas.}}}

\put(310,0){\makebox(68,24){$(r+s\le t)$}}
\end{picture}
\vspace{0.3cm}

\noindent
(Here we assume that the classical post-processing of the measurement outcomes is flawless.) 

Though we will not discuss the details, by following the principles of quantum fault tolerance, gadgets with these properties can be constructed for a stabilizer code that corrects $t$ errors. Property 0 says that a 1-EC with $r$ faults takes any input to an output that deviates from the code space by no more than distance $r$, and property 1 expresses that a 1-EC with $r$ faults takes an input with $s$ errors to an output that can still be successfully decoded. Property 2 says that a 1-Ga does not propagate errors too badly: if the 1-Ga has $r$ faults, then the number of errors in each output block is not more than $r$ plus the total number of errors in all input blocks; Property 3 says that these output blocks can be accurately decoded.

For the case $t=1$, these properties capture (syntactically) the content of the properties stated in Sec.~\ref{sec:level-1} and \ref{sec:recursive}. The new property 0 replaces the old properties 0 and $0'$, the new property 1 replaces the old properties 1 and 2, the new properties 2 and 3 replace the old properties 3 and 4. 

Using these new properties, we can establish:

\medskip
\noindent{\bf Lemma 5. exRec-Cor at level 1 (for a code that corrects $t$ errors)}. {\em Suppose that the level-1 gadgets obey properties 0--3. Then the 1-exRecs obey exRec-Cor.}

\medskip
\noindent {\bf Proof}: To simplify notation, we will consider the case of a single-qubit gate --- the argument is essentially the same for a multi-qubit gate. A 1-exRec is good if it contains no more than $t$ faults. Suppose there are $s$ faults in the leading 1-EC, $r$ faults in the 1-Ga, and $s'$ faults in the trailing 1-EC, with $s + r + s'\le t$. Then the 1-exRec followed by the ideal 1-decoder can be expressed as

\vspace{0.5cm}
\begin{picture}(242,24)
\put(0,12){\line(1,0){10}}
\put(10,0){\framebox(48,24){\shortstack{$s$-good\\$1$-EC}}}
\put(58,12){\line(1,0){10}}
\put(68,0){\framebox(48,24){\shortstack{$r$-good\\$1$-Ga}}}
\put(116,12){\line(1,0){10}}
\put(126,0){\framebox(48,24){\shortstack{$s'$-good\\$1$-EC}}}
\put(174,12){\line(1,0){10}}
\put(184,0){\framebox(48,24){\shortstack{ideal\\$1$-decoder}}}
\put(232,12){\line(1,0){10}}
\put(242,6){\makebox(20,12){.}}
\end{picture}
\vspace{0.3cm}

\noindent Using property 0 we can insert a filter after the leading 1-EC, and using property 2, we can insert a filter after the 1-Ga:

\vspace{0.5cm}
\begin{picture}(378,24)
\put(0,6){\makebox(20,12){=}}
\put(20,12){\line(1,0){10}}
\put(30,0){\framebox(48,24){\shortstack{$s$-good\\$1$-EC}}}
\put(78,12){\line(1,0){10}}
\put(88,0){\framebox(48,24){$s$-filter}}
\put(136,12){\line(1,0){10}}
\put(146,0){\framebox(48,24){\shortstack{$r$-good\\$1$-Ga}}}
\put(194,12){\line(1,0){10}}
\put(204,0){\framebox(48,24){\shortstack{$(s+r)$-\\filter}}}
\put(252,12){\line(1,0){10}}
\put(262,0){\framebox(48,24){\shortstack{$s'$-good\\$1$-EC}}}
\put(310,12){\line(1,0){10}}
\put(320,0){\framebox(48,24){\shortstack{ideal\\$1$-decoder}}}
\put(368,12){\line(1,0){10}}
\put(378,6){\makebox(20,12){.}}
\end{picture}
\vspace{0.3cm}

\noindent
Using property 1, we can omit the trailing 1-EC, then using property 2, we can omit the filter that precedes it:

\vspace{0.5cm}
\begin{picture}(362,24)
\put(0,6){\makebox(20,12){=}}
\put(20,12){\line(1,0){10}}
\put(30,0){\framebox(48,24){\shortstack{$s$-good\\$1$-EC}}}
\put(78,12){\line(1,0){10}}
\put(88,0){\framebox(48,24){$s$-filter}}
\put(136,12){\line(1,0){10}}
\put(146,0){\framebox(48,24){\shortstack{$r$-good\\$1$-Ga}}}
\put(194,12){\line(1,0){10}}
\put(204,0){\framebox(48,24){\shortstack{ideal\\$1$-decoder}}}
\put(252,12){\line(1,0){10}}
\put(262,6){\makebox(20,12){.}}
\end{picture}
\vspace{0.3cm}

\noindent Finally, using property 3, we can move the decoder to the left, converting the 1-Ga to the ideal 0-Ga, and using property 0, we can remove the filter that precedes it, thus obtaining

\vspace{0.5cm}
\begin{picture}(204,24)
\put(0,6){\makebox(20,12){=}}
\put(20,12){\line(1,0){10}}
\put(30,0){\framebox(48,24){\shortstack{$s$-good\\$1$-EC}}}
\put(78,12){\line(1,0){10}}
\put(88,0){\framebox(48,24){\shortstack{ideal\\$1$-decoder}}}
\put(136,12){\line(1,0){10}}
\put(146,0){\framebox(48,24){\shortstack{ideal\\$0$-Ga}}}
\put(194,12){\line(1,0){10}}
\put(204,6){\makebox(20,12){.}}
\end{picture}
\vspace{0.3cm}

\noindent This proves exRec-Cor.

\rightline{$\square$}

Our proof of the threshold theorem applies without much modification to codes that correct $t\ge 2$ errors. To define badness at level $k+1$, we need a notion of what it means for two successive overlapping $k$-exRecs to fail independently. Again, we are guided by the goal of proving exRec-Cor inductively via the threshold dance. As the ideal $k$-decoders sweep to the left through a $(k{+}1)$-exRec, they hop over a full $k$-exRec that contains more than $t$ bad $(k{-}1)$-exRecs. Then whether the preceding gate in the circuit is simulated accurately is determined by whether the $k$-ECGa contains more than $t$ faults. Thus, we may say that a $(k{+}1)$-exRec is bad if it contains $t+1$ independent bad $k$-exRecs, where two successive bad exRecs are said to be independent only if the earlier $k$-ECGa contains $t+1$ independent bad $(k{-}1)$-exRecs.

With this definition of goodness we can prove exRec-Cor at level $k$ inductively. Furthermore, since a $k$-exRec is bad only if it contains $t+1$ $(k{-}1)$-exRecs that fail independently, the probability $\varepsilon^{(k)}$ that a $k$-exRec is bad satisfies

\begin{equation}
\varepsilon^{(k)} \le A\left(\varepsilon^{(k{-}1)}\right)^{t+1}~,
\end{equation}
where $A$ is the number of ways to choose $t+1$ locations in the largest exRec.
This inequality can be rewritten as
\begin{equation}
\varepsilon^{(k)}/\varepsilon_0 \le \left(\varepsilon^{(k{-}1)}/\varepsilon_0\right)^{t+1}~,
\end{equation}
where 
\begin{equation}
\varepsilon_0=A^{-1/t}~,
\end{equation}
which can be iterated to find
\begin{equation}
\varepsilon^{(k)} \le \varepsilon_0\left(\varepsilon/\varepsilon_0\right)^{(t+1)^k}~.
\end{equation}
Arguing as in Sec.~\ref{sec:threshold-theorem} we arrive at a generalization of Theorem 1.

\medskip
\noindent {\bf Theorem 4. Quantum accuracy threshold for independent stochastic noise (using a code that corrects $t$ errors)}. {\em Suppose that fault-tolerant gadgets can be constructed such that all 1-exRecs obey the property exRec-Cor, where a 1-exRec that contains no more than $t$ faults is said to be good. Suppose that $\ell$ is the maximal number of locations in a 1-Rec, $d$ is the maximal depth of a 1-Rec, and $\varepsilon_0^{-t}$ is the maximal number of ways to choose $t+1$ locations in a 1-exRec. Suppose that independent stochastic faults occur with probability $\varepsilon < \varepsilon_0$ at each location in a noisy quantum circuit. Then for any fixed $\delta$, any ideal circuit with $L$ locations and depth $D$ can be simulated with error $\delta$ or better by a noisy circuit with $L^*$ locations and depth $D^*$, where
\begin{equation}
L^*=O\left(L(\log L)^{\alpha}\right)~,\quad D^*= O\left(D(\log L)^{\beta}\right)~,
\end{equation}
and
\begin{equation}
\alpha= {\log \ell \over \log (t+1)}~,\quad \beta= {\log d\over \log (t+1)}~.
\end{equation}
}
\medskip

\noindent As in Sec.~\ref{sec:malignant-pairs}, we can improve the threshold by counting {\em malignant} sets of $t+1$ locations, thus obtaining a generalization of Theorem 2.
%--------------------------------------------------------------------------------------------%
\section{An alternative proof}
\label{sec:distance-5}

The proof of Theorem 1 that we presented in Sec.~\ref{sec:threshold-theorem} applies to the case of concatenated distance-3 codes, and also, as extended in Theorem 4, to concatenation of codes with larger distance. Here we will present another proof, really an elaboration of the proof in \cite{ben-or}, which applies to concatenation of codes with distance 5 or more, but not to codes with distance 3.

As in our previous proof, the key is to define notions of goodness and correctness so that ``good implies correct'' is satisfied by fault-tolerant level-1 gadgets, and can also be established for higher-level rectangles by an inductive argument. Again, whether a rectangle simulates an ideal gate accurately depends on the context, on what precedes the rectangle in the circuit. In our previous argument, the appropriate context was established by considering the extended rectangle that contains the rectangle. Now the context will be provided by considering the output from the preceding rectangle. With this approach, we avoid the complication of dealing with extended rectangles that overlap with one another. But other complications arise instead. 

Since many of the same ideas are used in this proof as in the proof of Theorem 1, we will be sketchier than before, emphasizing the parts of the argument that require modification.

\begin{figure}
\begin{center}
\vspace{0.5cm}
\begin{picture}(160,54)
\put(0,19){\line(1,0){5}}
\put(0,35){\line(1,0){5}}
\put(5,13){\framebox(20,27){\footnotesize 0-Ga}}
\put(25,19){\line(1,0){5}}
\put(25,35){\line(1,0){5}}
\put(40,21){\makebox(30,12){$\Longrightarrow$}}

\put(78,12){\line(1,0){10}}
\put(88,0){\framebox(24,24){1-EC}}
\put(78,42){\line(1,0){10}}
\put(88,30){\framebox(24,24){1-EC}}
\put(112,12){\line(1,0){10}}
\put(112,42){\line(1,0){10}}
\put(122,0){\framebox(28,54){1-Ga}}
\put(150,12){\line(1,0){10}}
\put(150,42){\line(1,0){10}}
\end{picture}
\vspace{0.3cm}
\end{center}
\fcaption{Level-1 simulation for a distance-5 code. Each 0-Ga in the ideal circuit is replaced by a 1-Rec, which consists of the 1-Ga that simulates the 0-Ga, {\em preceded by} a 1-EC acting on each input 1-block of the 1-Ga.}
\label{fig:new-level-1-sim}
\end{figure}

\subsection{Properties of the level-1 gadgets}
In our level-1 simulation, a 0-Ga of the ideal circuit will be simulated by a 1-Rec that is composed of the appropriate 1-Ga {\em preceded by} 1-ECs acting on each input 1-block, as shown in Fig.~\ref{fig:new-level-1-sim}. (It is useful for the 1-ECs in the 1-Rec to act before the 1-Ga, so that we can formulate and use the property Rec-SemiCor, stated below.)

Because we are now using a quantum error-correcting code that can correct two errors, it is possible by following the principles of quantum fault tolerance to construct a universal set of 1-Recs with the property:
\begin{description}
\item {\bf Rec-Cor at level 1}. {\em If a 1-Rec contains no more than one fault, and its input has no more than one error in each input block, then its output has no more than one error in each output block.}
\end{description}
\noindent
We will say that a 1-Rec is {\em good} if it contains no more than one fault, and that a 1-Rec is {\em correct} if it takes an input with no more than one error per 1-block to an output with no more than one error per 1-block. (A preparation 1-Rec is correct if its output has no more than one error, and a measurement 1-Rec is correct if it successfully measures a 1-block that has no more than one error.) Then the property Rec-Cor can be stated more succinctly as
\begin{description}
\item {\bf Rec-Cor at level 1}. {\em A good 1-Rec is correct.} 
\end{description}

For independent stochastic faults that occur with probability $\varepsilon$ at each circuit location, a level-1 simulation of an ideal circuit with $L$ locations will be successful if every 1-Rec is good. Therefore, we may again bound the probability of failure of the simulation using eq.~(\ref{level1-pfail}), but 
where now $A$ is the number of pairs of locations in the largest 1-Rec.

\subsection{Recursive simulation: validity, goodness, and correctness}

The level of the simulation is advanced by one level by replacing each 0-Ga by a 1-Rec. We define goodness for a $k$-Rec in the obvious way --- it is good if it contains no more than one bad $(k{-}1)$-Rec. Thus the probability that a $k$-Rec is bad satisfies eq.~(\ref{p-bad-bound}), where $\varepsilon_0^{-1}$ is the number of pairs of locations in the largest 1-Rec.

Defining correctness is more subtle. In order to carry out the inductive step smoothly, we will define correctness using an ideal $k$-decoder, but we also wish to capture the idea that a correct $k$-Rec simulates an ideal gate accurately if each input $k$-block has no more than {\em one} $(k{-}1)$-block that is badly damaged. For this purpose, we will introduce a notion of {\em validity} for $k$-blocks. A 1-block is valid if it deviates by no more than distance 1 from the code space, and a $k$-block is valid if its deviation from the code space can be attributed to errors that are sparse in a certain sense. 

To make these notions precise, we need to define our ideal $k$-decoder properly. The ideal 1-decoder operates as follows: First, it measures and records the error syndrome. If the syndrome indicates a single error in the 1-block, that error is corrected, and then the 1-block is decoded to a qubit. But if the syndrome indicates more than one error, the 1-block is decoded as a random qubit (a maximally mixed state). Finally, the syndrome is discarded. Note that although the code is capable in principle of correcting two errors, the ideal decoder does not fully exploit the code's error-correcting power.

The level-$k$ decoder is defined recursively; it is realized by first applying the $(k{-}1)$-decoder to each of the $(k{-}1)$-subblocks, and then applying the 1-decoder to the resulting 1-block. 

We say that the state of a $k$-block is {\em valid} if, with probability one, the outcome of the syndrome measurement indicates no more than one error at the top level. That is, if the ideal $k$-decoder is viewed as $(k{-}1)$ decoders acting on all $(k{-}1)$-subblocks, followed by a final 1-decoder, then a $k$-block is valid if, with probability one, the final 1-decoder finds a syndrome pointing to no more than one error. Thus, validity is a property of a state that can be checked locally, at least in principle --- whether a state of a $k$-block is valid does not concern the nature of the entanglement of the $k$-block with other $k$-blocks. In fact, valid states form a linear subspace of the Hilbert space of the $k$-block, and we can define a {\em validity filter}, an orthogonal projector onto this subspace. 

With our definition of validity in hand, we can formulate what we mean by correctness. 

\medskip
\noindent {\bf Definition. Correctness}. {\em A $k$-Rec is {\em correct} if, acting on a valid input, the $k$-Rec followed by the ideal $k$-decoder is equivalent to the ideal $k$-decoder followed by the ideal 0-Ga that the $k$-Rec simulates:}

\vspace{0.5cm}
\begin{picture}(420,24)
\put(0,0){\makebox(68,24){valid input}}
\put(64,12){\line(1,0){10}}
\put(74,0){\framebox(48,24){\shortstack{correct\\$k$-Rec}}}
\put(122,12){\line(1,0){10}}
\put(132,0){\framebox(48,24){\shortstack{ideal\\$k$-decoder}}}
\put(180,12){\line(1,0){10}}
\put(190,6){\makebox(30,12){=}}
\put(210,0){\makebox(64,24){valid input}}
\put(274,12){\line(1,0){10}}
\put(284,0){\framebox(48,24){\shortstack{ideal\\$k$-decoder}}}
\put(332,12){\line(1,0){10}}
\put(342,0){\framebox(48,24){\shortstack{ideal\\$0$-Ga}}}
\put(390,12){\line(1,0){10}}
\put(400,6){\makebox(20,12){.}}
\end{picture}
\vspace{0.3cm}

\noindent (When we say that the input is valid, we mean that each input block is valid.)
Preparation and measurement are special cases. A $k$-preparation is correct if the ideal $k$-decoder maps its output to the ideally prepared state:

\vspace{0.5cm}
\begin{picture}(272,24)
\put(10,0){\framebox(48,24){\shortstack{correct\\$k$-prep.}}}
\put(58,12){\line(1,0){10}}
\put(68,0){\framebox(48,24){\shortstack{ideal\\$k$-decoder}}}
\put(116,12){\line(1,0){10}}
\put(126,6){\makebox(20,12){=}}

\put(146,0){\framebox(48,24){\shortstack{ideal\\$0$-prep.}}}
\put(194,12){\line(1,0){10}}
\put(204,6){\makebox(20,12){,}}
\end{picture}
\vspace{0.3cm}

\noindent and a $k$-measurement is correct if, acting on a valid input, it realizes the same POVM as the ideal $k$-decoder followed by the ideal 0-measurement:

\vspace{0.5cm}
\begin{picture}(420,24)
\put(0,0){\makebox(68,24){valid input}}
\put(64,12){\line(1,0){10}}
\put(74,0){\framebox(48,24){\shortstack{correct\\$k$-meas.}}}
\put(122,6){\makebox(30,12){=}}
\put(142,0){\makebox(64,24){valid input}}
\put(206,12){\line(1,0){10}}
\put(216,0){\framebox(48,24){\shortstack{ideal\\$k$-decoder}}}
\put(264,12){\line(1,0){10}}
\put(274,0){\framebox(48,24){\shortstack{ideal\\$0$-meas.}}}
\put(322,6){\makebox(20,12){.}}
\end{picture}
\vspace{0.3cm}

To prove the threshold theorem, we will need two properties of $k$-Recs:
\begin{description}
\item {\bf Rec-Cor}.  {\em A good $k$-Rec is correct.}
\item {\bf Rec-Val}. {\em For any input, the output of a good $k$-Rec is valid.}
\end{description}

\noindent (When we say that the output is valid, we mean that all output blocks are valid.)
If both properties hold, then the level-$k$ fault-tolerant simulation of an ideal quantum circuit will succeed if all $k$-Recs are good. To reach this conclusion, we first use Rec-Val to infer that every $k$-Rec has a valid output. Since the final measurements are correct and have valid inputs, we can replace them by ideal $k$-decoders followed by ideal 0-measurements. Now since all $k$-Recs simulating gates are correct and have valid inputs, we can move the $k$-decoders to the left through the circuit one step at a time until they reach the $k$-preparations. Since the $k$-preparations are correct, the ideal $k$-decoders transform them to ideal 0-preparations. 

For distance-5 codes, we can build fault-tolerant 1-gadgets such that Rec-Cor and Rec-Val are true. Now we need to show that they are true at level $k$ if they are true at level $k-1$. To carry out the inductive step, it is useful to introduce a third property. We define

\medskip
\noindent {\bf Definition. Semi-correctness}. {\em A $k$-Rec is {\em semi-correct} if, for any input, the $k$-Rec followed by the ideal $k$-decoder is equivalent to the $k$-EC contained in the $k$-Rec, followed next by the ideal $k$-decoder and then by the ideal 0-Ga that the $k$-Rec simulates:}

\vspace{0.5cm}
\begin{picture}(408,24)
\put(0,12){\line(1,0){10}}
\put(10,0){\framebox(48,24){$k$-EC}}
\put(58,12){\line(1,0){10}}
\put(68,0){\framebox(48,24){$k$-Ga}}
\put(116,12){\line(1,0){10}}
\put(126,0){\framebox(48,24){\shortstack{ideal\\$k$-decoder}}}
\put(174,12){\line(1,0){10}}
\put(184,6){\makebox(20,12){=}}
\put(204,12){\line(1,0){10}}
\put(214,0){\framebox(48,24){$k$-EC}}
\put(262,12){\line(1,0){10}}
\put(272,0){\framebox(48,24){\shortstack{ideal\\$k$-decoder}}}
\put(320,12){\line(1,0){10}}
\put(330,0){\framebox(48,24){\shortstack{ideal\\$0$-Ga}}}
\put(378,12){\line(1,0){10}}
\put(388,6){\makebox(20,12){.}}
\end{picture}
\vspace{0.3cm}

\medskip
\noindent And the third property is
\begin{description}
\item {\bf Rec-SemiCor}. {\em A good $k$-Rec is semi-correct.}
\end{description}

\medskip
\noindent Our fault-tolerant 1-gadgets can also be constructed to have the property Rec-SemiCor. Thus, to complete our alternative proof of the threshold theorem we need

\medskip
\noindent {\bf Lemma 6. Good implies correct}. {\em Suppose that all 1-Recs satisfy properties Rec-Cor, Rec-Val, and Rec-SemiCor. Then Rec-Cor, Rec-Val, and Rec-SemiCor  hold for all $k$-Recs at each level $k\ge 1$.} 
\medskip

We assume the induction hypothesis, that Rec-Cor, Rec-Val, and Rec-SemiCor hold at level-$k$. We are to prove all three properties at level-$(k{+}1)$. As in Sec.~\ref{good-implies-correct}, the proofs exploit the recursive construction of the ideal $(k{+}1)$-decoder --- it can be regarded as $k$-decoders acting on all $k$-subblocks, followed by a final 1-decoder. Using the induction hypothesis, we aim to move the $k$-decoders step-by-step to the left through the $(k{+}1)$-Rec, transforming it to a good 1-Rec, and then use the properties of the 1-Recs to complete the proof. 

By Rec-Val at level $k$, each good $k$-Rec has a valid output. Therefore, using Rec-Cor at level $k$, the $k$-decoder can be moved one step to the left past a good $k$-Rec that follows other good $k$-Recs, converting the $k$-Rec to an ideal 0-Ga. 

But as in Sec.~\ref{good-implies-correct}, the key to the argument is to show that a bad $k$-Rec gets transformed to a faulty 0-Ga as the $k$-decoder moves to the left.  A good $k$-Rec that follows a bad $k$-Rec may not have a valid input, so we cannot use Rec-Cor to justify moving a $k$-decoder to the left through the $k$-Rec. Instead, we use Rec-SemiCor, and move the $k$-decoder through the $k$-Ga, transforming it to an ideal 0-Ga. 

Now the bad $k$-Rec is accompanied by its trailing $k$-ECs, forming what we will call a bad $k$-RecEC. We would like to move the $k$-decoders left though the $k$-RecEC, transforming it to a faulty 0-Ga. But we encounter the same technical glitch that we confronted in Sec.~\ref{good-implies-correct}, and we overcome the difficulty in the same way, by replacing the $k$-decoders by $k$-$^*$decoders that retain their output syndromes.  Then, as in that previous discussion, moving $k$-$^*$decoders to the left past a good $k$-Rec that follows good $k$-Recs transforms the $k$-Rec to an ideal 0-Ga (using Rec-Val and Rec-Cor at level $k$), tensored with an operation that acts only on the syndrome. Similarly moving $k$-$^*$decoders to the left past the $k$-Ga contained in a good $k$-Rec transforms the $k$-Ga to an ideal 0-Ga (using Rec-SemiCor at level $k$), tensored with an operation that acts only on the syndrome. Furthermore, moving $k$-$^*$ decoders left past a bad $k$-RecEC transforms it to a faulty 0-Ga that acts jointly on its input qubits and on the syndrome.  

In the proof of Rec-Cor at level $k+1$, we are to assume that the input blocks of the $(k{+}1)$-Rec are valid. If a $(k{+}1)$-block is valid, that means that if $k$-decoders are applied to all the $k$-subblocks, then with probability one at most one of the $k$-subblocks has a syndrome indicating two or more errors at the top level. Therefore, a valid $(k{+}1)$-block can be expanded, such that for each term in the expansion no more than one $k$-subblock is invalid. In the proof of Rec-Cor below, we will glibly say that the valid $(k{+}1)$-block has no more than one invalid $k$-subblock. But what we are really doing is considering one fixed term in this expansion; since the proof works for each such term, it works for arbitrary valid input $(k{+}1)$-blocks.

\medskip
\noindent {\bf Proof of Rec-Cor}: The $(k{+}1)$-decoders placed behind the $(k{+}1)$-Rec can be realized as $k$-$^*$decoders applied to all $k$-subblocks of the output $(k{+}1)$-blocks, followed by final 1-decoders (where the output syndromes of the $k$-$^*$decoders are discarded). As described above, using Rec-Val, Rec-Cor, and Rec-SemiCor at level $k$, we can sweep the $k$-$^*$decoders that follow the $(k{+}1)$-Rec to the left one step at a time until they reach the front of the $(k{+}1)$-Rec; thereby we transform each good $k$-Rec to an ideal 0-Ga, and each bad $k$-Rec to a faulty 0-Ga.

The input to the $(k{+}1)$-Rec is valid; therefore, each input $(k{+}1)$-block has at most one invalid $k$-block. 
One of the $k$-Recs in the first time step of the $(k{+}1)$-Rec acts on the invalid input $k$-subblock. If this $k$-Rec is good, then we may use Rec-SemiCor at level $k$ to move the $k$-$^*$decoder to the left through the $k$-Ga, transforming the $k$-Ga to an ideal $0$-Ga. If it is bad, we can move the $k$-$^*$decoder to the left through the bad $k$-RecEC, transforming the RecEC to a faulty 0-Ga.

Since the $(k{+}1)$-Rec is good, after we have moved all the $k$ decoders to the left, they are followed by a good 1-Rec, and then by final 1-decoders. Furthermore, the input to the 1-Rec is valid.

 Using Rec-Cor at level 1, we can move the 1-decoders to the left past the 1-Rec, transforming it to an ideal 0-Ga. 
Now the $(k{+}1)$-decoders act on the input $(k{+}1)$-blocks, except for a possible $k$-EC acting on the one invalid $k$-subblock of each input $(k{+}1)$-block, preceding the action of the $(k{+}1)$-decoder on the input. But the input $(k{+}1)$-block is valid, and therefore an operation acting on its one invalid block does not interfere with the action of the decoder --- the $k$-EC is of no consequence and can be removed without altering the output of the $(k{+}1)$-decoder. Thus we have shown Rec-Cor for the $(k{+}1)$-Rec.

\rightline{$\square$}

\medskip
\noindent {\bf Proof of Rec-SemiCor}: The $(k{+}1)$-decoders placed behind the $(k{+}1)$-Rec can be realized as $k$-$^*$decoders applied to all $k$-subblocks of the output $(k{+}1)$-blocks, followed by final 1-decoders (where the output syndromes of the $k$-$^*$decoders are discarded). As in the above proof of Rec-Cor, we can sweep the $k$-$^*$decoders to the left through the good $(k{+}1)$-Rec, transforming it to a good 1-Rec, followed by final 1-decoders. 

Using Rec-SemiCor at level 1, we can move the 1-decoders to the left past the 1-Ga, transforming it to an ideal 0-Ga. Now the $k$-$^*$decoders are in front of the 1-EC and the 1-decoders are behind it. Again using Rec-Cor at level $k$, we can move the $k$-$^*$decoders back to the right, placing them behind the 1-ECs, while replacing the 1-ECs with the original $k$-ECs. This puts the $k$-$^*$decoders right in front of the final 1-decoders, reassembling the full $(k{+}1)$-decoders. Thus we have proven Rec-SemiCor for the $(k{+}1)$-Rec. 

\rightline{$\square$}

\medskip
\noindent {\bf Proof of Rec-Val}: We test the validity of the output from the $(k{+}1)$-EC by applying the $(k{+}1)$-$^*$decoder, realized as $k$-$^*$decoders acting on all $k$-subblocks, followed by a final 1-$^*$decoder. The output is valid if, with probability one, the final level-1 syndrome measurement records no more than one error. 

As in the above proof of Rec-Cor, we sweep the $k$-$^*$decoders to the left through the good $(k{+}1)$-Rec, transforming it to a good 1-Rec, followed by a final 1-$^*$decoder. Using Rec-Val at level 1, we see that the final 1-block is valid, no matter what state emerges from the $k$-$^*$decoders. Thus, the 1-$^*$decoder records no more than one error, and we conclude that the output of the $(k{+}1)$-Rec is valid.

\rightline{$\square$}

\medskip 
Straightforward modifications of this argument show that the $(k{+}1)$-preparation and $(k{+}1)$-measurement satisfy Rec-Cor, completing the proof of:

\medskip
\noindent {\bf Theorem 5. Quantum accuracy threshold for independent stochastic noise (alternative version)}. {\em Suppose that fault-tolerant gadgets can be constructed such that all 1-Recs obey the properties Rec-Cor, Rec-Val, and Rec-SemiCor. Suppose that $\ell$ is the maximal number of locations in a 1-Rec, $d$ is the maximal depth of a 1-Rec, and $\varepsilon_0^{-1}$ is the maximal number of pairs of locations in a 1-Rec. Suppose that independent stochastic faults occur with probability $\varepsilon < \varepsilon_0$ at each location in a noisy quantum circuit. Then for any fixed $\delta$, any ideal circuit with $L$ locations and depth $D$ can be simulated with error $\delta$ or better by a noisy circuit with $L^*$ locations and depth $D^*$, where
\begin{equation}
L^*=O\left(L(\log L)^{\log_2 \ell}\right)~,\quad D^*= O\left(D(\log L)^{\log_2 d}\right)~.
\end{equation}
}
\medskip

\noindent As for our previous version of the threshold theorem, we can improve the threshold estimate by counting malignant pairs of locations in the 1-Recs, where a pair of locations is benign if the 1-Rec satisfies the properties  Rec-Cor, Rec-Val, and Rec-SemiCor for arbitrary faults at that pair of locations. We can also extend the theorem to concatenation of codes that correct $t\ge 3 $ errors.

Thus for concatenation of a code of distance $2t+1 \ge 5$, we can use either Theorem 4 or (the extension of) Theorem 5 to estimate the threshold. Typically, Theorem 4 yields a more favorable estimate. For example, in the distance-5 case, we may either apply Theorem 4 to exRecs or Theorem 5 to Recs. But at level 1, exRec-Cor is more likely to hold than Rec-Cor for a given fault rate; at least three faults in a 1-exRec are needed for exRec-Cor to fail, while two faults in a 1-Rec can cause Rec-Cor to fail. 

\section{Local non-Markovian noise}
\label{sec:non-markovian}

\subsection{A more general noise model}
We have so far restricted our attention to a rather special noise model. We have assumed that faults are uncorrelated in space and in time, and we have assumed that each fault leaves a permanent record in the environment, attesting that the fault occured. 

Let us refer to a particular history, indicating all the locations in a circuit where faults occur, as a {\em fault path}. If the environment records the fault path --- that is, if different fault paths are labeled by perfectly distinguishable states of the environment --- then distinct fault paths decohere with one another. Therefore, we may assign a probability to each fault path. In our proof of the threshold theorem, we derived an upper bound on the sum of the probabilities of all the {\em bad} fault paths that contribute to an extended rectangle.

By adopting this model of independent stochastic noise, we were able to explain the key ideas underlying the proof of the threshold theorem in an especially simple setting. But the theorem we have proved so far has limited applicability. For example, it does not apply to the case of {\em unitary errors} in which each gate is a unitary transformation that deviates from the ideal gate by a slight over-rotation or under-rotation. Such errors leave no record. Furthermore, our threshold theorem does not apply if the faults are weakly correlated in space and/or in time. Such correlations are expected in realistic implementations. 

We will now broaden the setting, and extend our proof of the threshold theorem to a more general noise model, which we will call {\em local noise}. The defining characteristic of local noise is that the noise does not act collectively on many data qubits at once; rather the noise acts either on individual data qubits, or acts collectively on a set of data qubits that interact during the execution of a quantum gate. But the different noisy gates act on an environment or ``bath'' as well as on the data, and we allow the various noisy gates to share the same bath. In effect, the bath serves as a memory that can store information for an indefinitely long time --- the noise is {\em non-Markovian}. Although the noise acts locally on the data, the information recorded by the bath allows the faults to be correlated, both in time and in space. We will see that these correlations do not prevent us from proving a threshold theorem, provided the noise is sufficiently weak.

A threshold theorem for non-Markovian noise has also been formulated in an insightful paper by Terhal and Burkard \cite{terhal}. Our analysis differs from theirs, and has two distinct advantages. First, for the proof in \cite{terhal} to apply, a locality condition must be imposed not only on the interaction of the bath with the data, but also on interactions among the degrees of freedom within the bath. In contrast, we will not need any assumption about the locality of the bath. Second, the proof in \cite{terhal} applies when the fault-tolerant gadgets are nonoverlapping, but has no obvious extension to the case where the gadgets overlap. Therefore, it applies to concatenated distance-5 codes but not to concatenated distance-3 codes. In contrast, our approach is applicable even for the overlapping gadgets that arise in the analysis of concatenated distance-3 codes.

From a physics perspective, a non-Markovian noise model is most naturally formulated in terms of a Hamiltonian $H$ that governs the joint evolution of the ``system'' (the data qubits that are processed by the computation) and the ``bath'' (the environment whose interactions with the system drive the noise). We may express $H$ as
\begin{equation}
H=H_S+H_B +H_{SB}~,
\end{equation}
where $H_S$ is the time-dependent Hamiltonian of the system that realizes the ideal circuit, $H_B$ is an arbitrary Hamiltonian of the bath, and the Hamiltonian $H_{SB}$ coupling the system to the bath is
\begin{equation}
H_{SB}=\sum_a H_{SB,a}~;
\end{equation} 
here each $H_{SB,a}$ describes how a particular set of qubits $a$, which are acted upon by a particular gate in the ideal circuit, interact with the bath. The system-bath interaction $H_{SB}$ describes the noise, which is ``local'' in the sense that two (or more) data qubits are coupled by $H_{SB}$ in a given time step only if the ideal Hamiltonian $H_S$ also couples those data qubits in the same time step.

We can express the time evolution governed by the Hamiltonian $H$ in terms of a ``time-resolved fault path'' expansion \cite{terhal}.  That is, we divide the total time $T$ into $N$ time intervals each of width $\Delta=T/N$, where $N$ is sufficiently large that terms of order $\Delta^2$ can be safely neglected.  Then the time evolution operator for the time interval $(t,t+\Delta)$ can be accurately expressed as
\begin{equation}
U(t+\Delta,t) \approx e^{-i\Delta H_S} e^{-i\Delta H_B}\prod_a \left(I_{SB} -i\Delta H_{SB,a}\right)~~,
\end{equation}
and the time evolution operator $U(T,0)$ for interval $(0,T)$ is the product of $N$ such operators. For each summand in the fault path expansion, at each time resolved location labeled by $t$ and $a$, we insert either $I_{SB}$ (in which case we say there is no fault at that location) or $-i\Delta H_{SB,a}$ (in which case we say there is a fault at that location). 

Suppose that the time required to execute each gate in the quantum circuit is $t_0$, so that associated with each gate is a ``coarse-grained'' location consisting of $t_0/\Delta$ consecutive time-resolved locations. We say that a coarse-grained location is faulty if there is a fault at any one of the time-resolved locations that it contains. If the coarse-grained location is not faulty, then the noise is trivial there and the ideal gate is executed faithfully.

We will use the sup operator norm to characterize the strength of the noise --- the norm of an operator $A$ is denoted $\parallel A\parallel$ and defined by
\begin{equation}
\parallel A\parallel = {\rm sup}_{|\psi\rangle}\left(\parallel A|\psi\rangle\parallel/\parallel \psi\parallel\right) ~;
\end{equation}
this norm has the properties 
\begin{equation}
\parallel A + B\parallel~ \le ~\parallel A\parallel + \parallel B\parallel ~,\quad \parallel AB\parallel ~\le ~\parallel A\parallel \cdot \parallel B\parallel~, \quad \parallel A\otimes B\parallel ~=~ \parallel A\parallel\cdot \parallel B\parallel~.
\end{equation}
Suppose that the system-bath coupling at each time-resolved location satisfies
\begin{equation}
\parallel H_{SB,a}\parallel ~\le~ \lambda_0~.
\end{equation}
Specify $r$ particular coarse-grained locations in a quantum circuit, and let $E$ denote the sum over all time-resolved fault paths with faults at those $r$ locations, where the time-resolved fault path is completely unrestricted outside of the $r$ coarse-grained locations. Then, as shown in \cite{terhal}, $\parallel E\parallel ~\le~ \left(\lambda_0t_0\right)^r$.   This observation motivates the following definition:

\medskip
\noindent {\bf Definition. Local Noise}. {\em Consider a noisy quantum circuit, realized as a unitary transformation $U_{SB}$ acting on a system $S$ and a bath $B$, with $U_{SB}$ expressed as a fault-path expansion where each term is characterized by a set of fault locations. Let ${\cal I}_r$ denote a particular set of $r$ (coarse-grained) locations, and let $E({\cal I}_r)$ denote the sum over all fault paths that contain faults at all of the locations in ${\cal I}_r$. Then we say that the noise is {\em local with strength $\eta$} if for all ${\cal I}_r$
\begin{equation}
\label{local-noise}
\parallel E({\cal I}_r)\parallel \ \le \ \eta^r ~.
\end{equation}
}

\medskip

\noindent By this definition, our Hamiltonian noise model is local, with noise strength $\eta=\lambda_0 t_0$, for an arbitrary bath Hamiltonian $H_B$. Note that we are free to shift $H_{SB}$ by a constant in order to optimize $\eta$ \cite{terhal}.

Of course, it is also possible to formulate local noise models without direct reference to a Hamiltonian. Instead, we could characterize the noisy gates as unitary operators associated with circuit locations, where we allow the different noisy gates to act on a shared bath (and in fact arbitrary unitary transformations may be applied to the bath in between the successive quantum gates). To be specific, we will describe such a noise model first of all for a single-qubit gate. If the ideal gate applies the unitary transformation $U_{\rm ideal}$ to the qubit, the actual noisy gate instead applies a unitary transformation $\tilde U=U_{\rm fault} U_{\rm ideal}$ to the qubit and the bath, where
\begin{equation}
\label{U-fault}
U_{\rm fault}=\left(I\otimes A_0 + X\otimes A_1 + Y\otimes A_2 + Z\otimes A_3\right)~.
\end{equation}
Here $\{I,X,Y,Z\}$ are the Pauli operators, and the operators $\{A_0,A_1,A_2,A_3\}$ are arbitrary operators acting on the state of the bath (subject to the constraint that $U_{\rm fault}$ is unitary). Then we may say that the noise has strength $\eta$ if each noisy gate in the circuit obeys
\begin{equation}
\parallel X\otimes A_1 + Y\otimes A_2 + Z\otimes A_3\parallel ~\le ~\eta~.
\end{equation}
A measurement can be modeled as $U_{\rm fault}$ followed by an ideal measurement, and a preparation can be modeled as an ideal preparation followed by $U_{\rm fault}$. Similarly, a noisy two-qubit gate can be expressed as the ideal gate followed by an operator that can be expressed as an expansion in Pauli operators; the noise strength $\eta$ is an upper bound on the norm of the sum of the operators in this expansion, excluding the leading operator $I\otimes I\otimes A_{00}$. Note that, because gates that act in the same time step may have noncommuting actions on the bath, an operator ordering convention is needed to unambiguously specify the model, but the noise strength does not depend on how the ordering ambiguity is resolved. 

If each noisy gate acts on an independent bath, and the operators $I\otimes A_0$ and $(X\otimes A_1 + Y\otimes A_2 + Z\otimes A_3)$ map the environment to mutually orthogonal states, then our local noise model describes independent stochastic faults. In that case the probability of a fault is not $\eta$ but rather $\varepsilon=\eta^2$. We may regard $\eta$ as the {\em amplitude} weighting an error, whose square is a probability. If $A_0$, $A_1$, $A_2$, and $A_3$ are all scalar multiples of one another, then our local noise model describes unitary errors. For example, if $U_{\rm fault}$ acts only on a single data qubit and has eigenvalues $e^{\pm i\theta/2}$, then $\eta=|\sin\theta/2|$~.

Our proof of the threshold theorem for independent stochastic noise already applies to a restricted type of non-Markovian noise, in which distinct fault paths do not interfere. That proof works as long as each fault path with $r$ faulty locations can be assigned a probability no larger than $\varepsilon^r$; once a particular fault path is selected, we may allow the faulty locations to share a bath. We emphasized this point in the discussion of the inductive step in Sec.~\ref{subsubsec:nonindependent-pairs}, where we noted that a level-$(k{+}1)$ circuit built out of gates with independent stochastic faults can be regarded as a simulation of a level-1 circuit where the bad $k$-exRecs simulate faults that share a bath. The new analysis we will now describe is needed not just because we wish to consider faults that share a bath, but also because of the potential for quantum interference among the fault paths.

\subsection{Bounding the norm of the sum over bad fault paths}
The proof of the threshold theorem for local noise has two parts. First we will show that if the noise strength $\eta$ is sufficiently small, then the norm of the sum over bad fault paths becomes doubly exponentially small as the level $k$ of the simulation increases. Then in Sec.~\ref{sec:non-Markovian-accuracy} we will show that this small norm implies that the noisy circuit accurately simulates the ideal computation. 

The proof works whether or not overlapping rectangles are invoked in the definition of goodness. We will present the proof for nonoverlapping rectangles, and then elaborate on its applicability to overlapping rectangles in Sec.~\ref{subsec:overlap-nonMarkovian}. The result can be extended to the case where ancillas are rejected whenever a verification test fails, but we will ignore that complication in our proof. 

To begin, consider a noisy circuit with a total of $A$ level-0 locations, and let $F_s$ denote the sum over all fault paths with $s$ {\em or more} faults. Can we find an upper bound on $\parallel F_s\parallel$? We would like to express $F_s$ as a sum of terms of the form $E({\cal I}_r)$, since then we can bound the norm of the sum by the sum of the norms, and, using the definition of local noise, bound each norm in the sum by a power of $\eta$. But it is not completely straightforward to relate $F_s$ to the $E({\cal I}_r)$'s. In particular, if we sum $E({\cal I}_s)$ over all sets of $s$ locations, each fault path with more than $s$ locations will be counted multiple times. Let ${\cal I}$ denote the set of {\em all} locations in a circuit (with $|{\cal I}|=A$), and  let $E_r$ denote
\begin{equation}
E_r=\sum_{{\cal I}_r\subseteq {\cal I}}E({\cal I}_r) ~,
\label{level-0-bound}
\end{equation}
the sum of the $E({\cal I}_r)$'s over all sets of $r$ locations in the circuit. The crux of our analysis is this combinatoric lemma:

\medskip
\noindent {\bf Lemma 7. Counting of fault paths}. 
\begin{equation}
\label{fault-count}
F_s= \sum_{r=s}^{|{\cal I}|} (-1)^{r-s}{{r-1}\choose{s-1}} E_r~.
\end{equation}
\medskip

\noindent {\bf Proof}: Eq.~(\ref{fault-count}) can be derived from the inclusion-exclusion principle (see for example \cite{stanley}). For completeness, we provide a self-contained discussion of how the result can be obtained. To count correctly all the fault paths with $s$ or more faults, we first sum over all ways of choosing $s$ locations, and for each of these choices, we sum over all fault paths that are bad at those $s$ locations. But then we have overcounted the fault paths that are bad at (at least) $s+1$ locations, so we make a subtraction to correct for the overcounting. But now, because of this subtraction, we have undercounted the fault paths that are bad at (at least) $s+2$ locations, so we need to make an addition to correct for the undercounting, and so on. 

To determine the combinatoric factors in this expansion, we first choose an arbitrary ordering of the $|{\cal I}|$ circuit locations. This ordering is used just to facilitate the counting, and need not bear any relation to the time ordering in the circuit; nevertheless we will use temporal language to describe the ordering --- e.g., speaking of an ``earlier'' location or a ``later'' location. At each circuit location, we divide the sum over fault paths into a good part $G_i$ and a bad part $B_i$. (For example, in the Hamiltonian model the good part includes only the case where $I_{SB}$ is inserted at every time-resolved location contained in the coarse-grained location $i$, and the bad part includes every case in which $-i\Delta H_{SB,a}$ is inserted at at least one time-resolved location.) $F_s$ can be expressed as a sum of terms, where in each term the $s$ earliest fault locations are identified: $B_i$ is inserted at each of these $s$ locations, $G_i$ is inserted at all other locations prior to the last of the $s$ earliest bad locations, and $G_i+B_i$ is inserted at each location after the last of the $s$ earliest bad locations. With this scheme, every fault path with $s$ or more locations has been included exactly once. We refer to this expansion of $F_s$ as the ``original'' expansion; from it we are to obtain the ``derived'' expansion eq.~(\ref{fault-count}).

Next we rewrite each $G_i$ in each term of the original expansion as 
\begin{equation}
G_i= (G_i+B_i) - B_i~
\end{equation}
and expand in powers of $B_i$ to obtain the derived expansion. The resulting term with $B_i$ at each of a set of locations ${\cal I}_r$, and $G_i +B_i$ elsewhere, is $E({\cal I}_r)$. In how many ways can this term arise in the derived expansion? Of the $r$ $B_i$'s, $s$ are ``primary'' $B_i$'s that are already present in the original expansion, and the rest are ``secondary'' $B_i$'s that arose from expanding $G_i=(G_i+B_i)-B_i$. The $B_i$ at the latest bad location in the circuit must be a primary $B_i$, but each of the other $r-1$ could be either one of the remaining $s-1$ primary $B_i$'s or a secondary $B_i$. There are ${{r-1}\choose{s-1}}$ ways to choose which of the $B_i$ other than the last one are primary, and the number of minus signs is the number of secondary $B_i$'s; thus we find eq.~(\ref{fault-count}).

\rightline{$\square$}

\medskip
Using Lemma 7 and the definition of local noise, we obtain an upper bound on the norm of $F_s$:
\begin{eqnarray}
\parallel F_s\parallel &\le& \sum_{r=s}^A {{r-1}\choose{s-1}}{A\choose r}\eta^r= {A\choose s}\eta^s\sum_{r=s}^A\frac{s}{r}{A-s\choose r-s}\eta^{r-s}\nonumber\\
&\le & {A\choose s}\eta^s\sum_{t=0}^\infty \frac{(A-s)^{t}}{t!}\eta^{t}={A\choose s}\eta^s e^{(A-s)\eta}~.
\label{F-s-bound}
\end{eqnarray}
Now consider a level-1 simulation, in which each gate of an ideal circuit is replaced by the corresponding 1-Rec, and let ${\cal I}_r^{(1)}$ denote a particular set of $r$ 1-locations (locations of 1-Recs) in the 1-simulation. Let us say that a 1-Rec is bad if it contains $s$ or more faults, and consider $E({\cal I}_r^{(1)})$, the sum over all fault paths in the 1-simulation such that the 1-locations in ${\cal I}_r^{(1)}$ are bad. 

Now we can carry out the ``original'' expansion in each of the 1-Recs, and obtain a ``derived'' expansion by expanding $G_i=(G_i+B_i)-B_i$ at each good level-0 location. Label the $r$ bad 1-locations by the index $b\in \{1,2,\dots, r\}$, let ${\cal I}(b)$ denote the set of all 0-locations in bad 1-Rec $b$, and let ${\cal I}(b)_{\ell}$ denote a particular set of $\ell$ locations in bad 1-Rec $b$. Summing independently over the sets of bad locations inside each of the $r$ bad 1-Recs, and reasoning as in the derivation of Lemma 7, we find 
\begin{eqnarray}
E({\cal I}_r^{(1)})= \sum_{\ell_1=s}^{|{\cal I}(1)|}(-1)^{\ell_1-s}{\ell_1 -1\choose s-1}\cdots\sum_{\ell_r=s}^{|{\cal I}(r)|}(-1)^{\ell_r-s}{\ell_r -1\choose s-1}\nonumber\\
\sum_{{\cal I}(1)_{\ell_1}\subseteq {\cal I}(1)}\cdots \sum_{{\cal I}(r)_{\ell_r}\subseteq {\cal I}(r)} E(I(1)_{\ell_1} \cup \cdots \cup I(r)_{\ell_r})~.
\end{eqnarray}
For local noise, we have
\begin{equation}
\parallel E(I(1)_{\ell_1} \cup \cdots \cup I(r)_{\ell_r})\parallel ~\le~ \eta^{\left(\sum_{b=1}^r \ell_b\right)}=\prod_{b=1}^r\eta^{\ell_b}~.
\end{equation}
Therefore, our bound on $E({\cal I}_r^{(1)})$ factorizes into $r$ factors, each of the form in eq.~(\ref{F-s-bound}), and so we obtain:

\medskip
\noindent {\bf Lemma 8. Norm of the sum over level-1 fault paths}. 
{\em Define a 1-Rec to be bad if it contains $s$ or more faults, and let $E({\cal I}_r^{(1)})$ denote the sum over all fault paths in a 1-simulation such that the $r$ 1-locations in the set ${\cal I}_r^{(1)}$ are bad. Then for local noise with strength $\eta$
\begin{equation}
\label{level-1-bound}
\parallel E({\cal I}_r^{(1)})\parallel \ \le \ \left(\eta^{(1)}\right)^r~,
\end{equation}
where
\begin{equation}
\eta^{(1)} = C{A\choose s}\eta^s,
\end{equation}
$C\ge e^{(A-s)\eta}$, and $A$ is the maximal number of locations inside any 1-Rec. 
}
\medskip

Note that eq.~(\ref{level-1-bound}) has just the same form as eq.~(\ref{local-noise}), so we can use the same reasoning as in the derivation of Lemma 8 to take the inductive step. Now suppose that a $k$-Rec is bad if it contains $s$ or more bad $(k{-}1)$-Recs, let ${\cal I}_r^{(k)}$ denote a particular set of $r$ $k$-locations in a $k$-simulation, and let $E({\cal I}_r^{(k)})$ denote the sum over all fault paths such that the $r$ $k$-locations in ${\cal I}_r^{(k)}$ are bad. Suppose that $\parallel E({\cal I}_r^{(k)})\parallel \le \left(\eta^{(k)}\right)^r$; then we may argue that $\parallel E({\cal I}_r^{(k+1)})\parallel\  \le \ \left(\eta^{(k+1)}\right)^r$, where $\eta^{(k+1)} = C{A\choose s}\left(\eta^{(k)}\right)^s$ and $C\ge e^{(A-s)\eta}$. Therefore we have proved

\medskip
\noindent {\bf Lemma 9. Norm of the sum over bad level-$k$ fault paths}. 
{\em Define a $k$-Rec to be bad if it contains $s$ or more bad $(k{-}1)$-Recs, and let $E({\cal I}_r^{(k)})$ denote the sum over all fault paths in a $k$-simulation such that the $r$ $k$-locations in the set ${\cal I}_r^{(k)}$ are bad. Then for local noise with strength $\eta$
\begin{equation}\
\label{nonMarkovian-bound-level-k}
\parallel E({\cal I}_r^{(k)})\parallel \ \le \ \left(\eta^{(k)}\right)^r~,
\end{equation}
where
\begin{equation}
\eta^{(k)} = \eta_0\left(\eta/\eta_0\right)^{s^k}~,
\end{equation}
\begin{equation}
\label{solve-for-eta}
\eta_{0}^{-1} = \left(C{A\choose s}\right)^{1/(s-1)}~,
\end{equation}
$\eta \le \eta_0$, $C\ge e^{(A-s)\eta_0}$, and $A$ is the maximal number of locations inside any 1-Rec. 
}

\medskip
We can optimize our estimate of $\eta_0$ by setting $C= e^{(A-s)\eta_0}$ and solving eq.~(\ref{solve-for-eta}) for $\eta_0$, or more simply we may choose $C=e$, since the inequality $(A-s)^{s-1}\le \eta_0^{-(s-1)}=e {A\choose s}$ is satisfied for typical level-1 simulations of interest (e.g., for $s=2$).

\subsection{Overlapping rectangles}
\label{subsec:overlap-nonMarkovian}

Now let us note that Lemma 9 applies even when overlapping rectangles are used to define the badness of fault paths. 
At first, the overlaps may seem to complicate the discussion. We again wish to organize the sum over all the fault paths such that a particular $k$-location is bad into a sum of terms, where in each term a specified set of $r$ $(k{-}1)$-locations contained inside the $k$-location are bad. But now, depending on the context, a ``location'' can be either an exRec (if the locations immediately following are all good), or a truncated exRec (if a location immediately following is bad). 

Nevertheless, for any specified fault path, whether the location is truncated or not, and whether it is bad or not, is unambiguously determined, and in order for the location to be bad, it must contain at least $s$ bad locations at the next level down. For a $k$-simulation, we can, just as before, express the sum over all fault paths with $r$ or more bad $k$-exRecs as an inclusion-exclusion sum of the form in eq.~(\ref{fault-count}). The combinatorics, and therefore the coefficients in the sum, are the same as for the case of nonoverlapping rectangles. The only change is that for each fault path appearing in the sum, some exRecs are truncated and some are untruncated, and which ones are truncated varies from fault path to fault path. 

A truncated $k$-exRec has fewer locations than the full $k$-exRec. Therefore, at each level $k$, the upper bound eq.~(\ref{nonMarkovian-bound-level-k}) holds whether or not some of the $r$ bad $k$-exRecs are truncated, and the inductive proof of Lemma 9 goes through, where now $A$ is the maximal number of locations inside any 1-exRec.

\subsection{Accuracy}
\label{sec:non-Markovian-accuracy}

Now we have seen that if $\eta< \eta_0$, the norm of the bad part of a $k$-simulation  declines double-exponentially with the level $k$. What does this imply about the accuracy of the simulation?

The sum over all fault paths for a level-$k$ simulation of a circuit with $L$ locations can be divided into a good part $G^{(k)}_{\rm circuit}$ and a bad part $B^{(k)}_{\rm circuit}$, where in the good part every $k$-exRec is good. We can bound the norm of the bad part by combining eq.~(\ref{F-s-bound}) with Lemma 9, finding
\begin{equation}
\parallel B^{(k)}_{\rm circuit}\parallel~\le ~L \eta^{(k)} \exp\left((L-1)\eta^{(k)}\right)\le eL\eta^{(k)}~,
\end{equation}
where to obtain the last inequality we have assumed $(L-1)\eta^{(k)} <1$.

Now, the good part of the circuit is correct. Strictly speaking, we have proved this only when all of the 0-faults are physical operations. But we can always express the sum over the good fault paths such that in each term (except for an overall normalization) every 0-fault is a Pauli operator (which is physical) and then invoke linearity to conclude that the sum is correct because all of its terms are correct. 

The final outcome of the level-$k$ simulation is determined by measuring a quantum state. This state might be mixed but by considering its purification, we may imagine that the measured state is a pure state of the data and the environment (including ancillas), which can be expressed as a sum of good and bad parts as
\begin{equation}
\label{psi-level-k}
|\tilde\psi\rangle = G^{(k)}_{\rm circuit}|\psi_0\rangle + B^{(k)}_{\rm circuit}|\psi_0\rangle~.
\end{equation}
Here $|\psi_0\rangle$ is the ideal input to the circuit --- we may move any faults in the input state and any faults in the final measurement into the circuit, so we may imagine that the input and the final readout are ideal (recall that we are assuming that the classical decoding of the measurement outcomes is reliable). Up to normalization, the good part of the state, when measured, produces the same probability distribution of outcomes as the ideal circuit. Therefore, our bound on the norm of the $B^{(k)}_{\rm circuit}$ allows us to bound the accuracy of the simulation.

Consider a POVM with elements $\{E_a\}$ and let $\rho$ and $\tilde \rho$ be arbitrary density operators. Let $p$ denote the probability distribution for the outcomes when the POVM is performed on the state $\rho$ and let $\tilde p$ denote the probability distribution for the outcomes when the POVM is performed on the state $\tilde \rho$. Then the $L^1$ distance between the probability distributions is
\begin{equation}
\delta = \parallel p - \tilde p\parallel\equiv \sum_a|p_a-\tilde p_a|= \sum_a\left|{\rm tr}~E_a\left(\rho-\tilde\rho\right)\right| \le \sum_{a,i}|\lambda_i|\cdot|\langle i|E_a|i\rangle|=\parallel \rho-\tilde\rho\parallel_{\rm tr}~.
\end{equation}
Here $\{|i\rangle\}$ are the eigenstates of $\rho-\tilde\rho$, $\{\lambda_i\}$ are the corresponding eigenvalues, $\parallel\cdot\parallel_{\rm tr}$ denotes the trace norm, and in the last step we have used the completeness relation $\sum E_a=I$. If both states are pure, $\rho=|\psi\rangle\langle\psi|$, and $\tilde \rho=|\tilde\psi\rangle\langle \tilde\psi|$, then the two nonzero eigenvalues of $\rho-\tilde\rho$ are $\pm \sin\theta$, where $|\langle \psi|\tilde\psi\rangle|=\cos\theta
$, and therefore $\parallel \rho-\tilde\rho\parallel_{\rm tr}=2|\sin\theta|$.

To bound the accuracy of the level-$k$ simulation, then, we need to estimate the angle $\theta$ between $|\tilde\psi\rangle$ given by eq.~(\ref{psi-level-k}) and the (properly normalized) good part of this state. With the norm of the bad part of the state fixed, the angle is maximal when the good and bad parts are orthogonal, in which case $|\sin\theta|$ is the norm of the bad part. We conclude that the error of the simulation satisfies
\begin{equation}
\delta ~=~\parallel p - \tilde p\parallel ~\le ~ 2 \parallel B^{(k)}_{\rm circuit}\parallel~\le~  2e L \eta^{(k)}~\le~ 2eL \eta_0\left(\eta/\eta_0\right)^{2^k}~.
\end{equation}
This is essentially the same scaling of the accuracy with the level $k$ as found in eq.~(\ref{accuracy-scaling}), except that now the noise strength $\eta$ is a fault {\em amplitude}, while for independent stochastic errors, the fault probability $\varepsilon=\eta^2$ appeared instead.

Arguing as in Sec.~\ref{sec:threshold-theorem}, we then obtain

\medskip
\noindent {\bf Theorem 6. Quantum accuracy threshold for local non-Markovian noise}. {\em Suppose that fault-tolerant gadgets can be constructed such that all 1-exRecs obey the property exRec-Cor, where a 1-exRec that contains no more than 1 fault is said to be good. Suppose that $\ell$ is the maximal number of locations in a 1-Rec, $d$ is the maximal depth of a 1-Rec, and $\eta_0^{-1}/e$ is the maximal number of pairs of locations in a 1-exRec. Suppose that local noise occurs with strength $\eta < \eta_0$ at each location in a noisy quantum circuit. Then for any fixed $\delta$, any ideal circuit with $L$ locations and depth $D$ can be simulated with error $\delta$ or better by a noisy circuit with $L^*$ locations and depth $D^*$, where
\begin{equation}
L^*=O\left(L(\log L)^{\log_2 \ell}\right)~,\quad D^*= O\left(D(\log L)^{\log_2 d}\right)~.
\end{equation}
}
\medskip

\noindent The theorem specifies the same scaling of the overhead as the theorem for independent stochastic noise, and the value of the accuracy threshold is similar. The important difference is that for local noise the threshold condition must be satisfied by an error amplitude rather than an error probability. Of course, the theorem can be extended to the case where a 1-exRec that contains no more than $t$ faults is said to be good. 

Because distinct quantum gates are permitted to act on a shared bath, our local noise model incorporates correlations among faults, both in space and in time. The key to deriving an accuracy threshold is not strictly speaking the suppression of such correlations; rather it suffices for the coupling of the data qubits to the bath to be sufficiently weak. Though the noise acts locally on the data, we emphasize again that we did not need to impose any locality condition on the Hamiltonian of the bath; $H_B$ is completely arbitrary.

\subsection{Extension to counting malignant sets}

We can improve the estimate of the threshold by counting malignant sets of locations as in Sec.~\ref{sec:malignant-pairs}. Though the analysis also applies to overlapping rectangles, we will consider the case of nonoverlapping rectangles here, and speak of Recs rather than exRecs. 

For example, suppose that there are $B$ malignant pairs of locations in a 1-Rec. Then the 1-Rec can be bad if faults occur at a malignant pair of locations, or if faults occur at any set of three or more locations in the 1-Rec (where it may be that no two of these locations form a malignant pair). Again, we wish to bound the norm of the sum of all bad fault paths. 

The sum of all fault paths with three or more faults in the 1-Rec can be expressed as in Lemma 7 (with $s=3$). We need to add to this an additional contribution due to the fault paths with exactly two faults that form a malignant pair. For each malignant pair we place $B$ at each of the locations in the malignant pair, and $G$ at all other locations; then we expand each $G=(G+B) - B$. In this expansion, how many times does the term occur with $B$ inserted at each of the locations in a particular set ${\cal I}_r$ of $r$ locations? Any two of these $r$ locations could in principle be the ``primary'' locations, but only if these two locations form a malignant pair. Therefore the term occurs at most ${r\choose 2}$ times (if every pair of the $r$ locations is a malignant pair) and no fewer than zero times (if no pair of the $r$ locations is a malignant pair). On the other hand, each set of $r$ locations occurs $r-1 \choose 2$ times in the expansion arising from all sets of three or more locations, and furthermore, these terms are opposite in sign to the terms that arise from the malignant pairs. When we combine the two expansions, the absolute value of the coefficient for a term with $r$ bad locations is at most the larger of ${r-1\choose 2}$ and ${r\choose 2}- {r-1\choose 2}=r-1$. Invoking Lemma 7, and reasoning as in the proof of Lemma 8, we therefore have:

\medskip
\noindent {\bf Lemma 10. Norm of the sum over level-1 fault paths with malignant pairs}. 
{\em Define a 1-Rec to be bad if it contains faults at a malignant pair of locations, or at some set of three or more locations. Suppose that there are no more than $B$ malignant pairs of locations in any 1-Rec, and at most $A$ locations in any 1-Rec. Let $E({\cal I}_r^{(1)})$ denote the sum over all fault paths in a 1-simulation such that the $r$ 1-locations in the set ${\cal I}_r^{(1)}$ are bad. Then for local noise with strength $\eta$
\begin{equation}
\parallel E({\cal I}_r^{(1)})\parallel \ \le \ \left(\eta^{(1)}\right)^r~,
\end{equation}
where
\begin{equation}
\eta^{(1)} = B \eta^2 + (C+1){A\choose 3}\eta^3,
\end{equation}
and $C\ge e^{(A-3)\eta}$. 
}
\medskip

(The coefficient $C$ multiplying ${A\choose 3}\eta^3$ arises from estimating the sum over $r$ as in eq.~(\ref{F-s-bound}); the +1 arises from the $r=3$ term, because max$(r-1, {r-1\choose 2})=r-1=2$ for $r=3$.) A similar estimate applies at higher levels, so we can obtain an improved value for the accuracy threshold by counting malignant pairs. Furthermore, the argument also works for overlapping rectangles.

For the independent stochastic noise model, we estimated the threshold fault rate $\varepsilon_0$ by counting malignant pairs of faults, as reported in the proof of Theorem 3. It is not quite straightforward to adapt that analysis to obtain an estimate of the threshold noise strength $\eta_0$ for the local noise model. The circuits studied in Sec.~\ref{sec:explicit} include ancilla verification steps, and the threshold analysis involves an estimate of the effective fault rate in the ancillas that pass the verification test. Similarly, to estimate $\eta_0$ using the same circuits, we would need to bound the effective noise strength in postselected ancillas, and we have not performed this calculation.

\subsection{Decoherence of fault paths via syndrome measurement}
\label{subsec:decoherence-measurement}
Our estimate of the quantum accuracy threshold for general local noise is much worse than for stochastic errors, because the accuracy threshold for local noise specifies a value of a fault amplitude (the noise strength) rather than a fault probability. Expressed as a critical fault probability, the threshold for general local noise is in effect the square of the threshold for stochastic noise. 

This conclusion seems overly pessimistic, for several reasons. The crucial difference between the stochastic model and the local model is that in the stochastic model, each fault path in an exRec is associated with a perfectly distinguishable state of the environment, and therefore the fault paths decohere. In contrast, for local noise the states of the environment associated with distinct fault paths need not be distinguishable, and therefore the bad fault paths {\em might} interfere constructively. Thus we obtain a much more demanding condition on the noise.

In practice, though, it would seem to require a highly implausible conspiracy for many bad fault paths to add together with a common phase. Under a more plausible scenario in which the phases of distinct bad fault paths are only weakly correlated, we expect amplitudes to accumulate more like a random walk in the complex plane --- that is, it is more reasonable to add the probabilities of the fault paths than their amplitudes. 

However comforting in practice, this observation is of little help if we aspire to prove a rigorous theorem, since we are obligated to consider the worst possible case that is compatible with the assumptions of our noise model. But there is another observation that is more likely to point the way to a stronger rigorous result than found in Sec.~\ref{sec:non-Markovian-accuracy}. The ``environment'' includes not just the system needed to provide a dilation of our noise model, but also the ancillas that are used in the measurement of error syndromes. Therefore, for two different bad fault paths of a $k$-exRec to interfere, they must have identical error syndromes. 

We might, then, find an improved threshold estimate for local noise by carrying out the following stratagem. For the 1-exRec, suppose that we can divide the bad fault paths into $N$ classes, where fault paths in distinct classes are unable to interfere with one another. Suppose that for each class, the norm of the sum of the contributions to $B^{(1)}$ arising from that class can be expressed as a sum of no more than $M$ terms, each with norm no larger than $e\eta^2$. Since the norms of the decoherent contributions to $B^{(1)}$ add in quadrature, we then find that 
\begin{equation}
\parallel B^{(1)} \parallel^2 ~\le ~ N\left( M e\eta^2 \right)^2~;
\end{equation}
the condition $\parallel B^{(1)}\parallel \le \parallel B^{(0)}\parallel$ then implies a value of the threshold noise strength $\eta_0$ where
\begin{equation}
\eta_0^{-1} = eM\sqrt{N}~.
\end{equation}
At least roughly, this threshold estimate interpolates between our result for general local noise ($N=1$, $M=A$), and our result for independent stochastic noise $N=A$, $M=1$). 

For the seven-qubit Steane code analyzed in Sec.~\ref{sec:explicit}, there are $2^6=64$ possible values for the syndrome, and the \cnot exRec, which dominates the threshold estimate for the case of independent stochastic errors, contains four syndrome measurements. Deriving nontrivial bounds on $M$ and $N$, which we have not attempted, might lead to a substantial improvement in the accuracy threshold for local noise.

\section{Conclusions}
\label{sec:conclusions}

We have proven six versions of the quantum accuracy threshold theorem, all based on recursive simulations. Theorem 1 assumes an independent stochastic noise model, and establishes a threshold for a fault-tolerant simulation based on the concatenation of a distance-3 quantum code. Theorem 2 is a refinement of Theorem 1 that improves the estimate of the accuracy threshold. Theorem 3 derives a lower bound on the accuracy threshold, by applying Theorem 2 to Steane's [[7,1,3]] code, and Theorem 4 extends Theorem 1 to concatenation of codes that correct two or more errors. Theorem 6 generalizes the noise model considered in Theorem 1 to local non-Markovian noise. 

In the proofs of Theorems 1--4, we assess the accuracy of a recursive simulation by characterizing how faults are distributed in {\em extended rectangles} that overlap with one another. Theorem 5 establishes an accuracy threshold using a different approach in which accuracy is assessed by studying how faults are distributed in {\em rectangles} that do not overlap. Theorem 6 is applicable to both approaches.

Other versions of the quantum threshold theorem have been discussed previously \cite{ben-or,kitaev_threshold,knill}, but ours is the first rigorous proof that applies to the concatenation of a code that can correct only one error, and it provides the best lower bound on the accuracy threshold that has been rigorously established up to now. The proof uses novel ideas that we expect will have further applications in future studies of fault-tolerant quantum simulations. The crux of the proof of Theorems 1--4 is the inductive step based on the {\em threshold dance} depicted in Fig.~\ref{fig:dance}, which shows that if a level-$k$ rectangle is contained in a good extended rectangle, then the rectangle followed by an ideal decoder is equivalent to an ideal decoder followed by an ideal gate. 

Theorem 5 is really a reformulation and elaboration of the proof in \cite{ben-or} that, we hope, brings the essential elements of that argument into sharper focus. Theorem 6 shows that an accuracy threshold based on concatenated distance-3 codes can be established for noise models that allow faults to be spatially and temporally correlated, even if the environment can store quantum information for an indefinitely long time. The proof uses a different method, and applies under weaker assumptions, than the previous analysis of non-Markovian noise in \cite{terhal}.

In all proofs of the threshold theorem, including ours, certain assumptions are essential. The noise strength must be bounded above by a sufficiently small constant that is independent of the size of the computation. To flush from the computer the entropy arising from noise, we must have an inexhaustible supply of fresh ancilla qubits (or the ability to refresh and reuse our ancilla qubits an indefinite number of times) \cite{aharonov-ancilla}. To control memory noise that afflicts all parts of the computer simultaneously, we must be able to execute quantum gates in parallel. To combat errors effectively, we must assume a noise model in which faults that act collectively on many qubits are highly suppressed.

For our explicit gadget constructions in Sec.~\ref{sec:explicit}, and in the formulation of Theorems 1--6, we have made additional assumptions that are not necessary for the existence of an accuracy threshold, but are helpful for simplifying the analysis and improving the numerical value of the threshold. We have assumed that a qubit can be measured as quickly as a quantum gate can be executed, and we have assumed that the classical processing of measurement outcomes can be done instantaneously and flawlessly. We have assumed that two-qubit quantum gates can be executed with a fixed accuracy on any pair of qubits, irrespective of their spatial proximity. And we have assumed a noise model that excludes ``leakage errors'' in which qubits become inaccessible and must be replaced. It would be useful to repeat the threshold calculation with some or all of these assumptions relaxed. 

Under these same assumptions, it is also possible to derive a quantum accuracy threshold using two-dimensional topological codes \cite{kitaev-toric,kitaev-topo}. The lower bound on the threshold for quantum memory found in \cite{topological-jp}, based on topological codes, is comparable to the memory threshold that can be established for the concatenated [[7,1,3]] code using the method in Sec.~\ref{sec:explicit}. However, an optimized estimate using topological codes of the computation threshold has not been attempted, as the purification protocol for the ${\cal C}_3$ ancilla is rather complicated in this case \cite{bravyi}. It has also been shown \cite{topological-ahn} that an especially favorable overhead scaling can be established for fault-tolerant quantum computing using higher-dimensional topological codes.

Our detailed threshold estimate in Sec.~\ref{sec:explicit} was carried out for the [[7,1,3]] code. It would be interesting to analyze other codes that might conceivably yield a higher threshold, such as the [[23,1,7]] Golay code \cite{Steane02}, or polynomial codes \cite{dorit-daniel}.  It might also be useful to improve the threshold result for general local noise, by pursuing the program sketched in Sec.~\ref{subsec:decoherence-measurement}.

Another looming challenge is to put Knill's recent threshold estimates \cite{knill_detect} on a rigorous footing. Knill uses a more complex simulation method, in which concatenated {\em error-detecting} codes are used to prepare certain encoded ancilla states. Based on numerical studies, he reports a threshold value three orders of magnitude higher (!) than the value established by our Theorem 3. We hope that the methods we have introduced in this paper can be adapted to the analysis of Knill's scheme or related schemes, leading to tighter rigorous lower bounds on the quantum accuracy threshold.

\nonumsection{Acknowledgments}
\noindent
This work has been supported in part by: the Department of Energy under Grant No. DE-FG03-92-ER40701,  the National Science Foundation under Grant No. EIA-0086038, the Army Research Office under Grant No. DAAD19-00-1-0374 and Grant No. W911NF-05-1-0294,  NSERC of Canada, and CIAR. We thank Barbara Terhal for pointing out an error in an earlier version of this paper, and for helpful correspondence concerning non-Markovian noise. We also gratefully acknowledge useful discussions with Alexei Kitaev and Manny Knill. Some of our circuit diagrams were drawn using the Q-circuit macro package written by Steve Flammia and Bryan Eastin.

\nonumsection{References}

\end{document}